\documentclass[prd,eqsecnum,amsmath,amssymb,nofootinbib,11pt,superscriptaddress]{revtex4-1}
\usepackage{dcolumn}
\usepackage{bm}
\usepackage{xcolor}
\usepackage[applemac]{inputenc}
\usepackage[spanish,english]{babel}
\usepackage{amsmath,amssymb,amsfonts,latexsym,cancel}
\usepackage{color}
\usepackage{soul}
\usepackage{ulem}
\usepackage{slashed}
\usepackage{braket}
\usepackage{amsthm,hyperref,bbm}
\usepackage[mathscr]{euscript}
\usepackage{physics}
\usepackage{mathrsfs}
\usepackage{pgfplots}
\usepackage{graphicx, epsfig}
\usepackage[english]{babel}
\usepackage{float}
\usepackage{multirow}
\usepackage{xfrac}
\allowdisplaybreaks

\newcommand{\la}{\langle}
\newcommand{\ra}{\rangle}

\newcommand{\be}{\begin{equation}}
\newcommand{\ee}{\end{equation}}
\newcommand{\bea}{\begin{eqnarray}}
\newcommand{\eea}{\end{eqnarray}}
\newcommand{\bes}{\begin{subequations}}
\newcommand{\ees}{\end{subequations}}

\begin{document}

\title{\bf Pair production due to an electric field in 1+1 dimensions and the validity of the semiclassical approximation}

\author{Silvia Pla}\email{silvia.pla@uv.es}
\affiliation{Departamento de Fisica Teorica and IFIC, Centro Mixto Universidad de Valencia-CSIC. Facultad de Fisica, Universidad de Valencia, Burjassot-46100, Valencia, Spain.}
\affiliation{Department of Physics, Wake Forest University, Winston-Salem, North Carolina, 27109, USA}

\author{Ian M. Newsome}\email{newsim18@wfu.edu}
\affiliation{Department of Physics, Wake Forest University, Winston-Salem, North Carolina, 27109, USA}

\author{Robert S. Link}\email{linkrs15@wfu.edu}
\affiliation{Department of Physics, Wake Forest University, Winston-Salem, North Carolina, 27109, USA}

\author{Paul R. Anderson}\email{anderson@wfu.edu}
\affiliation{Department of Physics, Wake Forest University, Winston-Salem, North Carolina 27109, USA}

\author{Jose Navarro-Salas.}\email{jnavarro@ific.uv.es}
\affiliation{Departamento de Fisica Teorica and IFIC, Centro Mixto Universidad de Valencia-CSIC. Facultad de Fisica, Universidad de Valencia, Burjassot-46100, Valencia, Spain.}

%\affiliation[a]{Departamento de Fisica Teorica and IFIC, Centro Mixto Universidad de Valencia-CSIC. Facultad de Fisica, Universidad de Valencia, Burjassot-46100, Valencia, Spain.}
%\affiliation[b]{Department of Physics, Wake Forest University, Winston-Salem, North Carolina, 27109, USA}

\begin{abstract}

Solutions to the backreaction equation in 1+1-dimensional semiclassical electrodynamics are obtained and analyzed when considering a time-varying homogeneous electric field initially generated by a classical electric current, coupled to either a quantized scalar field or a quantized spin-$\frac{1}{2}$ field. Particle production by way of the Schwinger effect %will 
leads to backreaction effects that modulate the electric field strength. Details of the particle production process are investigated along with the transfer of energy between the electric field and the particles. The validity of the semiclassical approximation %for some of the solutions 
is also investigated using a criterion previously implemented for chaotic inflation and, in an earlier form, semiclassical gravity. The criterion states that the semiclassical approximation will break down if any linearized gauge-invariant quantity constructed from solutions to the linear response equation, with finite nonsingular data, grows rapidly for some period of time. Approximations to homogeneous solutions of the linear response equation are %considered 
computed and it is found that the criterion is violated when the maximum value, $E_{\rm max}$, obtained by the electric field is of the order of the critical scale for the Schwinger effect,  $E_{\rm max} \sim E_{\rm crit}\equiv m^2/q$, where $m$ is the
mass of the quantized field and $q$ is its electric charge. For these approximate solutions the criterion appears to be satisfied in the extreme limits  $\frac{qE_{\rm max}}{m^2} \ll 1$ and $\frac{qE_{\rm max}}{m^2} \gg 1$.
\end{abstract}

%%%%%%%%%%%%%%%%%%%%%%%%%%%%%%%%%%%%%%%%%%%%%%%%%%%%%%%%%%%%%%%%%%%%%%%%%%%%%%%%%%%%%%%%%%
\date{\today}
\maketitle

\section{Introduction}
The semiclassical approximation has been commonly used among a wide variety of physical scenarios where a quantized field on a classical background is investigated, with interesting phenomena emerging from such considerations including the decay of an electric field by the Schwinger effect \cite{schwinger}, particle creation in an expanding universe \cite{parker66}, and black hole evaporation via  the Hawking effect   \cite{hawking} (see also Refs.~\cite{parker-toms, birrell-davies} and references therein).
Consider for instance quantum electrodynamics, described in terms of an electromagnetic potential $A_\mu$ and a Dirac field $\psi$, with classical action $S[A_\mu, \bar \psi, \psi]$. The semiclassical theory can be formally described using the concept of the effective action $\Gamma[A_\mu]$, obtained by functional integration of the matter degrees of freedom \cite{qftbook}
\be
    \label{eff}\exp {i\Gamma[A_\mu]} =  \int D\bar \psi D\psi \exp {i S[A_\mu, \bar \psi, \psi]} \quad .
\ee
Within this framework the (semiclassical) Maxwell  field equations take  the form
\be
    \label{semiclassicalM} \partial_\mu F^{\mu\nu}  = q\langle 0_{A}| \bar \psi \gamma^\mu \psi | 0_{A} \rangle \quad ,
\ee
and replace the proper Maxwell equations of the full quantized theory in the Schwinger-Dyson form $\partial_\mu \langle F^{\mu\nu} \rangle  = q\langle \bar \psi \gamma^\mu \psi \rangle$. In Eqs.~(\ref{eff}) and (\ref{semiclassicalM}) the electromagnetic field  is treated as a purely classical entity. Moreover, the right-hand side of Eq.~(\ref{semiclassicalM}) is implicitly a function of $A_\mu$ in the sense that
the assumed vacuum depends on $A_\mu$. This is so because the modes of the charged Dirac field, defining the appropriate vacuum $|0_A \rangle$, satisfy equations involving the background field $A_\mu$. This semiclassical approach is usually regarded as a truncated and effective version of the fully quantized theory, with a limited range of validity. 

%Despite the expected limitations of the semiclassical approximation, it can nicely reproduce fundamental  aspects of exact quantum electrodynamics, such as the chiral anomaly where using the semiclassical approximation one obtains
%\be  \partial_\mu \langle 0_{A}| \bar \psi \gamma^\mu \gamma^5\psi | 0_{A} \rangle = 2im\langle 0_{A}| \bar \psi \gamma^5\psi | 0_{A} \rangle -\frac{\alpha}{2\pi} F_{\mu\nu}  {^{\star}F}^{\mu\nu} \quad . \ee The only change in this equation for the fully quantized theory is the replacement of $F_{\mu\nu}  {^{\star}F}^{\mu\nu}$ with $\la F_{\mu\nu}  {^{\star}F}^{\mu\nu} \ra$.

One advantage of the semiclassical viewpoint is that it provides a clear description of the spontaneous particle creation phenomena. The nonzero imaginary part of the effective action $\Gamma[A_\mu]$ indicates the quantum instability of the vacuum $| 0_{A} \rangle$ and the corresponding pair creation process \cite{schwinger}. This phenomena can be better understood in the canonical language: a positive-frequency solution of the Dirac equation $(i\slashed {D} - m)\psi=0$ at early times will evolve into a superposition of positive- and negative-frequency solutions at late times (this was first described for a gravitational background \cite{parker66}).  The semiclassical approach encapsulates in a  clear  way this very important  effect.
The original calculation by Schwinger~\cite{schwinger} involved a background field calculation in which the electric field $E$ is constant in both space and time.  A particle production rate was obtained.  The dependence on the coupling constant $q$ displayed an essential singularity $e^{-m^2/qE}$, showing the nonperturbative nature of the Schwinger effect. %{\color{purple} Moreover, it also points out a critical scale for the electric field $E_{crit}= m^2/q$  to have  significant particle production ~\cite{schwinger}.} 
The damping of the electric field can be deduced from this particle production rate. 
The real part of the (Heisenberg-Euler) effective action can also account for perturbative effects,  such as  light-by-light scattering, in agreement with the exact one-loop calculation  in the limit of low-frequency light, or the running of the effective coupling constant. 

Subsequently the semiclassical backreaction equation was solved for an electromagnetic field coupled to a massive scalar field or a massive spin-$\frac{1}{2}$ field in 1+1 dimensions (D)~\cite{Kluger91,Kluger92,Kluger93} and 3+1D~\cite{Kluger93,Tanji,stat-FT}.  The electric field was assumed to be homogeneous in space, but was allowed to vary in time in response to the electric current that occurs when the produced particles are accelerated by the electric field.  It was found the counter-electric field produced by this current initially starts to negate the original background electric field. Eventually the background field is completely canceled, but by this time there is a significant electric current due to particle production and the result is that the particles keep moving which generates an electric field in the opposite direction. The process continues and the particles end up undergoing plasma oscillations with an overall electric field oscillation in time. Similar studies have also been done by solving the Vlasov equation with a source term to account for particle production~\cite{Kluger91,Kluger92,Kluger93,Bloch}, using lattice simulations~\cite{lattice-1,lattice-2}, and classical statistical field theory techniques~\cite{stat-FT}.

In this paper we obtain and further study solutions to the semiclassical backreaction equation in 1+1D for both scalar and spin-$\frac{1}{2}$ fields coupled to an electromagnetic field initially generated by a homogeneous, classical current.  %In each case considered the electric field is generated by a classical current, is homogeneous, and is initially zero as it would be in a typical laboratory setting.  
We have  two primary goals.  The first is to study the details of the particle production process when backreaction effects have been taken into account, including also the transfer of energy between the electric field and the created particles. The second goal is to estimate the importance of certain types of quantum fluctuations and use the results to assess the validity of the semiclassical approximation.

We study three classical current profiles %that are used 
which generate an electric field that is initially zero. The first is similar to the previous cases in that the current is proportional to a delta function potential and the electric field goes from zero to a constant value instantaneously. A second profile involves a sudden turn on of the classical current but a gradual turn on of the electric field.  The third profile is that of the Sauter pulse~\cite{Sauter} in which the current is in the form of a smooth pulse that has a significant value only for a finite period of time.  For the Sauter pulse the turn on and, if quantum effects are ignored, the turn off of both the current and the electric field are very gradual.  %Unlike the previously studied cases in which the electric field is nonzero initially, 
In all three cases there is a well-defined vacuum state for the quantum fields since the electric field is initially zero. 
%For each of the three profiles 
The semiclassical backreaction equation is solved numerically both in the case of semiclassical scalar and spinor electrodynamics.  %, which includes a massive charged scalar field, and in the case of semiclassical quantum electrodynamics.%, which includes a charged spin $\frac{1}{2}$ field. 
To our knowledge the semiclassical backreaction equations have not been %previously solved 
generically studied for the second and third classical current profiles. The first one has been considered in Refs.~\cite{Tanji,stat-FT,lattice-1,lattice-2,Bloch}.

The particle production process  for individual modes of the quantum field has previously been studied in background field calculations where the electric field is either constant~\cite{and-mot-I,dunne-part-prod} or is gradually turned on and then off~\cite{and-mot-san}.  It was found that a single particle creation event occurs for many modes  when the electric field is either constant or approximately constant. Here, we consider particle production when backreaction effects are taken into account.  Because of the plasma oscillations, there is a richer evolution for some modes that involves multiple particle creation events and can also involve particle destruction events.  We do this for individual modes for the delta function classical current profile. 

For completeness, and to give better insight into the particle creation process,  
we also compute the total number of particles produced for all three profiles and  
the energy density of the produced particles  for the delta function current profile.   The energy density of the particles is compared with the energy density of the electric field.  Similar calculations have  been done previously in $1+1$D using lattice simulations 
~\cite{lattice-1} and in $3+1$D using canonical quantization~\cite{Tanji} and classical statistical field theory techniques~\cite{stat-FT}.
  %Similar analysis have been done using lattice field theory techniques \cite{lattice}, 

We compute the energy density of the quantum field using the continuous adiabatic regularization prescription, obtaining compatible results. The agreement between both approaches for Dirac massless fermions can be easily understood since the full QED$_2$ model is integrable \cite{Schwingermass} and particle production can be well described within the semiclassical framework. The presence of a nonzero mass breaks integrability and hence one could expect it to also break the accuracy of the semiclassical picture.

The validity of the semiclassical approximation is studied here by estimating the importance of some of the quantum fluctuations.  The semiclassical approximation breaks down if quantum fluctuations are too large. We use a criterion for the validity of the semiclassical approximation that has been previously applied to the process of preheating in models of chaotic inflation~\cite{validitypreheating}.  An earlier version of the criterion has also been used to study the validity of the semiclassical approximation for free scalar fields in flat space when the fields are in the Minkowski vacuum state~\cite{validitygravity} and for the conformally invariant scalar field in the Bunch-Davies state in de Sitter space in the usual spatially flat cosmological coordinates~\cite{validitydeSitter}.  To our knowledge no similar study of the validity of the semiclassical approximation has been done previously for scalar electrodynamics or quantum electrodynamics when particle creation occurs due to the presence of a strong electric field. 

The method we  use to study the validity of the semiclassical approximation involves an analysis of solutions to the linear response equation which can be obtained by perturbing the semiclassical backreaction equation.  In general, the linear response equation obtained in this way is an integro-differential equation which involves an integral over the retarded two-point correlation function for the source term in the semiclassical backreaction equation.  In this case, that is the two-point correlation function for the electric current.  While the general form is known, the specific forms for the case of a homogeneous electric field in 1+1D coupled to either a massive scalar field or a spin-$\frac{1}{2}$ field has not previously been derived.  We do so in the Appendix for both of these cases.  

Although the linear response equation can be solved directly, there is a simpler method which can be used to obtain an approximate solution which should be valid at early times if the exact solution is relatively small.  The method involves computing the difference $\Delta E$ between two solutions to the semiclassical backreaction equation which have similar starting values at a given time. This method was used to investigate the validity of the semiclassical approximation during the preheating phase of chaotic inflation in Ref.~\cite{validitypreheating}.  It works for the homogeneous solutions to the linear response equation that we consider here.

The paper is organized as follows. In Sec.~\ref{sec:model} brief reviews are given of the quantization of complex charged scalar and spin-$\frac{1}{2}$ fields in electrodynamics.  The semiclassical backreaction equations are also discussed  along with the renormalization techniques used. In Sec.~\ref{sec:energy} the details of the particle production process are investigated for the case of a classical current profile proportional to a delta function.  Also discussed is the transfer of energy between the electric field and the created particles. 
The criterion for the validity of the semiclassical approximation that we use is discussed in Sec.~\ref{sec:criterion} where both the general form and the specific form of the linear response equation are displayed for the separate cases of a scalar field and spin-$\frac{1}{2}$ field coupled to the electromagnetic field.
In Sec.~\ref{sec:numerical}
some of the results of numerical calculations we have made related to the validity of the semiclassical approximation are presented and discussed.  A summary of our results and some conclusions are given in Sec.~\ref{sec:conclusions}.  The Appendix contains derivations of the specific contributions to the linear response equations from the current-current commutators when scalar fields and spin $\frac{1}{2}$ fields are coupled to the electromagnetic field.

%%%%%%%%%%%%%%%%%%%%%%%%%%%%%%%%%%%%%%%%%%%%%%%%%%%%%%%%%%%%%%%%%%%%%%%%%%%%%%%%%%%%%%%%%%

\newpage

\section{Quantization and renormalization of complex scalar and \\ spin-$\frac{1}{2}$ fields}
\label{sec:model}
In this section we will briefly describe the models under consideration: a quantized complex scalar field and a quantized Dirac field, both interacting with a background electromagnetic field generated by a prescribed classical source.  For the two systems under investigation, we restrict our analysis to a $1+1$D Minkowski space and assume that the background electric field is spatially homogeneous so that
$E=E(t)$ in a given reference frame. We use units such that $\hbar = c = 1$ and our convention for the metric signature is  $(-,+)$.

\subsection{Scalar field} \label{sec:model-scalar}

The classical action representing a %quantized
scalar field $\phi(t,x)$ coupled to a background electromagnetic field is
\be
S=\int d^{2}x\bigg[-\frac{1}{4}F_{\mu \nu}F^{\mu \nu}+ A_\mu J_C^\mu -D_{\mu}\phi^{\dagger}D^{\mu}\phi-m^{2}\phi^{\dagger}\phi\bigg] \label{S} \quad ,
\ee
where $F_{\mu \nu}=\partial_\mu A_\nu - \partial_\nu A_\mu$ is the electromagnetic field-strength tensor, the mass of scalar field excitations is given by $m$, and $D_{\mu}=\partial_{\mu}-iqA_{\mu}$ is the gauge-covariant derivative required to make the action gauge invariant. $J_C^\mu$ is a classical and conserved external source. Variation of Eq.  \eqref{S} with respect to the vector potential yields the  classical Maxwell equations
\be
-\Box A^{\mu}+ \partial^\mu \partial_\nu A^\nu  = J^\mu_C+ J^{\mu}_Q \label{Max} \quad ,
\ee
where the source term $J^\mu_Q$ induced by the scalar field is given by
\be
J^{\mu}_Q =  \eta^{\mu \nu}\bigg[-iq\bigg(\phi^{\dagger}\partial_{\nu}\phi-\left(\partial_{\nu}\phi^{\dagger}\right)\phi\bigg)-2q^{2}A_{\nu}\left(\phi^{\dagger}\phi\right)\bigg] \label{Jsource} \quad .
\ee
The field equation for $\phi(t,x)$ is $
\left(D^{\mu}D_{\mu}-m^{2}\right)\phi(t,x)=0$. % \label{mode} \quad .
%Note that \eqref{Max}, \eqref{Jsource}, and \eqref{mode} are manifestly gauge invariant.
We choose the   Lorentz gauge $\partial_{\mu}A^{\mu}=0$, and fix the vector potential in the convenient form
\be
    A^{\mu}=(0,A(t)) \;, \label{gauge}
\ee
which therefore yields $F_{01} = \partial_0 A_1 = \dot A = -E$.
%\be F_{01} = \partial_0 A_1 = \dot A = -E \;.  \ee
The field equation reduces to
\be
    \bigg[-\partial_{t}^{2}+\partial_{x}^{2}-2iqA(t)\partial_{x}-q^{2}A^{2}(t)-m^{2}\bigg]\phi(t,x)=0 \quad . \label{mode3}
\ee
Quantizing the scalar field and expanding it in terms of modes yields
\begin{equation}
    \phi(t,x)=\frac{1}{\sqrt{2\pi}}\int_{-\infty}^\infty dk \bigg[a_{k}U_{k}(t,x)+b_{k}^{\dagger}V_{k}(t,x)\bigg] \ . \label{phi1}
\end{equation}
where $a_k,a_{k}^{\dagger},b_{k}, \textnormal{and} \, b^\dagger_k$ are the usual creation and annihilation operators obeying the commutation relations $[a_{k},a_{k^{'}}^{\dagger}]=[b_{k},b_{k^{'}}^{\dagger}]=\delta (k-k^{'})$.
Due to spatial homogeneity we can write the modes $U_{k}(t,x)$ and $V_{k}(t,x)$ in the convenient form $U_{k}(t,x)=f_{k}(t)e^{ikx}$, $V_{k}(t,x)=f_{-k}^{*}(t)e^{-ikx}$,
%\begin{equation}U_{k}(t,x)=f_{k}(t)e^{ikx} \qquad , \qquad V_{k}(t,x)=f_{-k}^{*}(t)e^{-ikx} \ , \label{func}\end{equation}
where $f_k(t)$ satisfies the ordinary differential equation
\begin{equation}\label{fmode}
   \ddot f_{k}(t) + \bigg[ \left(k-q A\right)^{2}+m^{2} \bigg]f_{k}(t) = 0
\end{equation}
and is normalized using the Wronskian condition
\be
   f_{k}\dot f_{k}^{*} - f_{k}^{*}\dot f_{k} = i.\label{scalar-wronskian}
 \ee
This allows us to recast the scalar field mode decomposition as
\begin{equation}
    \phi(t,x)=\frac{1}{\sqrt{2\pi}}\int dk \bigg[a_{k}f_{k}(t)+b_{-k}^{\dagger}f_{k}^{*}(t)\bigg]e^{ikx} \label{phi3} \; .
\end{equation}

%%%%%%%%%%%%%%

\subsection{Spin-$\frac{1}{2}$ field}
\label{sec:spin12}
The classical action representing a spin-$\frac{1}{2}$ field $\psi(t,x)$ coupled to a background electric field is
\be
S=\int d^{2}x \, \bigg[-\frac{1}{4}F_{\mu \nu}F^{\mu \nu} + A_\mu J_C^\mu +i\Bar{\psi}\gamma^{\mu}D_{\mu}\psi-m\Bar{\psi}\psi\bigg] \label{SS} \quad .
\ee
where $\Bar{\psi}=\psi^{\dagger}\gamma^{0}$, with $F^{\mu \nu}$ and $D_{\mu}$ defined the same as for the scalar field case. The Dirac matrices $\gamma^{\mu}$ satisfy the anticommutation relations $\{\gamma^{\mu},\gamma^{\nu}\}=-2\eta^{\mu \nu}$.
As for the scalar field, $J_C$ is an external classical source.
The Maxwell equations include the source term induced by the field $\psi$
\be
J^{\mu}_Q = q \, \Bar{\psi}\, \gamma^{\mu} \, \psi \label{J2} \quad .
\ee
The field equation for $\psi(t,x)$ is the Dirac equation $\big(i \, \gamma^{\mu}D_{\mu}-m\, \big)\psi(t,x) = 0$.
%\be\big(i \, \gamma^{\mu}D_{\mu}-m\, \big)\psi(x) = 0 \label{mode2} \quad .\ee
%Both~\eqref{J2} and~\eqref{mode2} are also gauge invariant.
With the gauge choice~\eqref{gauge} the explicit form of the Dirac equation is
\be
\bigg[i \, \gamma^{t}\partial_{t}+i \, \gamma^{x}\partial_{x}+q \, \gamma^{x}A(t)-m \bigg]\psi(t,x) =0 \quad . \label{modefermi}
\ee
Quantizing the spin-$\frac{1}{2}$ field and expanding it in terms of modes yields
\bea
\psi(t,x)&=&\int_{-\infty}^\infty dk \, \bigg[B_{k}u_{k}(t,x)+D_{k}^{\dagger}v_{k}(t,x)\bigg] \;, \label{psi1}
\eea
where here $B_{k},B_{k}^{\dagger},D_{k}, \, \textnormal{and} \, D_{k}^{\dagger}$ are the usual creation and annihilation operators obeying the anticommutation relations $\{B_{k}, B^{\dagger}_{k^{'}}\} = \{D_{k}, D^{\dagger}_{k^{'}}\} = \delta(k-k^{'})$. Using the formalism introduced in Refs.~\cite{FN, BNP}, we can construct two independent spinor solutions as follows
\be
u_{k}(t,x)=\frac{e^{ikx}}{\sqrt{2\pi}}
{\small \left({\begin{array}{c}
h_{k}^{I}(t) \\
-h_{k}^{II}(t)
\end{array}}\right)}
\quad , \quad
v_{k}(t,x)=\frac{e^{-ikx}}{\sqrt{2\pi}}{\small
\left({\begin{array}{c}
h_{-k}^{II *}(t) \\
h_{-k}^{I *}(t)
\end{array}}\right)}. \label{v}
\ee
Utilizing the Weyl representation of the Dirac matrices $\gamma^{\mu}$
\be \label{weylM}
\gamma^{t} =
\left( {\begin{array}{cc}
 0 & 1  \\
 1& 0  \\
 \end{array} } \right) \quad , \quad
\gamma^{x} =
 \left( {\begin{array}{cc}
 0 & 1  \\
 -1& 0  \\
 \end{array} } \right) \quad , \quad
\gamma^{5} =\gamma^{t}\gamma^{x}=
\left( {\begin{array}{cc}
 -1 & 0  \\
 0& 1 \\
 \end{array} } \right)
\ , \ee
one can show that $h_{k}^{I}(t)$ and  $h_{k}^{II}(t)$ are solutions of the mode equations
\bes \bea
\Dot{h}_{k}^{I}-i\left(k-qA\right)h_{k}^{I}-i\,m\,h_{k}^{II} &=& 0\, ,\label{modeh1} \\
\Dot{h}_{k}^{II}+i\left(k-qA\right)h_{k}^{II}-i\,m\,h_{k}^{I} &=& 0.\label{modeh2}
 \eea \ees
The normalization condition $|h_k^{I}|^{2}+|h_k^{II}|^{2}=1$ ensures that the standard anticommutation relations between the creation and annihilation operators are satisfied.

%%%%%%%%%%%%%%%%%%%%%%%%%%%%%%%%%%%%%%%%%%%%%%%%%%%%%%%%%%%%%%%%%%%%%%%%%%%%%%%%%%%%%%%%%%

\subsection{Semiclassical backreaction equation and renormalization}%THE SEMICLASSICAL BACKREACTION EQUATION AND RENORMALIZATION}
\label{sec:semiclass}

A simple way to obtain the semiclassical backreaction equation is to replace $J^\mu_Q$ in Eq.~\eqref{Max} with $\la J^\mu_Q \ra$ and then use Eq.~\eqref{gauge} and either~\eqref{phi3} or Eq.~\eqref{psi1}, with the result
\begin{equation}
    \frac{d^{2}}{dt^{2}} A(t) =-\frac{d }{d t}E(t)=J_{C} + \langle J_{Q} \rangle \quad . \label{sb}
\end{equation}
Here we have simplified the notation by omitting the superscript $x$ on $J_C$ and $J_Q$ since in this case the $t$ component of these vectors vanishes.  When particle production occurs the background electric field accelerates the produced particles creating a current which then reacts back on this electric field. In the semiclassical approximation this current is $\langle J_{Q} \rangle$.  The net electric field $E(t)$ is then generated by both the classical current $J_{C}$ and the current from the created particles $\langle J_{Q} \rangle$.

%In this section 
We now obtain the generic forms of the finite, physical expression of $\langle J_Q \rangle$
for both scalar and fermion fields.
This is nontrivial since the formal expressions for the current are quadratic in the quantized fields. %This produces one of the well-know ultraviolet divergences that plague every quantum field theory.  
Here we will explain how the ultraviolet divergences can be tamed by using the so-called adiabatic regularization method. The method was originally proposed to obtain finite expectation values for the stress-energy tensors of scalar fields in expanding universes \cite{parker-fulling, Birrell78, Anderson-Parker} (see also Refs.~\cite{parker-toms, birrell-davies} for
scalar fields and Refs.~\cite{rio1, rio2, rio3, ghosh1, ghosh2, BFNV}
for fermion fields).  The adiabatic method has been
adapted to treat spatially homogeneous electric backgrounds in Refs.~\cite{Cooper-Mottola89, CMRA, Kluger92}, 
%The adiabatic regularization scheme for semiclassical electrodynamics was further 
 and it has been improved to make it consistent with gravity in 
%generalized to cosmological spacetimes in
Refs.~\cite{FN, FNP, FNP2} 
and connected to the DeWitt-Schwinger proper-time expansion in Ref.~\cite{BNP20}.  
Here we follow the procedure 
proposed in Ref.~\cite{FN, FNP, FNP2}. 
%For completenessa brief account is given below.

\subsubsection{Scalar field}
\label{sec:scalar-2}

It is useful to symmetrize the current operator for the scalar field %\footnote{Symmetrization offers a theoretically favored way to yield particle production from both particles and antiparticles upon evaluation in the vacuum state. Also, the benefit of not having to renormalize certain terms is afforded by the symmetrization process, a case being the $\mu=0$ component of the current operator. Antisymmetrization is used for the current operator in the spin $\frac{1}{2}$ case for the same reason.} 
with the result
\be
    J^{\mu}_Q =  \frac{1}{2}\eta^{\mu \nu}\bigg[-iq\bigg(\phi^{\dagger}\partial_{\nu}\phi-\left(\partial_{\nu}\phi^{\dagger}\right)\phi\bigg)+iq\bigg(\phi\partial_{\nu}\phi^{\dagger}-\left(\partial_{\nu}\phi\right)\phi^{\dagger}\bigg)-2q^{2}A_{\nu}\bigg(\phi^{\dagger}\phi+\phi \phi^{\dagger}\bigg)\bigg] \quad . \label{jsymm1}
\ee
Using Eq.~\eqref{gauge} and evaluating Eq.~\eqref{jsymm1} in the vacuum state gives for the nontrivial spatial component
\be
    \la J_{Q} \ra = \frac{q}{\pi} \int_{-\infty}^{\infty} dk \, \bigg(k-qA(t)\bigg)|f_{k}(t)|^{2} \quad . \label{Jq}
\ee
Note that the $\mu = 0$ component of the current is identically zero, meaning that no net charge is created. The integral \eqref{Jq} contains ultraviolet divergences and hence must be renormalized. Since the external electric field is assumed to be spatially homogeneous, it is  especially convenient to use an extension of the adiabatic regularization method. For scalar fields the procedure is based on the standard WKB-type expansion of the field modes. In our case one writes the ansatz
\bea
f_{k}(t)=\frac{1}{\sqrt{2\Omega_{k}(t)}}e^{-i\int^t \Omega_{k}(t')dt'} \quad , \label{WKB-ansatz-scalar}
\eea
where $\Omega_k$ is expanded in powers of derivatives of $A(t)$, as $\Omega_{k}= \omega^{(0)} + \omega^{(1)}+ \omega^{(2)} + \cdots$. The leading term $\omega^{(0)}$  is assumed to be of  zeroth adiabatic order, while $\omega^{(1)}$ is of adiabatic order one, etc. The choice of the leading-order term $\omega^{(0)}$ determines univocally the subsequent orders. A natural possibility \cite{Cooper-Mottola89}  is  $\omega^{(0)}=\sqrt{(k - qA)^2+m^2}$, which assumes that $A(t)$ should be considered as a variable of adiabatic order %$0$,
zero, $\dot A$ of adiabatic order %$1$
one, etc. %This parallels the conventional choice made in the original application of the method for scalar fields existing in expanding universes. The scale factor $a(t)$ is of adiabatic order $0$ and $\dot a$ of adiabatic order $1$, etc.

However, $A(t)$ is intrinsically a dimensionful  quantity and this suggests an alternative possibility. As proposed in Refs.~\cite{FN, FNP}, one can also choose   $\omega^{(0)}\equiv \omega= \sqrt{k^2+m^2}$. This choice  is attached to the adiabatic assignment of %$1$ 
one for $A(t)$, while $\dot A$ is considered to be of adiabatic order %$2$,
two, etc.  This second possibility is actually the only consistent possibility in the presence of both electromagnetic and gravitational backgrounds. % Furthermore,  the adiabatic expansion induced for the Feynman two-point function  is equivalent to the DeWitt-Schwinger proper time expansion only if $A(t)$ is assumed of adiabatic order $1$~\cite{BNP20}.
We then obtain
%The procedure to subtract off the required zeroth and first adiabatic orders, assuming the gauge field $A(t)$ is of adiabatic order 1 and  $\omega \equiv \sqrt{k^{2}+m^{2}}$, yields the following renormalized expression for the current \cite{FNP}
\be
    \la J_{Q} \ra_{\textnormal{ren}} =  \frac{q}{\pi} \int_{-\infty}^{\infty} dk \, \bigg[ \bigg(k-qA(t)\bigg)|f_{k}(t)|^{2} -\frac{k}{2\omega}+ \frac{q\, m^{2}}{2\omega^{3}}A(t)\bigg] \quad . \label{Jrenorm}
\ee
Similarly, one can also determine the renormalized energy density $\la T_{00} \ra=\la \rho \ra$ induced by the quantized field%. One obtains %\cite{FNP}
\bea
\langle \rho \rangle_{\rm ren} = \frac{1}{2\pi} \int_{-\infty}^{\infty}  dk \left[|\dot{f}_k(t)|^2+ \bigg(m^2+\big(k-q A(t)\big)^2 \bigg)|f_k(t)|^2- \omega +\frac{ k q}{\omega}A(t)-\frac{m^2 q^2}{2\omega^3} A^2(t)\right] \quad . \label{engsc} \nonumber \\
\eea

\subsubsection{Spin-$\frac{1}{2}$ field}
\label{sec:spin12-2}

For the spin-$\frac{1}{2}$ field the appropriate antisymmetrized term is \cite{parker-toms}
\be J^\mu_Q = \frac{q}{2} [ \bar \psi, \gamma^\mu \psi ]. \label{JsymDirac} \ee 
The expression for $\mu=0$ corresponds to the induced electric charge and, as expected, $\langle J^0_Q\rangle$ is identically zero, i.e. no net charge is created. 
The renormalized expression for the spatial component of  the spin-$\frac{1}{2}$ current evaluated in the vacuum state is \cite{FN, FNP2}
\be
\la J_{Q} \ra_{\textnormal{ren}} = \frac{q}{2\pi}\int_{-\infty}^{\infty} dk \, \bigg[ |h_{k}^{I}(t)|^{2}-|h_{k}^{II}(t)|^{2}+\frac{k}{\omega}-\frac{q\, m^{2}}{\omega^{3}}A(t)\bigg] \label{Jrenorm2}
\quad .
\ee
It is particularly interesting to consider the massless case where the first two terms in the above integral  cancel and the expression for the current becomes
\be
\la J_{Q} \ra_{\textnormal{ren}} = -\frac{q^2}{\pi}A(t) \label{Jmassless}
\quad .
\ee
This result is consistent with the two-dimensional axial anomaly
\be
    \partial_\mu \la J^{\mu}_{5} \ra_{\textnormal{ren}} = \frac{q}{\pi} \epsilon^{\mu\nu}F_{\mu\nu} \quad ,
\ee
where $J^\mu_5 = \bar \psi \gamma^\mu \gamma^5 \psi $ and
$J^\mu_Q =- q\epsilon^{\mu\nu}J_{\nu 5}$. Furthermore, the renormalized energy density is given by %It is also useful to evaluate the renormalized expression for the energy density \cite{FNP2}
\bea
    \langle \rho \rangle_{\rm ren}=\frac{1}{2\pi}\int_{-\infty}^{\infty}  dk \bigg[ i \left[h_k^{II}(t)\dot{h}_k^{II*}(t)+h_k^{I}(t)\dot{h}_k^{I*}(t)\right]+\omega -\frac{k q}{\omega}A(t)+\frac{m^2 q^2}{2\omega^3}A^2(t)\bigg] \quad . \label{fermden}
\eea

%%%%%%%%%%%%%%%%%%%%%%%%%%%%%%%%%%%%%%%%%%%%%%%%%%%%%%%%%%%%%%%%%%%%%%%%%%%%%%%%%%%%

\section{Particle production  and energy conservation in the semiclassical framework}
\label{sec:energy}

In this section we study both the details of the particle production process and the transfer of energy between the electric field and the produced particles for some solutions to the semiclassical backreaction equation for the delta function current profile mentioned in the Introduction.

The vacuum instability due to pair production was first realized by Heisenberg and Euler \cite{Heisenberg1936}, who predicted, on the basis of 
an effective action for a constant and homogeneous electromagnetic background, a pair production rate in an electric field of order $\sim q^2E^2e^{-\frac{m^2\pi}{qE}} $.  Schwinger, using the modern language of QED, computed the imaginary part of the one-loop effective action, also for a homogeneous and constant electric field, to evaluate the vacuum persistence amplitude (for a historical perspective, see Ref.~\cite{dunne}).
%from the formula \be |\langle out | in \rangle|^2 = \exp(-2\operatorname{Im} \Gamma(A_\mu))  \;. \ee The result, per unit volume and unit time, and in an arbitrary space dimension $d$, is \cite{schwinger} (see also \cite{kim})  \be \frac{2\operatorname{Im}\Gamma (A_\mu)}{VT}= \frac{2}{(2\pi)^d} \sum_{n=1}^{\infty} \Big(\frac{qE}{n}\Big)^{(d+1)/2} \exp \frac{-n\pi m^2}{q E} \ . \ee
From the exponential factor one notes immediately that the order of the critical scale for pair production can be defined to be   \be E_{\rm crit}\sim m^2/q \;. \label{def-Ecrit}\ee 
For some of the numerical work
described in the following sections we compare the classical electric
field to $E_{\rm crit}$ and for those comparisons we take $E_{\rm crit}$ to be
equal to $m^2/q$, %as this is also 
as is customary in the literature on the Schwinger effect.

While particle production in quantum field theory is a nonlocal process, for free quantum fields such as the ones we are considering, it is possible to define a time-dependent particle number that is based on the WKB approximation for the modes of the quantum field.  This has been done previously in the electric field case in Refs.~\cite{and-mot-I,dunne-part-prod,and-mot-san} where background electric fields were considered.  While there was some variation in the details depending on the order of the WKB approximation used, it was found for a constant electric field that when a given mode starts out in an adiabatic vacuum state, as the vector potential $A(t)$ increases in time, there is a particle creation event that occurs when $|k - q A|\sim m$ and lasts for a relatively short period of time.  After which the particle number for that mode approaches a constant value.
%To better understand the process of particle creation it is
%useful to define a time-dependent particle number~\cite{and-mot-I,dunne-part-prod,and-mot-san} (see also \cite{parker66, rio2} for $\langle N(t)\rangle$ in the cosmological setting). Recall the particle number is only unambiguously defined during intervals of time for which the potential $A(t)$ is time-independent, which occurs only if $E$ and its first time derivative are zero.  As an example, assume that the electric field is zero for times $t > t_f$.  Defining $\Omega_k(t) \equiv \sqrt{[k-q A(t)]^2+m^2} $,
%\be   \Omega_k(t) \equiv \sqrt{[k-q A(t)]^2+m^2} \;, \label{Omega-def}\ee
%the exact modes for a massive scalar field are for $t \ge t_f$
%\bea \label{f}
%f_k(t) &=& \alpha_k \frac{1}{\sqrt{2\Omega_k(t_f)}}e^{-i \int_{t_0}^t \Omega_k(t_f) dt_1}+\beta_k \frac{1}{\sqrt{2 \Omega_k(t_f)}}e^{+i \int_{t_0}^t \Omega_k(t_f) dt_1},\\ \nonumber \\
%\dot f_k(t) &=& -i \alpha_k \sqrt{\frac{\Omega_k(t_f)}{2}}e^{-i \int_{t_0}^t\Omega_k(t_f) dt_1} +i\beta_k  \sqrt{\frac{\Omega_k(t_f) }{2}}e^{+i \int_{t_0}^t \Omega_k(t_f) dt_1  } \ , \label{dotf}\eea
%where $t_0$ is an arbitrary constant and $\alpha_k$ and $\beta_k$ are the (time-independent) Bogolubov coefficients, obeying $|\alpha_k |^2 - |\beta_k|^2 = 1$.
%{\color{red} The  positive quantity $|\beta_k|^2$ determines univocally the number density of created particles.}
%\bea
%\langle N \rangle=2\int_{-\infty}^{\infty}\frac{dk}{2\pi}|\beta_k|^2 \ ,
%\eea
%where the factor of $2$ accounts for antiparticles.
Here we use the zeroth-order WKB approximation, used in Refs.  \cite{and-mot-I, dunne-part-prod, and-mot-san}: 
\bes 
\bea  
    g_k(t) &\equiv& \frac{1}{\sqrt{2\Omega_k(t)}}e^{-i \int_{t_0}^t \Omega_k(t_1) dt_1} \;, \label{wkb-0} \\ \nonumber \\
    \dot{g}_k(t) &\equiv& -i \sqrt{\frac{\Omega(t)}{2}} e^{-i \int_{t_0}^t \Omega_k(t_1) dt_1} \;, \label{wkb-dot-0} \\ \nonumber \\
    \Omega_k(t) &\equiv& \sqrt{[k-q A(t)]^2+m^2}   \;.  \label{Omega-def}
\eea 
\ees
Writing the exact mode functions as
\bes 
\bea 
    f_k(t) = \alpha_k(t) g_k(t) + \beta_k(t) g_k^{*}(t)  \quad    \label{f-alpha-beta-1} ,  \\ \nonumber \\
    \dot{f}_k(t) = \alpha_k(t) \dot{g}_k(t) + \beta_k(t) \dot{g}_k^{*}(t) \quad ,
    \label{f-alpha-beta} 
\eea
\ees
and substituting these expressions into Eq.~\eqref{fmode} converts the mode equation into two first-order coupled differential equations for $\alpha_k(t)$ and $\beta_k(t)$.  Substitution into the Wronskian condition~\eqref{scalar-wronskian} 
gives the condition $|\alpha_k(t)|^2 - |\beta_k(t)|^2 = 1$. 
%\be |\alpha_k(t)|^2 - |\beta_k(t)|^2 = 1.\ee  
Note that if the vector potential stops varying in time then the zeroth-order WKB approximation becomes exact and $\alpha_k$ and $\beta_k$ become Bogoliubov coefficients which relate the {\it in} vacuum state to the {\it out} vacuum state.  With this motivation one can define the time-dependent particle number for a given mode $k$ to be
\be 
    N_k(t) \equiv |\beta_k(t)|^2 \quad , \label{Nk-def} 
\ee
%During periods of time
%when the potential $A(t)$ is
%time-dependent the number of particles is intrinsically ambiguous. However, one can introduce an approximate definition of particle number by
%replacing $\Omega_k(t_f)$ by $\Omega_k(t)$ and $\Omega_k(t_1)$ in the  expressions (\ref{f}) and (\ref{dotf}). The mode function $f_k$ continues to be an exact solution to the mode equation if the Bogolubov coefficients $\alpha$ and $\beta$
%become time-dependent. It is straight-forward to derive a coupled set of differential equations for them, but that is not necessary for the analysis shown here.
%The time-dependent particle number is defined to be
with the total number of created particles at time $t$ given by
\bea
\langle N (t) \rangle \equiv 2\int_{-\infty}^{\infty}\frac{dk}{2\pi}|\beta_k (t) |^2 \quad . \label{Ntot-def}
\eea
Inverting Eqs. \eqref{f-alpha-beta-1} and \eqref{f-alpha-beta} gives
\bea\beta_k (t)=\frac{1}{i}(g_k\dot f_k  - \dot g_k f_k)  \;. \label{beta-soln} \eea
%and
%\bea
%g_k\equiv \frac{1}{\sqrt{2\Omega_k(t)}}e^{-i\int_{t_0}^{t} \Omega(t_1) dt_1},  \hspace{0.5cm} \dot g_k\equiv-i \sqrt{\frac{\Omega_k(t)}{2}} e^{-i \int_{t_0}^{t} \Omega_k(t_1) dt_1} \quad .
%\eea

A similar analysis can be done for spin-$\frac{1}{2}$ particles.  Time-dependent Bogoliubov coefficients can be obtained by first defining
\bea
g_k^I\equiv \sqrt{\frac{\Omega_k-(k-qA)}{2\Omega_k}} e^{-i\int_{t_0}^{t}\Omega_k(t_1) dt_1} , \hspace{0.5cm}
g_k^{II}\equiv-\sqrt{\frac{\Omega_k+(k-qA)}{2\Omega_k}} e^{-i\int_{t_0}^{t}\Omega_k(t_1) dt_1}
\label{definiciones2f} \ ,
\eea	
and then imposing the relations
\bes 
\bea
    h_k^I(t)=\alpha_k(t) g_k^I(t)+\beta_k(t) g_k^{II*}(t) \quad ,\\ \nonumber \\
    h_k^{II}(t)=\alpha_k(t) g_k^{II}(t)-\beta_k(t) g_k^{I*}(t) \quad , 
\eea
\ees
with the result that 
\be \beta_k(t) = [g_k^{I}(t)h_k^{II}(t)-g_k^{II}(t) h_k^{I}(t)] \;. \label{beta-soln-fermion} \ee
%\bea\beta_k(t) = [g_k^{I}(t)h_k^{II}(t)-g_k^{II}(t) h_k^{I}(t)] \quad .\eea
%Note that, if $A = A(t_f)$ for $t \ge t_f$, then the coefficient $\beta_k$ is time-independent and determines univocally the number of created particles.

%It is also interesting to remark that after the time for which the external classical source $J_C$ is zero, 
A classical current adds energy to the electric field and, if particle production occurs, then some of the electric field's energy is used for this process.  If the classical current shuts off at some point then, since the calculations are being done in flat space, energy is conserved but can still be transferred between the electric field and the produced particles.  
To see %that the particle creation phenomena is fully consistent with energy conservation even when quantum effects are included and the semiclassical backreaction equations are solved. 
this, note that the energy density of the electric field is $\rho_{\rm elec}=\frac{1}{2}E^2$.  A formula for the energy density of a scalar field in the case of a homogeneous electric field in $1+1$D is given in Eq.~\eqref{engsc} and one for the energy density of a spin-$\frac{1}{2}$ field is given in Eq.~\eqref{fermden}.  With these definitions it is easy to check that 
%It is not difficult to prove that~\cite {FNP}
\be
\frac{d}{d t} \Big(\rho_{\rm elec}+\langle \rho \rangle_{\rm ren} \Big)=\frac{d A}{dt}\left(\frac{d^2 A}{d t^2}-\langle J_Q \rangle \right)=0 \label{conservationEn} \quad ,
\ee
where the last term in parentheses is precisely the semiclassical Maxwell equation for the electric field \eqref{sb}.
Thus one can investigate the time dependence of the transfer of energy between the electric field and the particles by simply plotting $\rho_{\rm elec}$ and $\la \rho \ra_{\rm ren}$.  
  We note that in our approach energy conservation is a rigorous  consequence of the adiabatic renormalization prescription.

To study the effects of both particle production and the transfer of energy we consider models in which %, in this section only, 
the electric field is initially generated by a classical current of the form 
\be J_C =- E_0 \,\delta(t) \;. \label{Jc-delta} \ee
%If quantum effects are neglected then this gives \be   E = E_0 \Theta(t) \quad , \quad A = - E_0 \, t \, \Theta(t) \quad ,\eewith $\Theta(t)$ the step function.  
Since the electric field is zero for $t < 0$, there is a natural initial vacuum state which for a scalar field is
\bea
    f_k(t=0) = \frac{1}{\sqrt{2 \omega}} \quad , \quad \dot{f}_k(t=0) = -i\sqrt{\frac{\omega}{2}}  \label{delta-initial-vacuum-scalar} \quad . \label{initial-state-delta-scalar}
\eea
For a spin-$\frac{1}{2}$ field the initial vacuum state is
\bea
    h_k^{I}(t=0) = \sqrt{\frac{\omega-k}{2\omega}} \quad , \quad h_k^{II}(t=0) = -\sqrt{\frac{\omega+k}{2\omega}} \quad . \label{initial-state-delta-fermi}
\eea
Since the classical current %~\eqref{Jc-delta} 
is zero %, as it is 
for $t > 0$, the total energy density of the system is constant for both the scalar and spin-$\frac{1}{2}$ cases.

To solve the semiclassical backreaction equations numerically we have used dimensionless variables and parameters. We have scaled the mode equations, \eqref{fmode} for scalars, \eqref{modeh1}, \eqref{modeh2} for spin-$\frac{1}{2}$ fields, and also the semiclassical Maxwell equation \eqref{sb} in terms of the electric charge $q$. The new scaled parameters are %$k \to k/q,  \omega \to \omega/q,  t \to qt,  m \to m/q$.
\bea k \to k/q, \qquad \omega \to \omega/q, \qquad t \to qt, \qquad m \to m/q     \;. \label{scaled-parameters} \eea
For the mode functions for the scalar field% $f(t) \to \sqrt{q}f(t)$.
\be  f(t) \to \sqrt{q}f(t) \;.\ee
%To follow the conventions used in \cite{Kluger91,Kluger92}  
We also use the definitions% $\tilde E \equiv \frac{E}{E_{\rm crit}}$, $\tilde J \equiv  \frac{J}{qE_{\rm crit}}$, with 
\be 
    \tilde E \equiv \frac{E}{E_{\rm crit}} \quad , \quad \tilde J \equiv \frac{J}{qE_{\rm crit}}  \quad , \quad \tilde \rho=\frac{\rho}{E_{\rm crit}^2} \quad , \quad \langle \tilde N  \rangle=\frac{\langle  N  \rangle}{E_{\rm crit}} \quad ,\label{Etilde} 
\ee
% $E_{\rm crit} = \frac{m^2}{q}$ and $\tilde \rho=\rho/E_{crit}^2$, $\langle \tilde N  \rangle=q^2\langle  N  \rangle/m^2$.
%\be \tilde \rho=\rho/E_{\rm crit}^2 \quad , \quad \langle \tilde N  \rangle=q^2\langle  N  \rangle/m^2 \;,  \ee
where $E_{\rm crit}$ is the critical scale for pair production defined in Eq. \eqref{def-Ecrit}.

%To directly compare solutions for different masses, in Section \ref{sec:numerical} we will use the scalings $ E \to E/q$ and $J \to J/q^2$. %  \label{Eq-Jq} \ee
%From the exponential factor one notes immediately that the order of the critical scale for pair production can be defined to be   \be E_{\rm crit} \equiv m^2/q \;.  \label{Ecrit-def} \ee

%\subsection{Details of the Particle Production Process} %\label{subsecsec:scalar-1}
%\label{subsec:details-part-prod}
%Here, we show the solutions of the
%semiclassical backreaction equations a scalar field for two different initial conditions of the electric field, $E_0=E_{\rm crit}$ and $E_0=5 E_{\rm crit}$, where $E_{\rm crit}$ is the critical Schwinger value  discussed above.
%We have fixed the mass of the created particles to be $\frac{m^2}{q^2}=10$. The results are displayed in Fig. \ref{figscalars}. % where we show (i) the time evolution of the scaled electric field $\tilde E$ and the induced electric current $\langle \tilde J_Q \rangle $ as functions of the dimensionless time $qt$, (ii) the time evolution of the scaled energy density $\tilde \rho$ for both the electric field and the created particles, and (iii) the time evolution of the scaled particle number $\langle \tilde N \rangle$.

\subsection{Particle production and energy transfer}

Here we investigate some of the details of the particle production process including the transfer of energy between the electric field and the particles for solutions to the semiclassical backreaction equation when either a scalar field or a spin-$\frac{1}{2}$ field is coupled to the electric field and the classical current is given by Eq.~\eqref{Jc-delta}.  The specific solutions considered have $E_{\rm crit} =\frac{m^2}{q} = 10$ and either $E_0=E_{\rm crit}$ or $E_0=5 E_{\rm crit}$.      

\begin{figure}[htbp]
\begin{center}
\begin{tabular}{c}
\hspace{-1.9cm}\includegraphics[width=70mm]{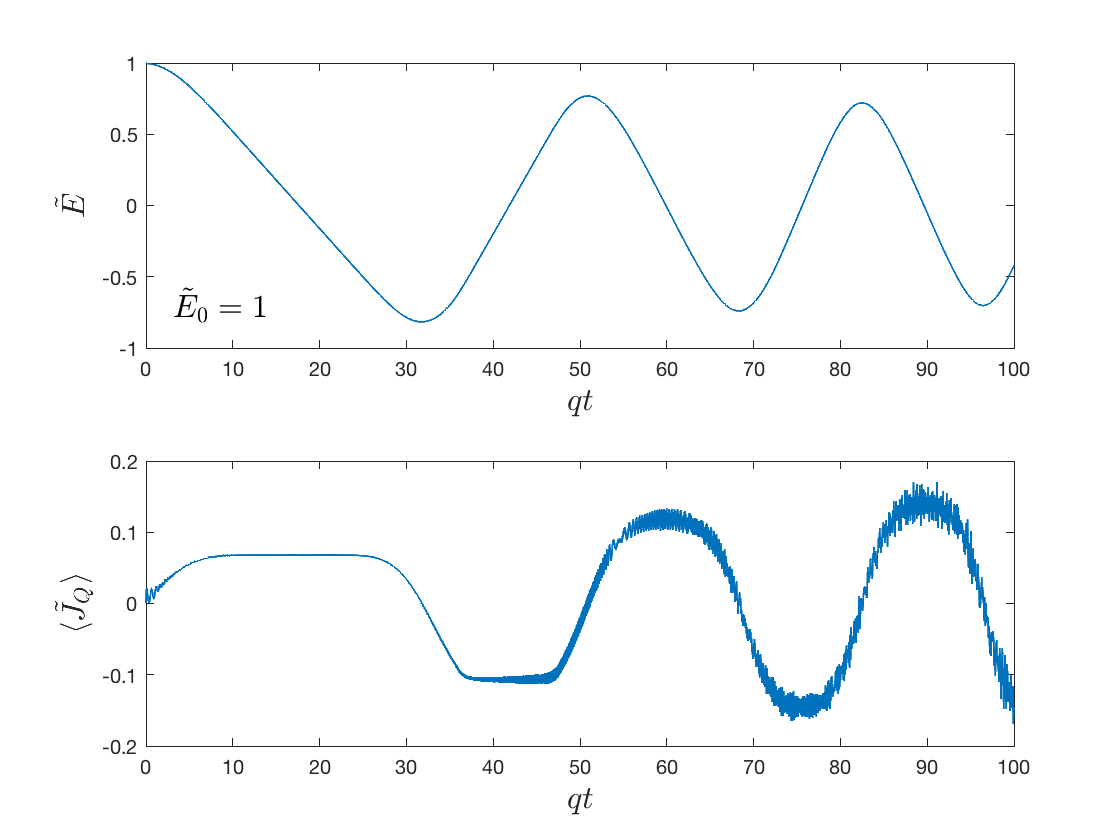}\hspace{-0.7cm}
\includegraphics[width=70mm]{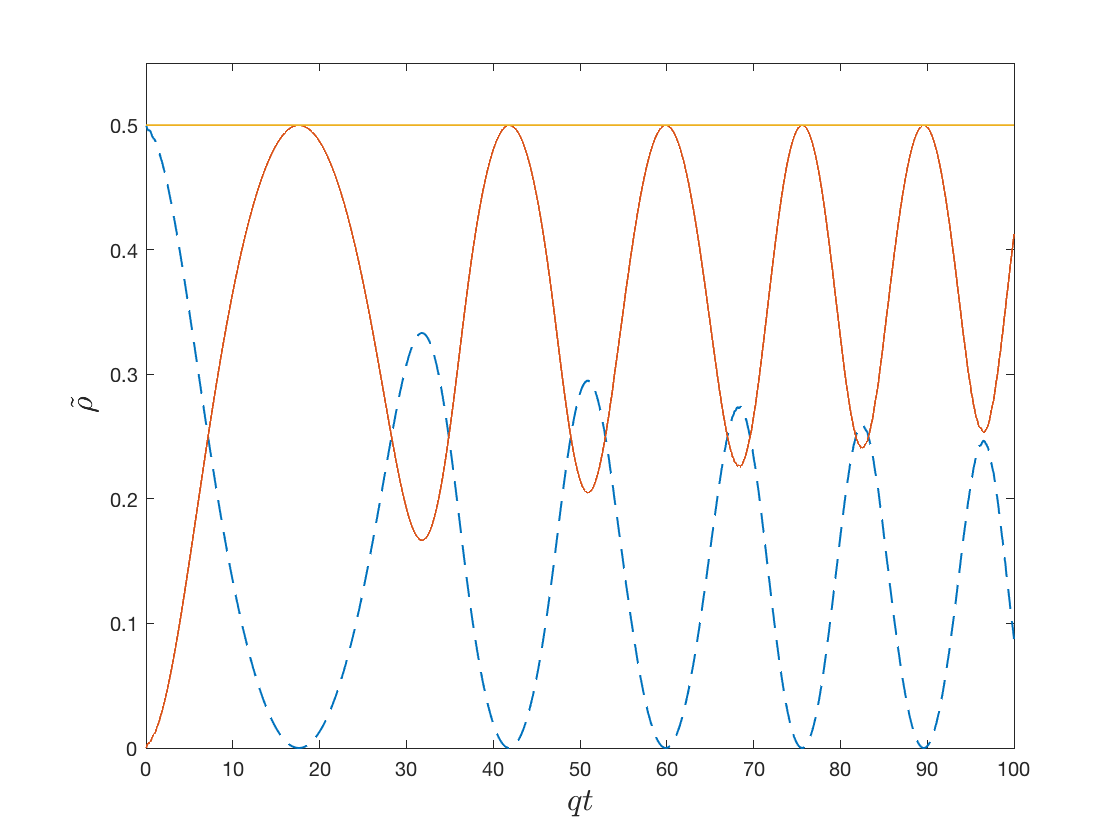}\hspace{-0.7cm}
\includegraphics[width=70mm]{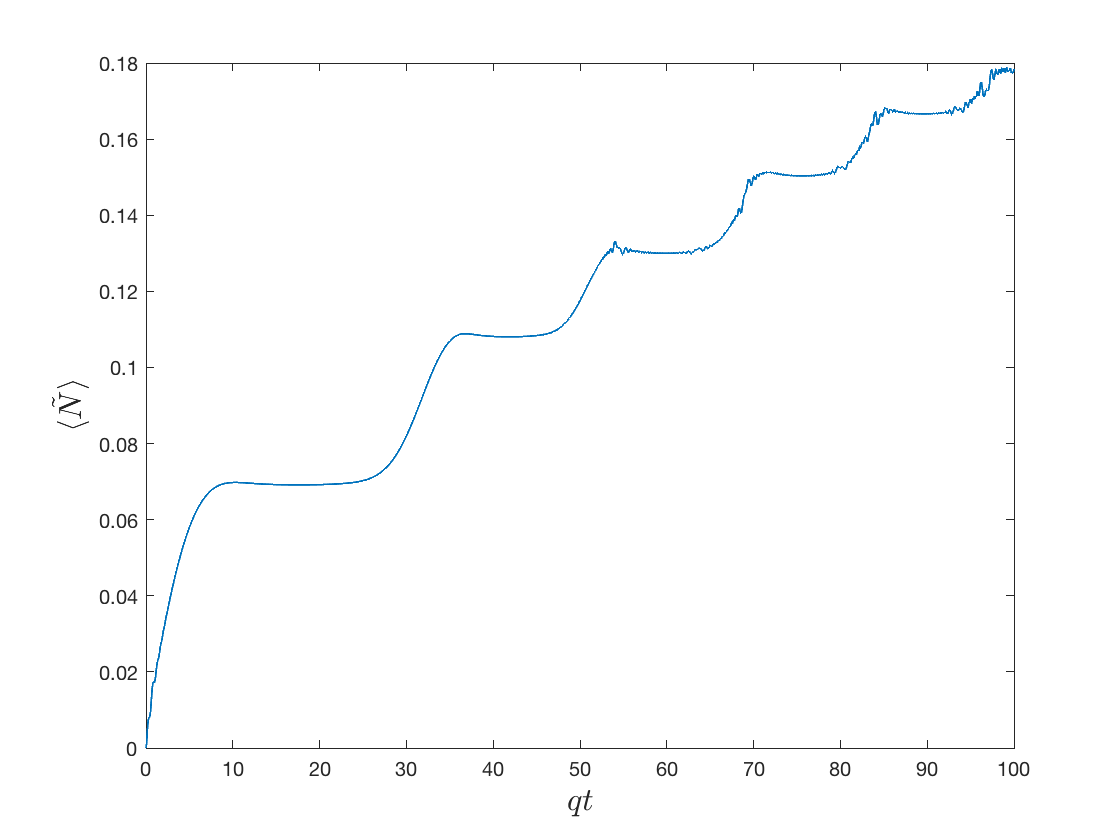}\\
\hspace{-1.9cm}\includegraphics[width=70mm]{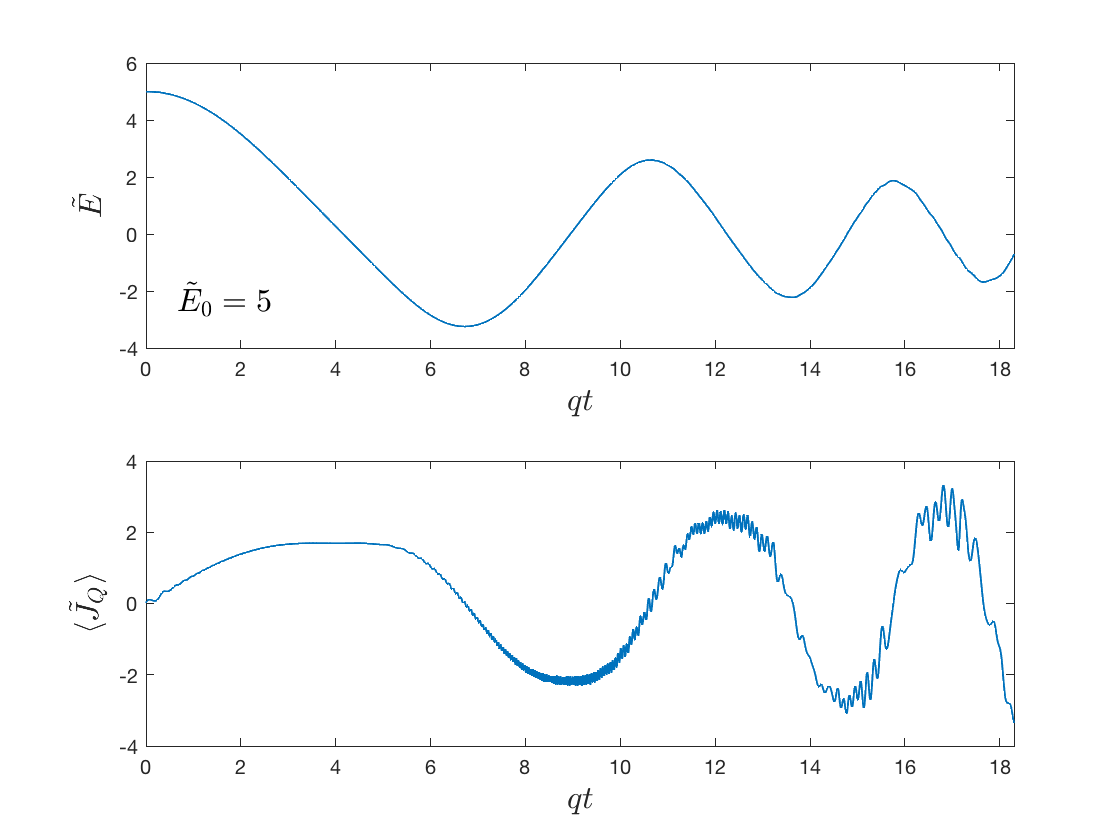}\hspace{-0.7cm}
\includegraphics[width=70mm]{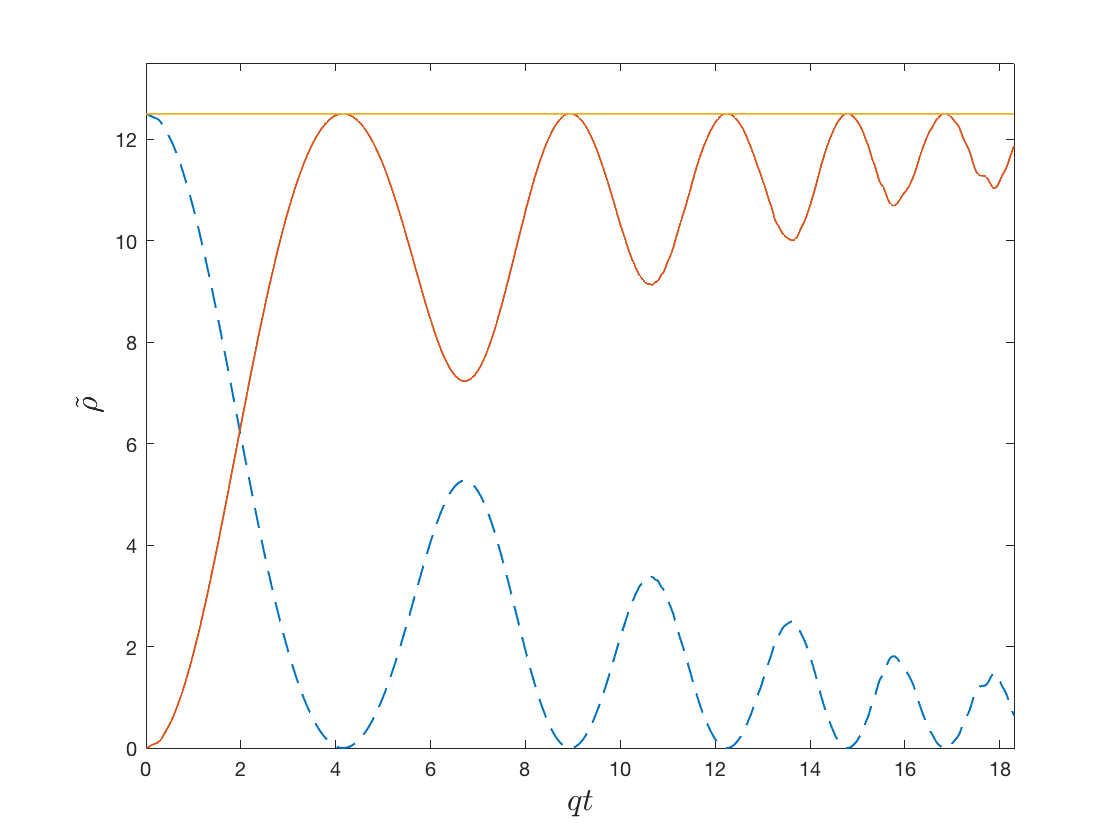}\hspace{-0.7cm}
\includegraphics[width=70mm]{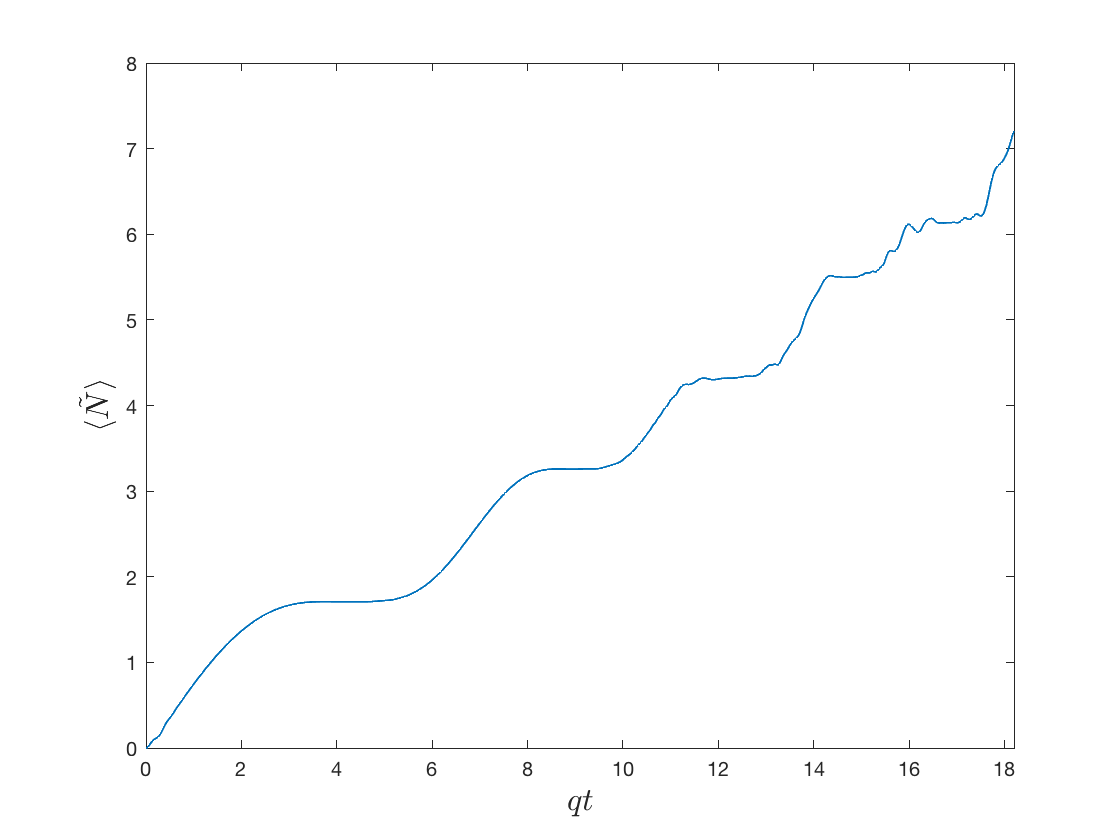}
\end{tabular}
\end{center}
\caption{\small{Various quantities are plotted for solutions to the semiclassical backreaction equations for a quantized scalar field with the classical current profile $J_C =- E_0 \,\delta(t)$. %~\eqref{Jc-delta}.  
The solutions for $\tilde E_0=1$ are shown across the top row of panels and those for $\tilde E_0=5$ are shown across the bottom row.  The mass of the scalar field is $ \frac{m^2}{q^2}=10$ and thus $\frac{E_0}{q} = 10$ and $50$ respectively. In the left panels the electric field $\tilde E$ and the electric current $\langle \tilde J_Q\rangle_{\rm ren}$ are plotted.  For each of the middle panels the blue dashed curve corresponds to the energy density of the electric field $\rho_{elec}$, the orange solid curve represents the energy density of the created particles $\langle \tilde{\rho} \rangle_{\rm ren}$, and the straight yellow line is the total energy density of the system.  The total particle number $\langle \tilde{N} \rangle$ is plotted in the right panels.
}}
\label{figscalars}
\end{figure}

In Fig.~\ref{figscalars}, some of our results for a scalar field coupled to the electric field are shown for $E_0=E_{\rm crit}$ in the top panels and $E_0=5 E_{\rm crit}$ in the bottom ones.  
cIt is apparent that as soon as particle production starts to occur, the initial electric field decays and the electric current increases as a consequence of the created particles. When the electric field has been reduced significantly the current reaches a plateau and the particle creation saturates. Furthermore, when the electric field changes sign  and
its magnitude again becomes large, the particle creation
rate is enhanced while the current is slowed and then reversed.   This results in plasma oscillations.  %As discussed above the  process is always exactly consistent with energy conservation.
%The transfer of energy between the particles and the field can be seen in the middle panels of Fig.~\ref{figscalars}.
Note also that the duration of the initial growth of the electric current $\langle \tilde J_Q \rangle$ is of the same order as the duration of the initial growth in the particle number $\langle \tilde N \rangle$.

\begin{figure}[htbp]
\begin{center}
\begin{tabular}{c}
\hspace{-1.9cm}
\includegraphics[width=70mm]{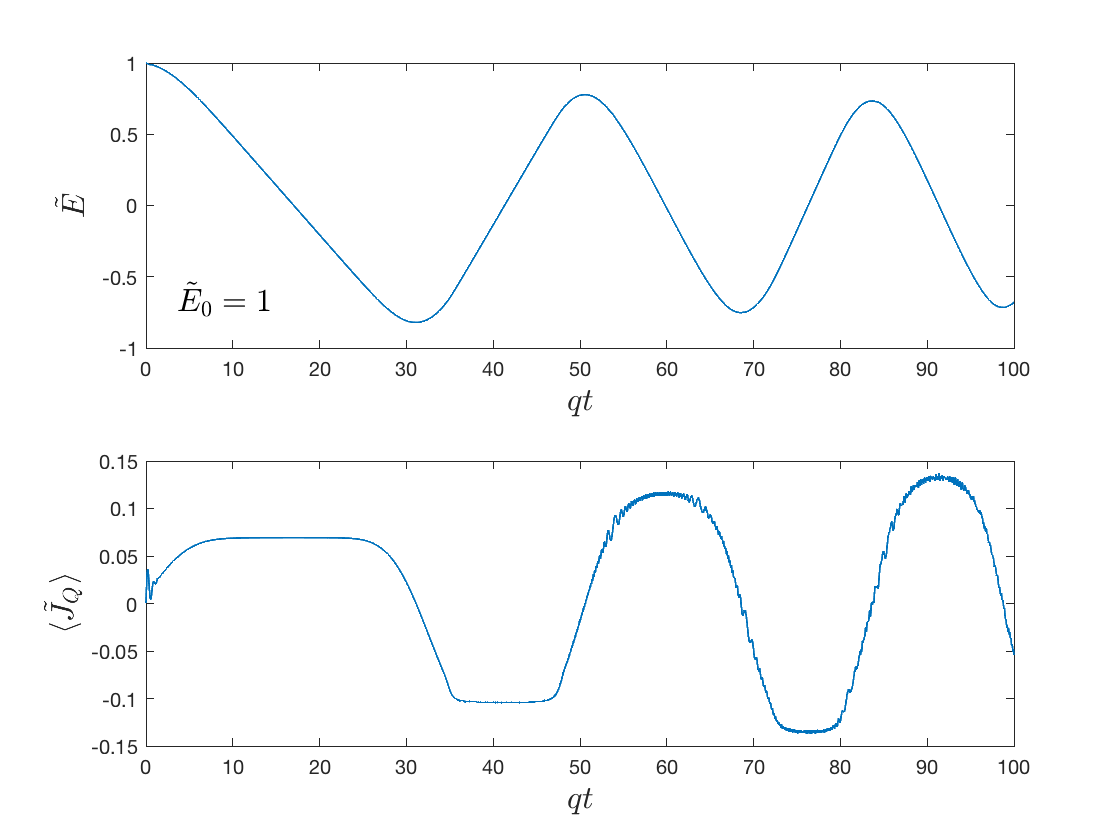}\hspace{-0.7cm}
\includegraphics[width=70mm]{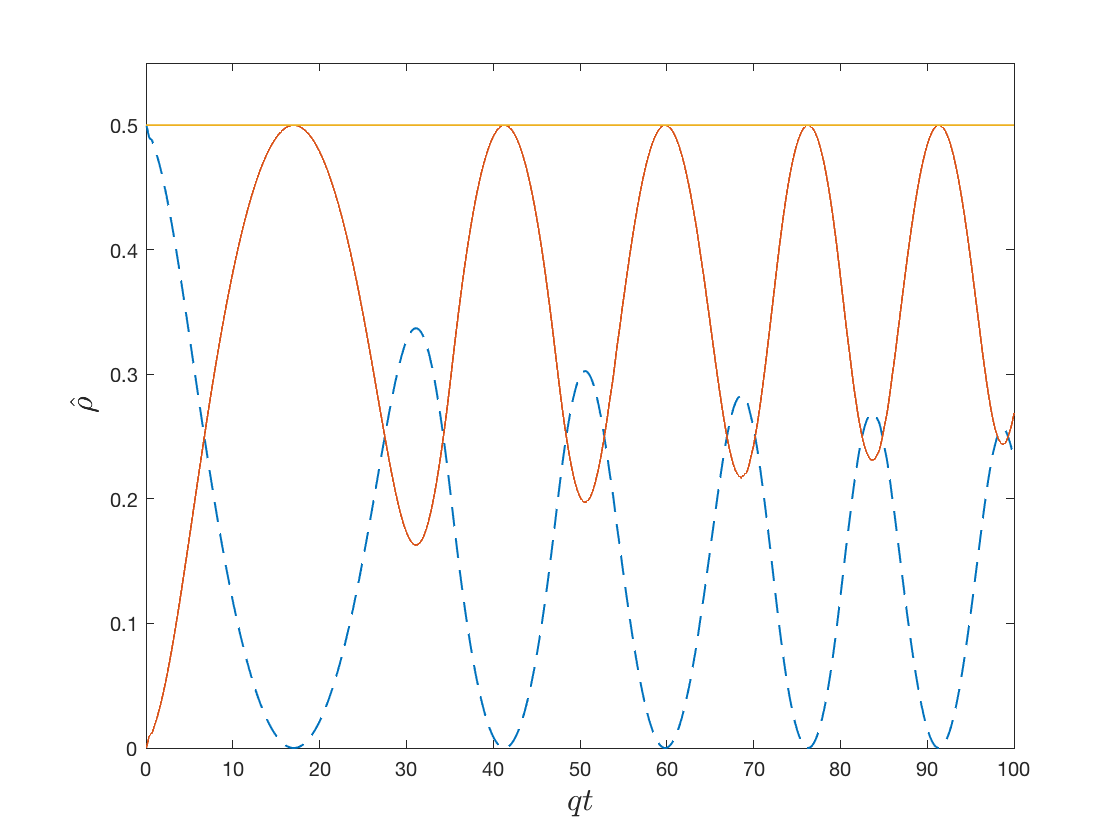}\hspace{-0.7cm}
\includegraphics[width=70mm]{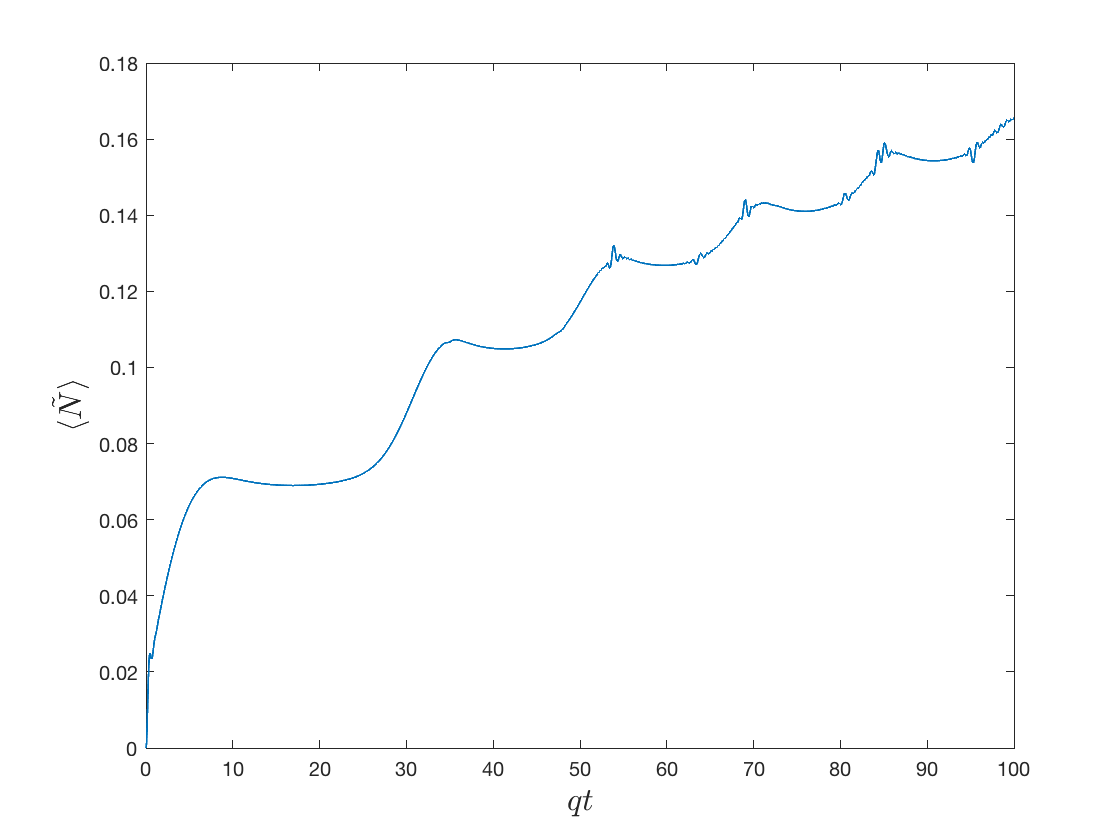}\\
\hspace{-1.9cm}\includegraphics[width=70mm]{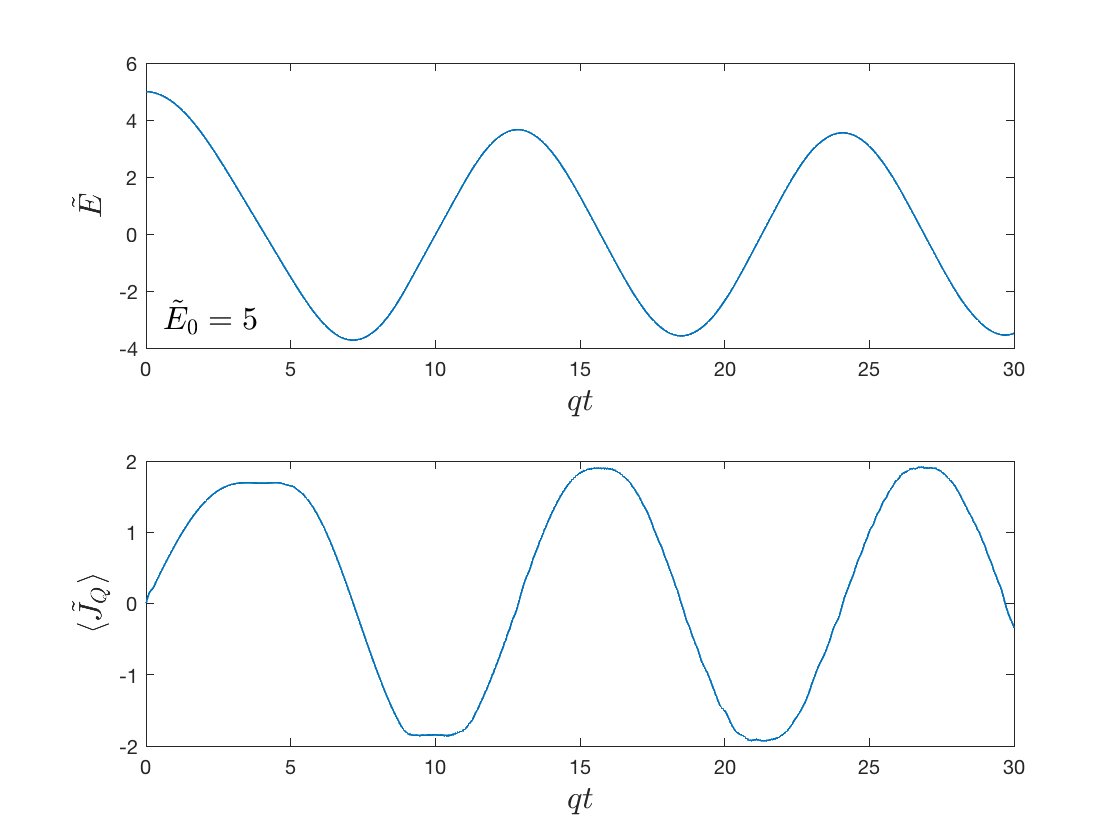}\hspace{-0.7cm}
\includegraphics[width=70mm]{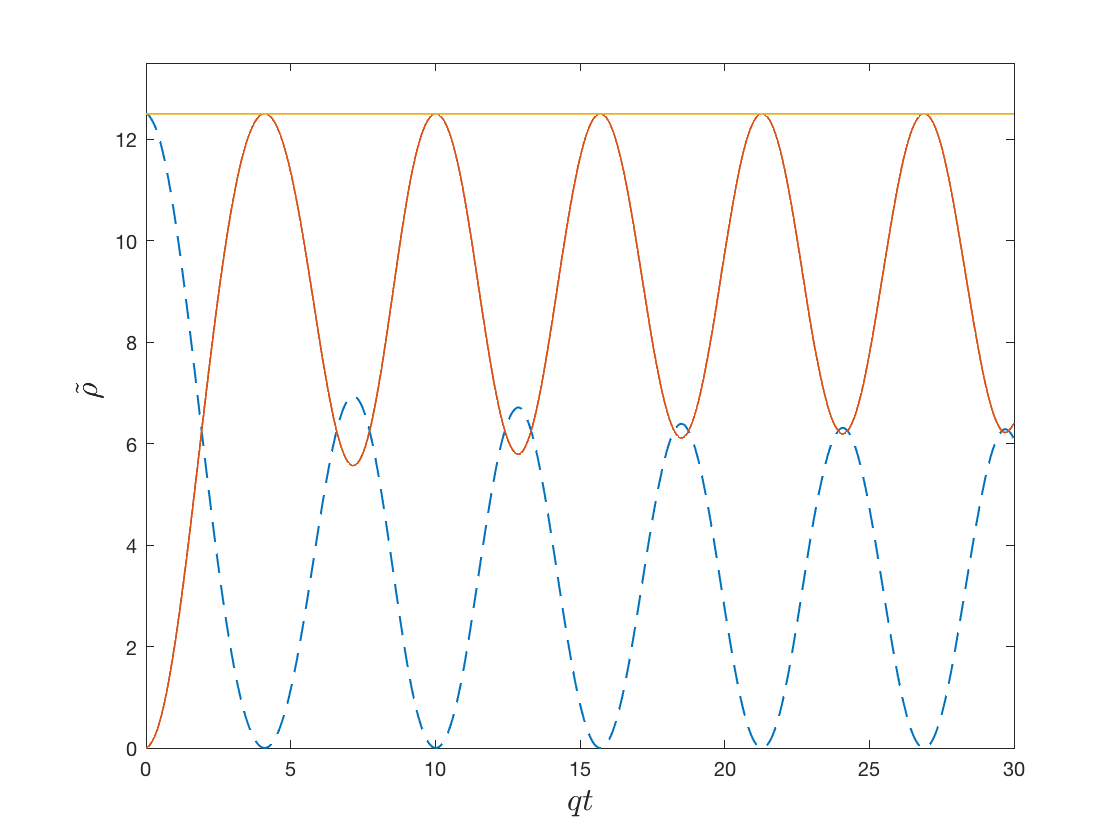}\hspace{-0.7cm}
\includegraphics[width=70mm]{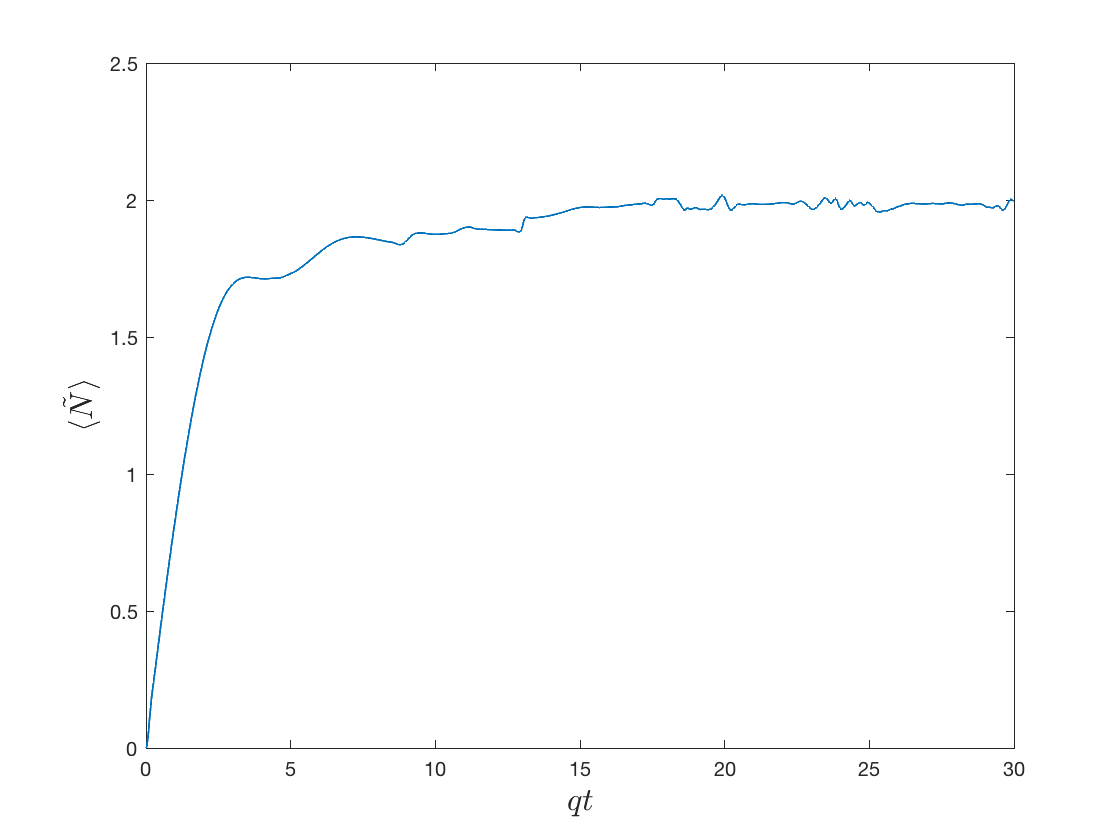}
\end{tabular}
\end{center}
\caption{\small{Various quantities are plotted for solutions to the semiclassical backreaction equations for a quantized spin-$\frac{1}{2}$ field with the classical current profile $J_C =- E_0 \,\delta(t)$. %~\eqref{Jc-delta}. 
The structure of the figure, and also the initial conditions for the electric profile, are the same than in Figure \ref{figscalars}.
%The solutions for $\tilde E_0=1$ are shown across the top row and those for $\tilde E_0=5$ are shown across the bottom row.  The mass of the spin $\frac{1}{2}$ field is $\frac{m^2}{q^2}=10$ and thus $\frac{E_0}{q} = 10$ and $50$ respectively. In the left panels the electric field $\tilde E$ and the electric current $\langle \tilde J_Q\rangle_{\rm ren}$ are plotted.For each of the middle panels the blue curve corresponds to the energy density of the electric field $\rho_{\rm elec}$, the orange curve represents the energy density of the created particles $\langle \tilde{\rho} \rangle_{\rm ren}$, and the yellow line is the total energy density of the system.  The total particle number $\langle \tilde{N} \rangle$ is plotted in the right panels.
}}
\label{figfermions}
\end{figure}

%\subsection{Spin $\frac{1}{2}$ Field} \label{subsec:fermions-1}
In Fig.~\ref{figfermions}, some of our results for a spin-$\frac{1}{2}$ field coupled to the electric field are shown for $E_0=E_{\rm crit}$ in the top panels and $E_0=5 E_{\rm crit}$ in the bottom ones. 
Comparing Fig.~\ref{figfermions} with Fig.~\ref{figscalars}, one finds that for the smaller value of the initial electric field, $E_0 = E_{\rm crit}$,  all of the details are very similar to the scalar field case.   For the larger initial value of the electric field many of the general features are also similar including the initial damping of the electric field and subsequent plasma oscillations.  However, some of the details differ significantly.  Due to Pauli blocking the particle production for the spin-$\frac{1}{2}$ field effectively shuts off fairly early in the process. %maximum number of created particles is bounded\textcolor{orange}{(at a lower level than it would be due to the conservation of energy})  and the particle production process for the spin $\frac{1}{2}$ field effectively shuts off fairly early in the process.  
One result is that there is less energy permanently transferred to the particles than in the scalar field case.    
%This fact is also reflected in the energy density of the electric field whose envelope rapidly reaches a constant value, meaning that no new particles are created. 

%Due to Pauli blocking, .
%Again, the growth in the electric current $\langle \tilde J_Q \rangle$ at early times coincides with the growth in the particle number $\langle \tilde N \rangle$.

There are some differences in both the scalar field and spin-$\frac{1}{2}$ cases between the solution for which the electric field is at the critical value initially and the solution for which it is initially much larger.  As would be expected there is significantly more particle production and a significantly faster initial damping for the larger field.  Once the plasma oscillations begin there also appears to be a much faster approach of the amplitude of the electric field and the total number of particles to their asymptotic values when the initial electric field is larger.  Further, examination of the energy density shows that a significant amount of the initial energy of the larger electric field is permanently transferred to the particles during the first damping phase and this increases during the plasma oscillation phase.  For the smaller field less energy is transferred initially to the particles during the first damping phase and the permanent transfer of energy to the particles upon each plasma oscillation is smaller.   

For both the scalar and spin-$\frac{1}{2}$ fields, a clear correlation is found between the maxima of the energy density of the created particles and the maxima and minima of the current due to the created particles.  For cases in which the total number of particles continues to increase significantly after the first burst of particle production, the maxima in the energy density of the created particles correlate with the middles of the time periods when the total number of particles is approximately constant.  As expected, the minima of the energy densities of the created particles correspond to times when a new round of significant particle production is just beginning in cases where there is significant particle production after the first burst.
In general the periods of significant particle production correspond to periods when energy is being transferred to the particles.  It is interesting to note that the above results, obtained within the adiabatic renormalization prescription in the continuous limit,  are  compatible with the results obtained using a similar method in $3+1$D~\cite{Tanji} as well as those obtained in $1+1$D and/or $3+1$D 
using lattice simulations~\cite{lattice-1,lattice-2} and classical statistical field theory techniques \cite{stat-FT}.

\begin{figure}[htbp]
\begin{center}
\begin{tabular}{c}
\hspace{-1.9cm}\includegraphics[width=70mm]{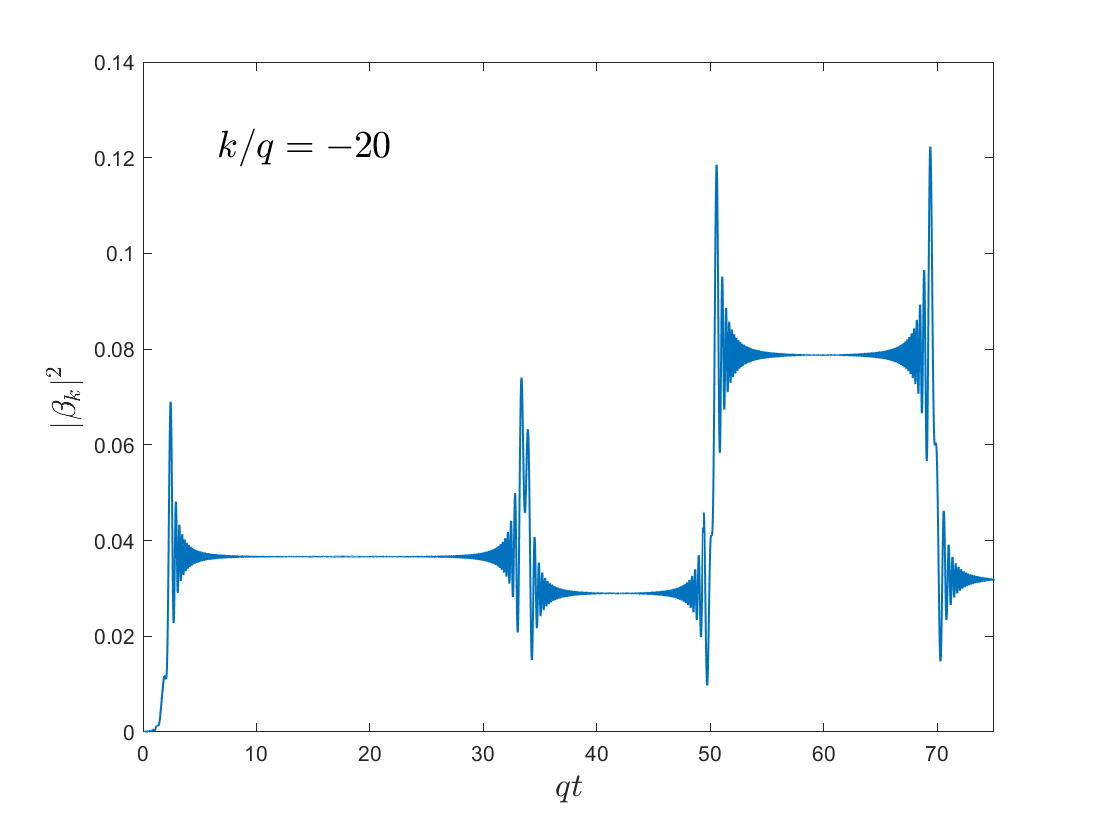}\hspace{-0.7cm}
\includegraphics[width=70mm]{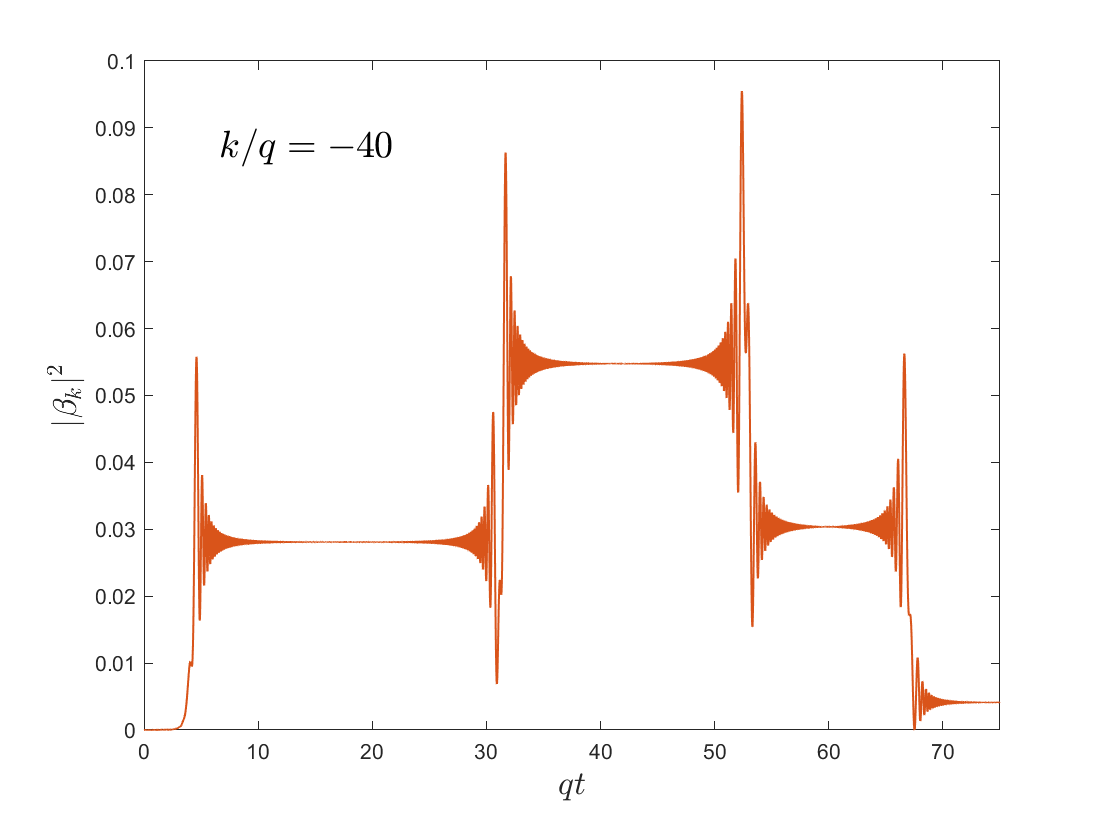}\hspace{-0.7cm}
\includegraphics[width=70mm]{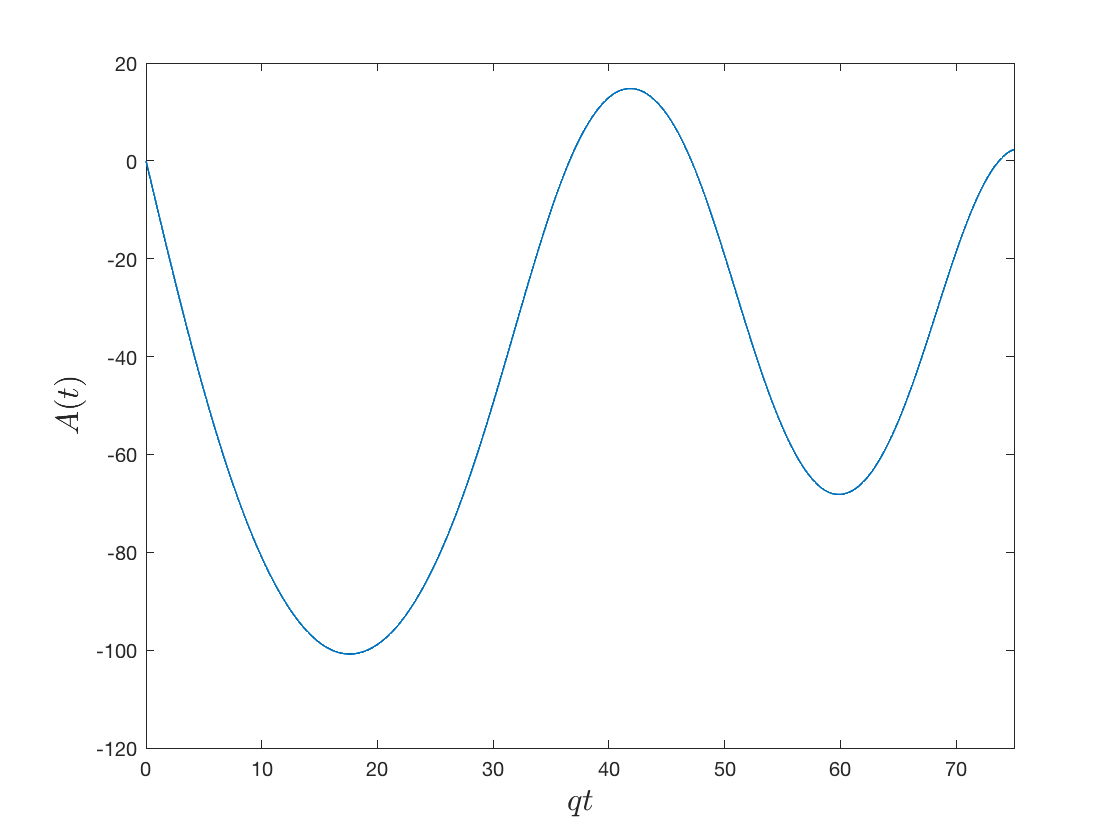}\\
\hspace{-1.9cm}\includegraphics[width=70mm]{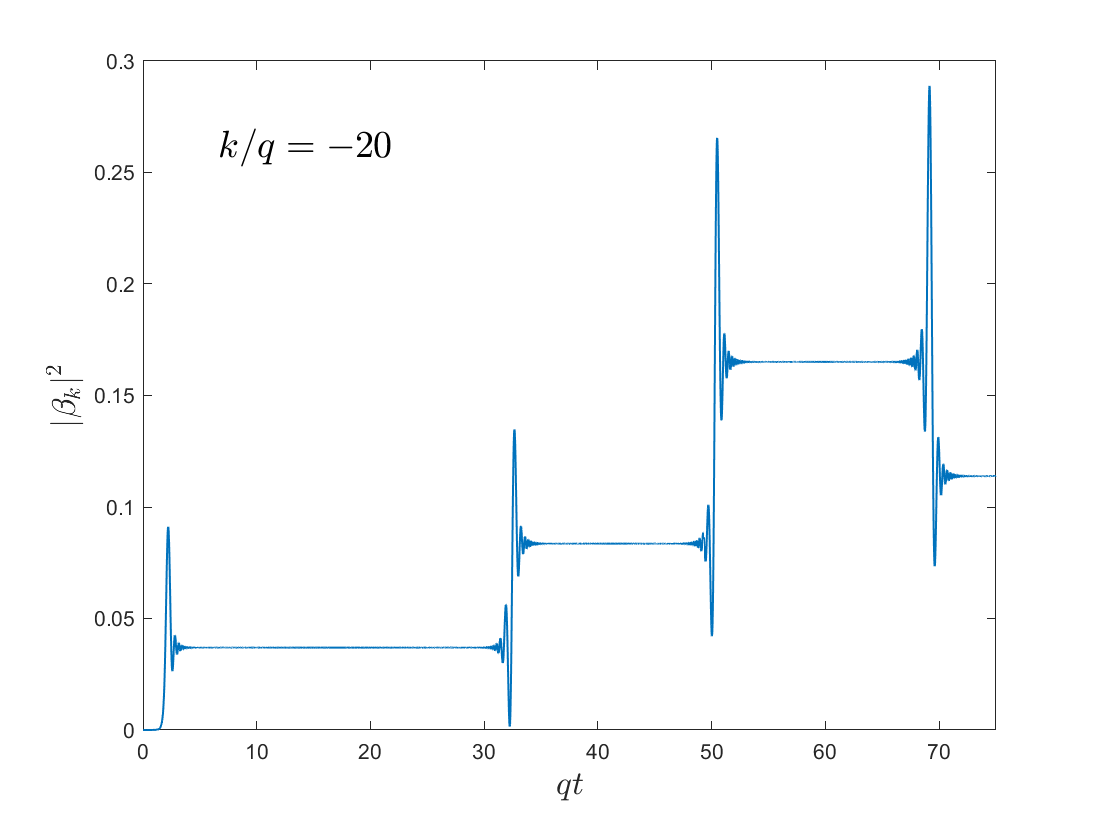}\hspace{-0.7cm}
\includegraphics[width=70mm]{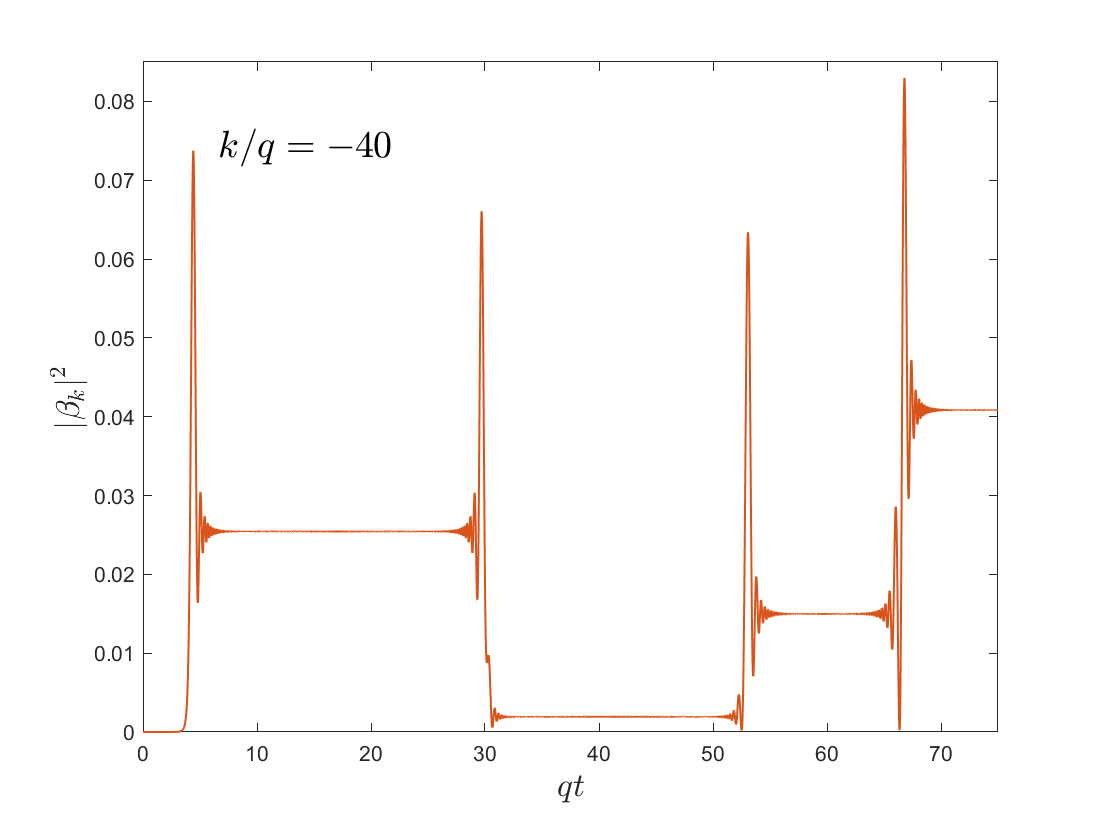}\hspace{-0.7cm}
\includegraphics[width=70mm]{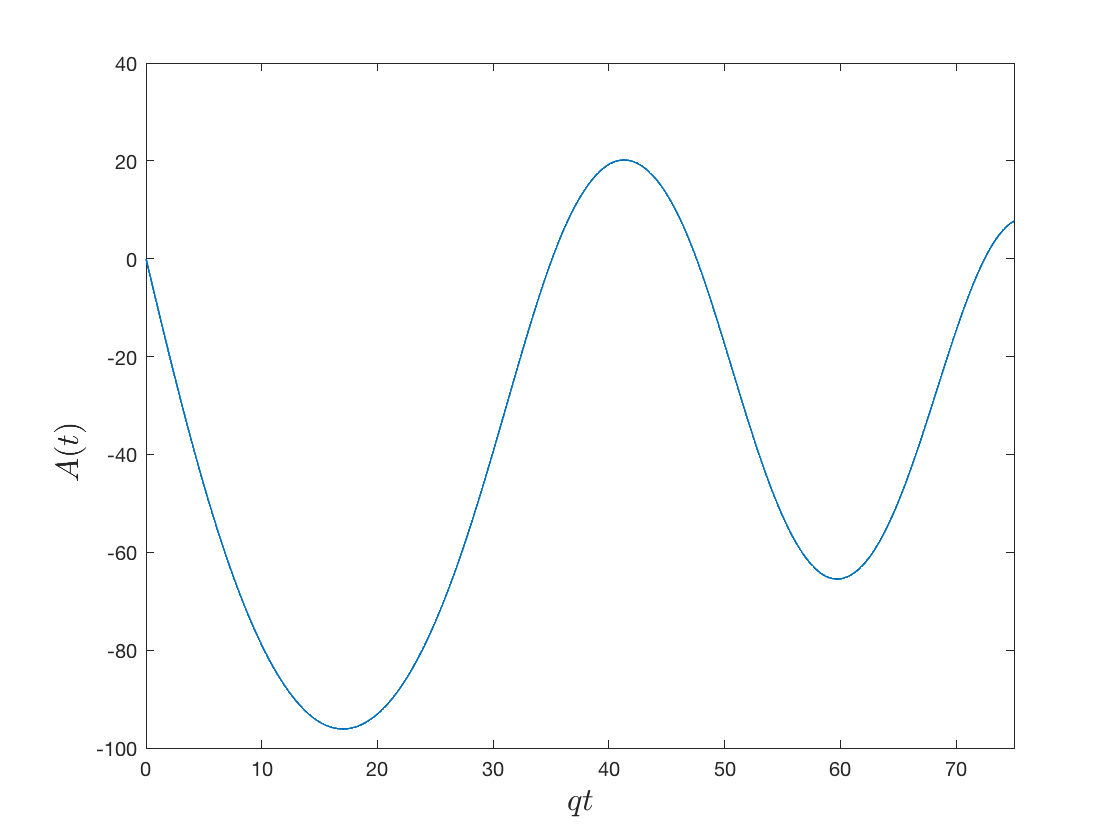}
\end{tabular}
\end{center}
\caption{\small{ The time-dependent particle number is shown for individual modes when $\tilde{E}_0 = 1$ for the scalar field (top row) and spin-$\frac{1}{2}$ (bottom row) cases for the classical current profile $J_C =- E_0 \,\delta(t)$. %~\eqref{Jc-delta}.
The mass of the scalar field is $ \frac{m^2}{q^2}=10$ and thus $\frac{E_0}{q} = 10$. The vector potential is plotted in the far right panels.  For each row the first panel on the left shows the particle number for $\frac{k}{q} = -20$ and the middle panel shows the particle number for $\frac{k}{q} = -40$.
}}
\label{mode-mode}
\end{figure}

It was shown in Refs.~\cite{and-mot-I,dunne-part-prod,and-mot-san} that a single particle creation event occurs for an individual mode if the background electric field is either constant or approximately constant. What is different here is that the backreaction of the produced particles produces plasma oscillations.  The resulting oscillations of the electric field lead to some modes undergoing multiple particle creation events and sometimes also particle destruction events.  This can be seen in Fig.~\ref{mode-mode} where the time evolution of the function $|\beta_k|^2$ for $\tilde E_0=1$ is shown for both the scalar field and spin-$\frac{1}{2}$ field cases. Comparison with the plot of the vector potential $A(t)$ shows that the creation, or destruction, process for an individual mode $k$ happens when $k-qA(t)\approx m$.

\newpage

\subsection{Massless limit for the spin-$\frac{1}{2}$ field}\label{masslessLim}
\label{sec:massless}

For completeness we extend our analysis to the massless limit for the spin-$\frac{1}{2}$ field. In this case, the mode equations \eqref{modeh1} and \eqref{modeh2} decouple, and with the initial conditions given in Eq.~\eqref{initial-state-delta-fermi}, their solutions are given by
\be 
    h^{I,II}_k(t)=\pm \theta (\mp k)e^{\pm i\int^t_{t_0}(k-qA(t'))dt'} \quad, \label{masslessmode}
\ee
where $\theta(x)$ is the Heaviside step function. The electric current $\la J_Q \ra_{\rm ren}$ has the simple form given in Eq.~\eqref{Jmassless}, and hence, the semiclassical Maxwell equation~\eqref{sb} turns out to be the equation of a harmonic oscillator $\ddot A (t)+\frac{q^2}{\pi}A (t)=0$.
%\bea\ddot A (t)+\frac{q^2}{\pi}A (t)=0 \quad .\eea
With the initial conditions $E(0)=E_0$ and $A(0)=0$, we immediately find the analytic solution
$E(t)=E_0\cos(\frac{|q|}{\sqrt{\pi}}t)$. 
The energy density~\eqref{fermden} and the number of the created particles are $\la \rho(t) \rangle_{\rm ren}=\frac{q^2}{2\pi}A^2(t)$ and $\langle N(t) \rangle =\frac{|qA(t)|}{\pi}$.
%\bes \bea\langle \rho(t) \rangle_{ren}&=&\frac{q^2}{2\pi}A^2(t) \;, \\\langle N(t) \rangle &=&\frac{|qA(t)|}{\pi}\;.\eea \ees
 For a detailed analysis of the adiabatic invariance of the particle number see Ref.~\cite{BFNP}. As in the general case, the total energy of the system is conserved. 
 %In Fig.~\ref{figfermionsm0} these quantities are plotted for the case $E_0/q=2$. 
 We note the exact analytic solubility of the case $m=0$ is due entirely to the axial anomaly in $1+1$D.  In fact, the constant $\frac{|q|}{\sqrt{\pi}}$ is the mass of the ``photon'' in the Schwinger model generated by radiative corrections \cite{Schwingermass}. In the massless case the (nonlocal) effective action $\Gamma[A_\mu, J_C]$ can be obtained exactly and it describes a gauge-invariant vector field with mass $\frac{|q|}{\sqrt{\pi}}$ (see, for instance, Ref.~\cite{Dash}). The semiclassical calculation of the produced energy due to the external source provides an accurate  result. In the massive case the effective action does not describe an integrable model \cite{coleman-1, gross} and  the semiclassical picture is expected to break down at some point.  The validity of the semiclassical approximation for massless and  massive spin-$\frac{1}{2}$ fields for the asymptotically constant classical profile is addressed in Sec. \ref{sec:validity-asym-class-profile}.   %using $\Gamma[A_\mu, J_C]$ as a classical action (i.e., the so-called semiclassical approximation)  coincides with the calculation within the exact quantum theory.}

\section{Validity criterion for the semiclassical approximation}
\label{sec:criterion}

The semiclassical backreaction equation can be derived from Eq.~\eqref{eff} via a loop expansion \cite{qftbook}.
In this case when solving the semiclassical backreaction equation, the semiclassical approximation breaks down if contributions from the quantum terms to the equations become comparable to that of the classical background field and any other classical fields.  The reason is that one expects higher-order terms in the loop expansion to be important in that limit. However, there is a different way to derive the semiclassical backreaction equation called the large-$N$ expansion.  In this expansion one considers $N$ identical quantum fields coupled to the background field, which to leading order is treated as a classical field.  At next-to-leading order in the large-$N$ expansion, quantum effects due to the background field first appear \cite{large-N-1994, hu-verdaguer}.  Thus in this expansion it is consistent to consider solutions to the semiclassical backreaction equation for which the quantum fields have a significant effect on the classical background field.
%{\color{red} Here we will take $N=1$ and consider a priori  strong electric fields. However, in the second part of the paper we will reanalize this issue and consider a wider range of  electric field backgrounds by comparing it with the (Schwinger) critical electric field scale $E_{crit}\equiv m^2/q$.} 
 Here we will take $N=1$ and consider a wide range of
situations ranging from those where the background electric field is small compared with the (Schwinger) critical scale $E_{\rm crit}\equiv m^2/q$ and quantum effects are correspondingly small to those where the background electric field is large compared to the critical value and quantum effects are correspondingly large.  The critical value is the threshold for which a significant amount of particle production is expected to occur.  %It is discussed in more detail in Sec. \ref{sec:energy}.

%As discussed in Sec.~\ref{sec:semiclass}, 
The large-$N$ expansion provides a formal framework for the semiclassical backreaction equation when quantum effects are significant.  However, it does not guarantee that the semiclassical approximation is valid.  There are three reasons.  The first is that interactions of the quantum fields which are coupled to the classical background field are ignored in most cases, including those considered here.  This works if the interactions are small over the time scales relevant to the problem. The second is that even if the next-to-leading order terms in the large-$N$ expansion are initially small in size, it has been shown in certain quantum mechanics calculations that they undergo secular growth~\cite{fred1} and there is evidence that secular growth also occurs for such terms in quantum field theory~\cite{fred-emil-private}.  However, there is also evidence that partial resummations of certain classes of Feynman diagrams eliminate this problem~\cite{fred2, fred3}. The third is that the semiclassical backreaction equation involves an expectation value of some quantity such as the electric current or stress-energy tensor that is constructed from the quantum fields. For an expectation value to be a good approximation to what one would measure in quantum theory, it is necessary that quantum fluctuations are small.

A natural way to estimate the size of quantum fluctuations is to evaluate the two-point correlation function for the current. There are several different two-point correlation functions including (i) $\langle J(t,x) J(t',x') \rangle $, (ii) the connected part, i.e., $\langle J(t,x) J(t',x') \rangle - \langle J(t,x)\rangle \langle J(t',x') \rangle $, (iii) the time-ordered correlation function $\langle T( J(t,x) J(t',x')) \rangle $, etc. There are problems associated with some of these, as described in Refs.~\cite{wu-ford, phillips-hu, validitygravity}. For example, it has been shown for the symmetric part of the stress-energy tensor two-point correlation function that there can be state-dependent divergences in the limit that the points come together~\cite{wu-ford}.  A related issue is that it has been shown in at least one case in the limit that the points come together that different renormalization schemes can give different results for a particular quantity made from one component of the stress-energy tensor two-point correlation function~\cite{phillips-hu}.  There can also be covariance issues with some of the quantities made from the stress-energy tensor two-point correlation function~\cite{validitygravity}.

There is a correlation function that is free of these problems and which emerges naturally from the semiclassical theory itself and that is $\la [J(t,x),J(t',x')]\ra$.  By perturbing the semiclassical backreaction equation one is led to the so-called linear response equation which contains this correlation function and which describes the time evolution of perturbations about a given semiclassical solution. A criterion was developed in Ref.~\cite{validitygravity} for the validity of the semiclassical approximation in gravity which states that a necessary condition for the validity of the %large $N$ 
semiclassical approximation to be valid is that any linearized, gauge-invariant scalar quantity constructed from solutions to the linear response equations with finite nonsingular initial data should not grow without bound.  It is important to emphasize that this is not a sufficient condition for the validity of the semiclassical approximation.  %Thus if the criterion is satisfied the semiclassical approximation may be valid but if it is not satisfied then the semiclassical approximation is not valid. 
The criterion was adapted to cover preheating during chaotic inflation~\cite{validitypreheating} where a significant amount of particle production occurs and quantum effects are large. If the criterion is applied to semiclassical quantum electrodynamics then it would state that the semiclassical approximation breaks down if any linearized gauge-invariant quantity constructed from solutions to the linear response equation with finite nonsingular initial data grows rapidly for some period of time. %The modified version of the criterion states that ``The {\color{red}(large $N$) we think it is preferable to omit here the large $N$ to avoid any apparent contradiction with the assumption of taking $N=1$} semiclassical approximation will break down if any linearized gauge invariant quantity constructed from solutions to the linear response equation with finite non-singular initial data, grows rapidly for some period of time''. 

\subsection{Linear response equation}
\label{sec:linear-response-eq}

The linear response equation for semiclassical electrodynamics can be obtained by perturbing Eq.~\eqref{sb} about a background solution to the semiclassical equation with the result
\be
\frac{d^{2}}{dt^{2}}\delta A(t)=-\frac{d}{d t}\delta  E= \delta J_{C}+\delta \langle J_{Q}\rangle \quad . \label{LRE}
\ee
It can be seen from Eq.~\eqref{LRE} that a first integral of the linear response equation gives the perturbed electric field, which is gauge invariant.

To analyze the behaviors of solutions to this equation, particularly at early times, it is useful to break the solutions to the semiclassical backreaction equation into two parts with
\bes \bea E_Q &\equiv& E  - E_C \quad , \label{Eq-def} \\
          E_C &\equiv& - \int_{t_0}^t dt_1 \,J_C(t_1) \quad .  \label{Ec-def}
\eea \ees
From the structure of the linear response equation it is clear that its solutions $\delta E$ can be broken up in exactly the same way.  Then, the criterion for the validity of the semiclassical approximation can be modified to state that if the quantity $\delta E_Q$ grows significantly during some period of time then the semiclassical approximation is invalid.
It is worth noting that because $\la J_Q \ra$ and $\delta \la J_Q \ra$ are constructed from solutions to the mode equation which depend on the vector potential $A$, and therefore indirectly on $E$, then $E_Q$ depends on $E_C$ and $\delta E_Q$ depends on $\delta E_C$.

In Appendix A it is shown for  both the scalar and spin-$\frac{1}{2}$ coupled systems that for homogeneous perturbations, $\delta \la J_{Q}\ra$ depends upon the two-point correlation function for the current. A more general derivation is given in Ref.~\cite{kubostuff}.
For scalar fields the result is
\be
    \delta \la J_{Q} \ra_{\rm ren} = -\frac{q^{2}}{\pi}\delta A(t)\int_{-\infty}^{\infty} dk \, \bigg(|f_{k}(t)|^{2}-\frac{m^{2}}{2\omega^{3}}\bigg) +  \, i \int_{-\infty}^{\infty} dx' \int_{-\infty}^{t} dt' \, \la [J_{Q}(t, x),J_{Q}(t', x')] \ra \, \, \delta A(t') \quad ,  \label{delta-jq-scalar}
\ee
where
\be
    \omega \equiv \sqrt{m^2 + k^2} \quad , \label{omega-def}
\ee
and
\be
    \int_{-\infty}^{\infty} dx' \langle [J_{Q}(t, x),J_{Q}(t', x')] \rangle =\frac{4 \, i \, q^{2}}{\pi}  \int_{-\infty}^{\infty} dk \bigg(k-qA(t)\bigg)\bigg(k-qA(t')\bigg)\textnormal{Im}\bigg\{f_{k}(t)^{2}f_{k}^{*}(t')^{2}\bigg\} \quad .
\ee
It can be shown, using the point-splitting technique, that the divergence structure %introduced by
in the first integral is conveniently compensated for by the divergence structure that is inherent in the second integral. \footnote{It is not obvious that there is a divergence in the second integral because the commutator vanishes in the limit that the points come together.  However, a careful analysis shows it to be there.}  Therefore, $\delta \la J_Q \ra$ is finite and the overall equation is  well defined.

For spin-$\frac{1}{2}$ fields the renormalized perturbation of the quantum current in~\eqref{LRE} is
\be \label{deltajf}
    \delta \la J_{Q} \ra_{\rm ren} = -\frac{q^{2}m^{2}}{2\pi}\delta A(t)\int_{-\infty}^{\infty}\frac{dk}{\omega^{3}}+ \, i\int_{-\infty}^{\infty}dx^{'}\int_{-\infty}^{t}dt^{'} \la [ J_{Q}(t,x) , J_{Q}(t^{'},x^{'}) ] \ra \, \delta A(t^{'}) \quad ,
\ee
with
\be \label{fcomm}
    \int_{-\infty}^{\infty}dx' \la [ J_{Q}(t,x) , J_{Q}(t^{'},x^{'}) ] \ra =\frac{4 i \, q^{2}}{\pi}\int_{-\infty}^{\infty}dk \, \textnormal{Im}\bigg\{h_{k}^{I}(t)h_{k}^{II}(t)h_{k}^{I*}(t')h_{k}^{II*}(t') \bigg\}  \;.
\ee

Recall that in  the massless limit we find that the mode equations decouple and the solutions are given in Eq.~\eqref{masslessmode}. %$h^{I,II}_k(t)=\pm \theta(\mp k)e^{\pm i\int(k-qA)dt}$.
Thus, for a given value of $k$ either $h^I_k$ or $h^{II}_k$ is zero,  and hence $h^I_k h^{II}_k=0$ for any value of $k$.
Therefore in the massless limit the current-current commutator in Eq.~\eqref{fcomm} is zero.

%%%%%%%%%%%%%%%%%%%%%%%%%%%%%%%%%%%%%%%%%%%%%%%%%%%%%%%%%%%%%%%%%%%%%%%%%%%%%%%%%%%%%%%%%%

\subsection{Approximate solutions to the linear response equation}
\label{sec:approx-solutions-LSR}

From Eqs.~\eqref{LRE},~\eqref{delta-jq-scalar}, and~\eqref{deltajf}, it is clear that the linear response equation is an integro-differential equation.  This makes it significantly more difficult to solve numerically compared to an ordinary differential equation.
A useful way to approximate the solutions to the linear response equation for the case of homogeneous perturbations was given in Ref.~\cite{validitypreheating}. It involves solving the semiclassical backreaction equation for two sets of initial conditions which differ from each other by only a small amount. At early times we expect these two solutions to be an approximate solution to the linear response equation so long as the difference does not grow too large. If this difference grows significantly, then the corresponding solution to the linear response equation should also grow substantially. Hence, our criterion for the validity of the semiclassical approximation is considered to be violated.

%As can be seen in Fig.~\ref{figLateTimes}, 
As has been mentioned previously, solutions to the semiclassical backreaction equation tend to oscillate over long periods of time due to plasma oscillations
and it is possible that solutions to the linear response equation could  oscillate over shorter periods of time.  While there is no problem in comparing the absolute difference between two solutions to the semiclassical backreaction equation, it is more problematic when one considers the relative difference because the denominator will vanish at certain points.  For this reason we introduce a modified version of the relative difference which is guaranteed to be no smaller than zero and no larger than one. Consider two solutions to either the classical or semiclassical backreaction equation in 1+1D, $\vec{E}_1 = E_1 \hat{x}$ and $\vec{E}_2 = E_2 \hat{x}$ (or just $E_1$ and $E_2$ since we are only considering one spatial dimension).  Then the absolute and relative differences are respectively
\bes
\bea
    \Delta E &\equiv& E_2 - E_1  \quad , \label{Delta-def} \\ \nonumber \\
    R &\equiv& \frac{|\Delta E|}{|E_1|+|E_2|} \quad . \label{R-def}
\eea   \label{R-Delta-def}
\ees
We note that $R$ can be easily reexpressed as a Lorentz-invariant quantity.
%It is interesting to find a Lorentz-invariant form for $R$. To this end we take advantage of the fact that we have assumed that the electric field is spatially homogeneous. This means that there is an inertial observer, with covariant velocity $u^\mu$ for which $E^\mu= F^{\mu\nu}u_\nu$ takes the form $(0, E(t))$ when $u^\mu=(1,0)$. Furthermore, the tangent vector to the one-dimension spatial surface is $n^\mu=(0,1)$. With these definitions \be R =  \frac{|(F_{\mu\nu})_2 u^\nu n^\mu - (F_{\mu\nu})_1 u^\nu n^\mu |}{|(F_{\mu\nu})_2 u^\nu n^\mu|+|(F_{\mu\nu})_1 u^\nu n^\mu| } \;. \label{RLI-def0} \  \ee

%A similar measure to $R$ that is also Lorentz invariant is \be \bar R = \frac{|(F_{ab})_2 (F^{ab})_2 - (F_{ab})_1 (F^{ab})_1 |}{|(F_{ab})_2(F^{ab})_2|+|(F_{ab})_1 (F^{ab})_1| } \;, \label{RLI-def} \ee with \be F_{ab} F^{ab} = - 2 E^2   \;.  \label{Fab2} \ee It is straight-forward to show that \be \bar R = R + O(R^2)  \;. \label{RLI-R} \eeTherefore since $R$ is a simpler quantity, we will use it to analyze the issue of the validity of the semiclassical approximation.

It is useful to apply the relative difference $R$ for two solutions to the classical backreaction equation which, as can be seen in Eq.~\eqref{Ec-def}, are simply integrals over the classical current $J_C$.
Consider a classical current of the form
\be
    J_C=-E_0 \, \dot{g}(t) \quad . \label{Jc-g}
\ee
Here $\dot{g}(t)$ is the time derivative of some well-behaved, dimensionless function $g(t)$, and the solution to the classical Maxwell equation is $E_{C} = E_0 g(t)$.  In the following sections we will consider the cases $g(t) = \frac{qt}{1+qt}$ and
$g(t) = \sech^2(qt)$, with the latter being the Sauter pulse. %and $t$ is a dimensionless time coordinate.
The solutions are parametrized by the constant $E_0$. For two solutions to Eqs.~\eqref{Ec-def} with~\eqref{Jc-g},  $E_{C1}$ and $E_{C2}$, with $E_{0}=E_{01}$ and $E_{0}=E_{02}$ respectively, we have for the absolute and relative difference
\bes
\bea
    \Delta E_{0} &\equiv& E_{02}-E_{01} \label{Delta-E0} \quad , \\ \nonumber \\
    R_C &=& \frac{| \Delta E_{C} |}{|E_{C1}| + |E_{C2}|} = \frac{| \Delta E_{0} |}{|E_{01}| + |E_{02}|} \quad . \label{Rc}
\eea  \label{sahh}
\ees

Next, consider two solutions to the semiclassical backreaction equation.  Since we are considering classical currents, which are zero initially, and an electric field that is zero initially, there is no ambiguity in the choice of vacuum state. Therefore these solutions are also parametrized by the value of $E_0$ for a given function $g(t)$. Using the subscripts $1$ and $2$ to denote quantities computed for these solutions, it is clear that the difference $\Delta E$ is an exact solution to the equation
\be
    -\frac{d \Delta E}{d t}=\Delta J_{C} + \Delta \langle J_{Q} \rangle \quad .  \label{Delta-E}
\ee
with $\Delta J_C = J_{C2} - J_{C1}$ and $\Delta \langle J_{Q} \rangle = \langle J_{Q2} \rangle - \langle J_{Q1} \rangle $.

Suppose at some early time $t_1$, when $E_C$ is still very small with no significant amount of particle production, that
$ R_C(t_1) \ll 1 $. One can then arrange the initial conditions for the perturbation $\delta E$ such that $\delta E(t_1) = \Delta E(t_1)$.  It is also obvious that one can set for all times $\delta J_C(t) = \Delta J_C(t)$.  Then Eq.~\eqref{Delta-E} is approximately equivalent to the linear response equation \eqref{LRE} so long as $\Delta \langle J_{Q} \rangle \approx \delta \langle J_Q \rangle$, which one would certainly expect to be the case at times near $t_1$.

As discussed in the previous subsection [see Eq.~\eqref{Eq-def}], it is more useful at early times to consider the quantity $\Delta E_Q \approx \delta E_Q$. To measure the relative growth of $\Delta E_Q$ we compute the relative difference
\be R_Q = \frac{|\Delta E_Q|}{|E_{Q1}|+|E_{Q2}|}  \;. \label{Rq-def} \ee
 This difference can then be compared to the relative difference between the corresponding classical solutions $R_C$ in Eq.~\eqref{Rc}, which does not change in time.

Consider two times $t_2 > t_1$ where $t_1$ is the initial time discussed above when one imagines fixing the starting values for the linear response equation and $t_2$ is a relatively early time after that.  Then the possibilities are as follows. 
 (i) If $R_Q(t) \lesssim R_C$
            then the criterion for the validity of the semiclassical approximation will be satisfied by the approximate homogeneous solutions that we consider up to the time $t_2$.
            (ii) If for any times between $t_1$ and $t_2$, $R_Q(t) \gg R_C$,
             then the solution to the linear response equation, $\delta E$, grows rapidly during at least some part of the period $t_1 \le t \le t_2$  and the criterion for validity of the semiclassical approximation is not satisfied.
            Note that once the semiclassical approximation has broken down, one can no longer trust its solutions even if for later times $R_Q \lesssim R_C$. (iii) Finally, the intermediate case when $R_Q$ is larger than $R_C$ but still of the same order of magnitude is ambiguous. Perhaps the best that can be said is in this case quantum fluctuations are increasing and so the accuracy of the semiclassical approximation is decreasing in proportion to this increase.

\section{Numerical Results}
\label{sec:numerical}
\begin{figure}[htbp]
\begin{center}
\begin{tabular}{c}
\includegraphics[width=70mm]{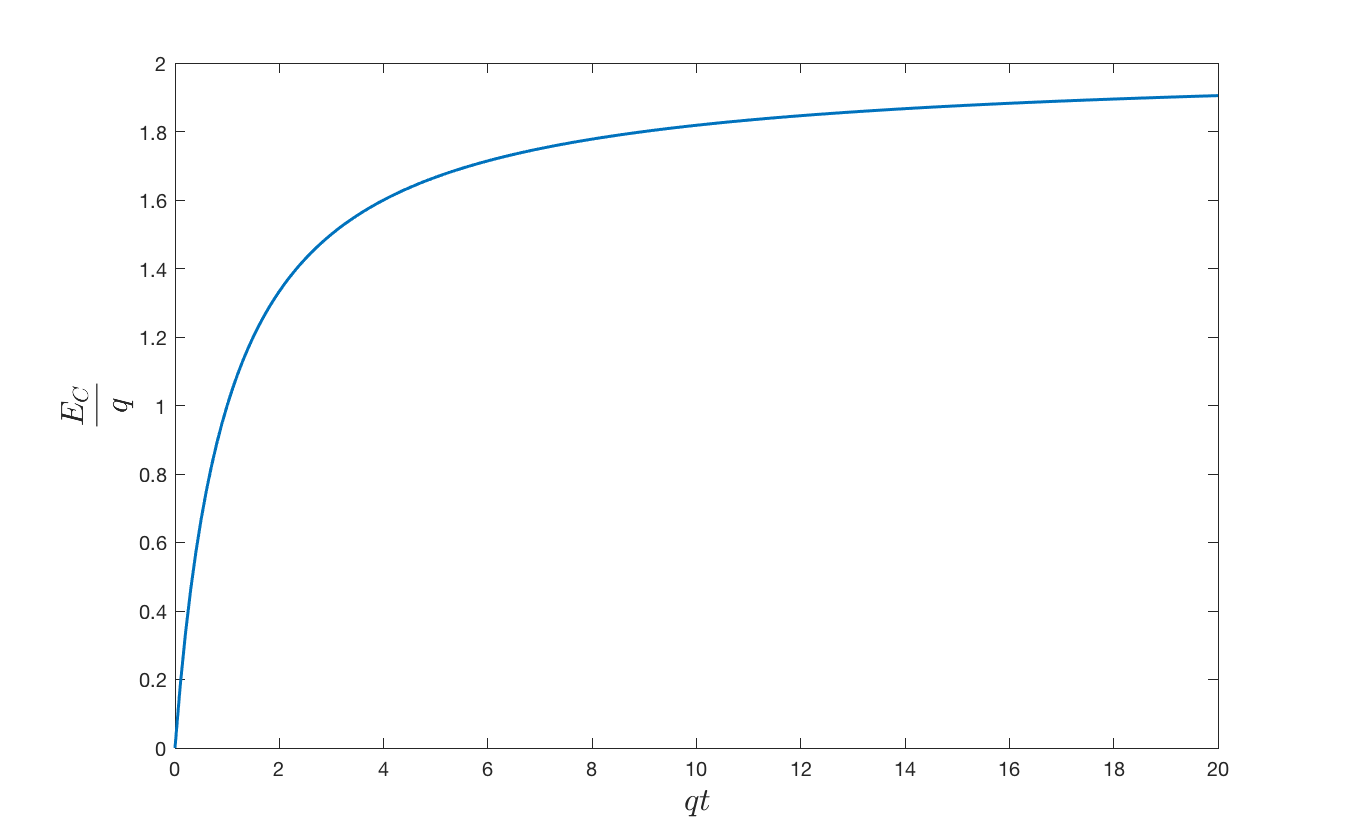}
\includegraphics[width=70mm]{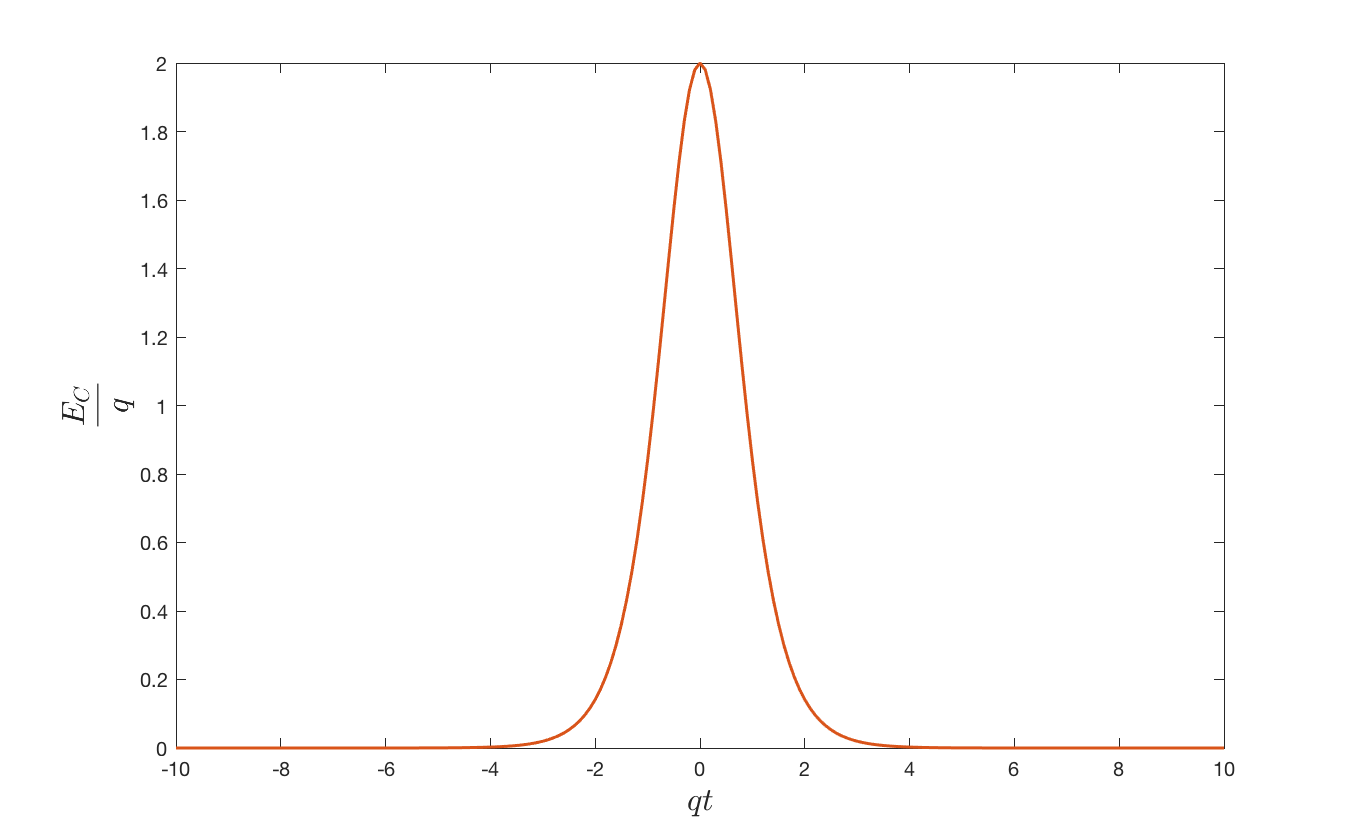}\\
\includegraphics[width=70mm]{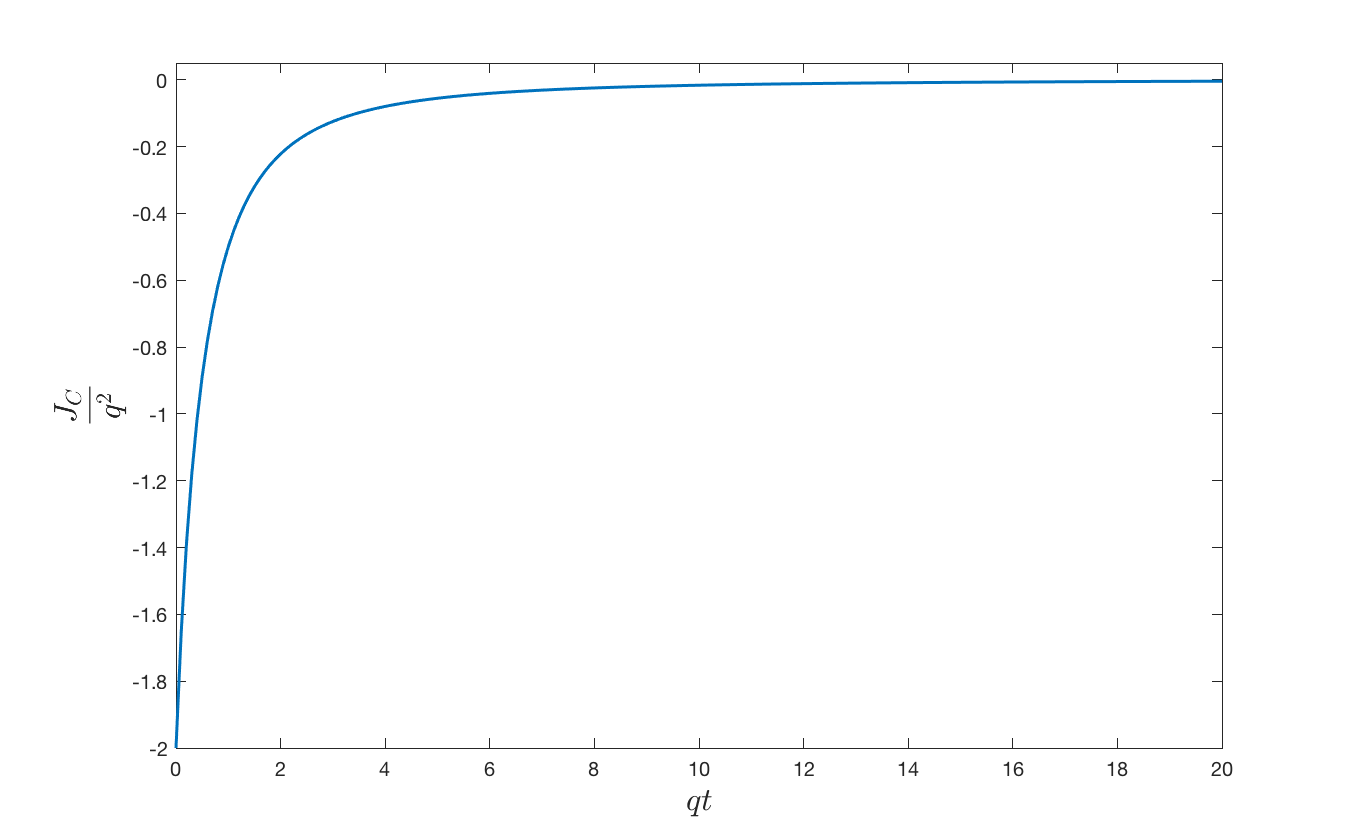}
\includegraphics[width=70mm]{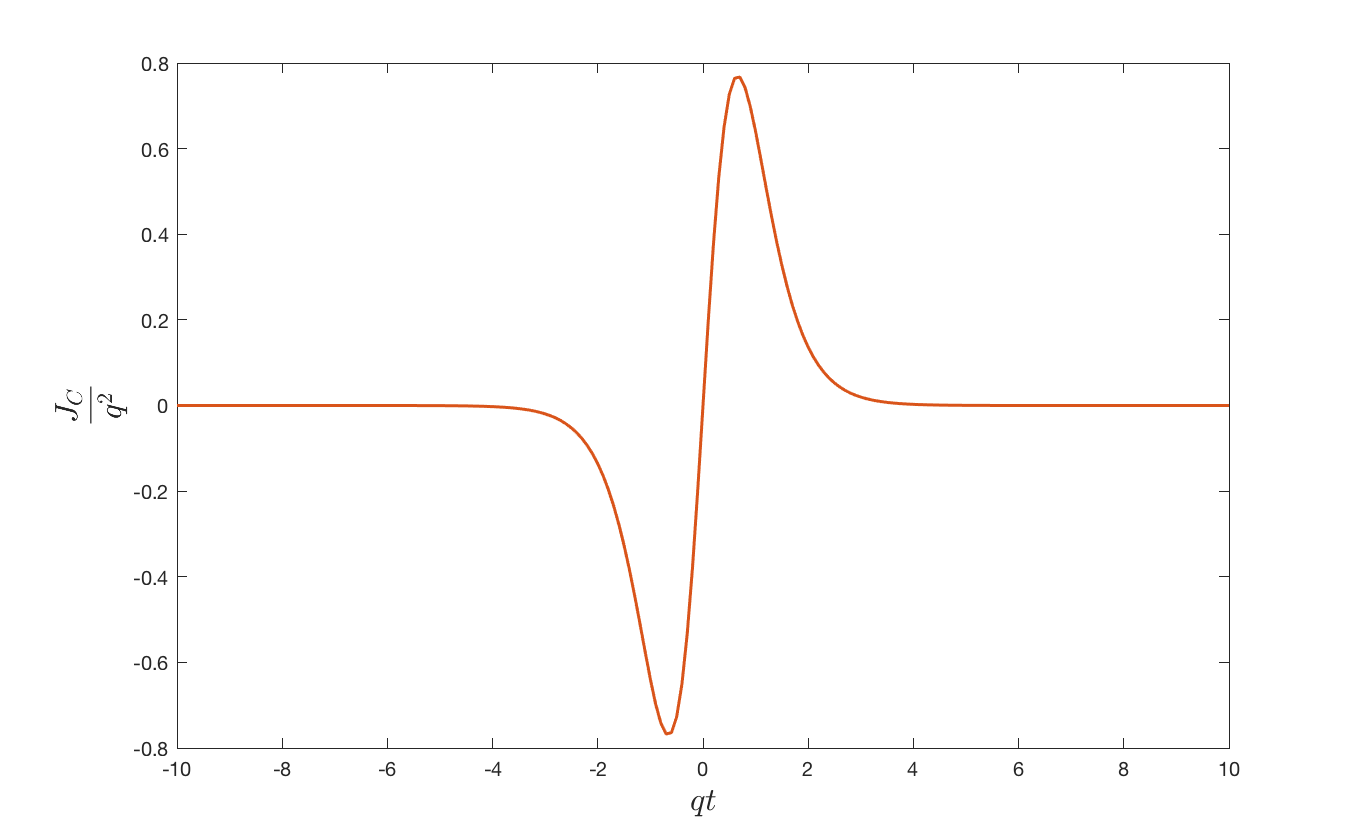}
\end{tabular}
\end{center}
\caption{\small{Electric profiles for $E_0/q=2$. In the left (top) panel we show the asymptotically constant profile. In the right (top) panel we show the Sauter pulse. In both bottom panels, the classical current generating the respective electric field profiles is plotted. For the asymptotically constant profile we choose an initial time $t_0=0$, while for the Sauter pulse the initial time has to be fixed as $t_0 = -\infty$.}}
\label{electric_profiles2.0}
\end{figure}

In this section we implement a numerical analysis to study the validity of the semiclassical approximation for two different classical source profiles. To do so, we
use the method described in the previous section to compare the numerical solutions of the semiclassical backreaction equation for two distinct, but very close values of the external source amplitude $E_0$.
The first profile considered has a classical source current given by
\be
     J_{C}=-\frac{ q E_{0}}{(1+qt)^{2}} \quad , \label{Jclass}
\ee
for $t\geq 0$ and $J_{C}=0$ for $t<0$.  The classical solution of the Maxwell equation ($-\dot E_C=J_C$) gives rise to the asymptotically constant electric field profile for $t\geq 0$
\be
E_{C}(t)=E_{0}\left(\frac{qt}{1+qt}\right) \quad . \label{Eclass}
\ee
The second profile considered is the Sauter pulse with source current given by
\be
    J_{C} = 2qE_{0}\textnormal{sech}^2(qt)\textnormal{tanh}(qt) \quad , \label{Sautercurrent}
\ee
and corresponding classical electric field
\be
    E_{C}(t)=E_{0}\textnormal{sech}^{2}(qt) \quad . \label{SauterE}
\ee
In Fig.~\ref{electric_profiles2.0} we show the classical behavior of both profiles. For the first profile, one can easily see that at late times the electric field approaches the constant value $E_0$. The Sauter pulse models a possibly more realistic scenario for the detection of the Schwinger effect, in which both the initial and the final values of the classical electric field tend to zero. Note that, for the first profile we choose an initial time $t_0=0$, while for the Sauter pulse the initial time has to be fixed as $t_0 = -\infty$.
As discussed in Sec.~\ref{sec:approx-solutions-LSR}, it is useful, particularly at early times, to work with the quantity $E_Q$ in Eq.~\eqref{Eq-def} which is the difference between the net electric field and the electric field $E_C$ that would be present if there were no quantum effects.  Therefore the natural quantity to consider is the relative difference $R_Q$ in Eq.~\eqref{Rq-def} which is constructed from two solutions to the semiclassical backreaction equation with values of $E_0$ that differ by some small amount.  This can be compared to the relative difference $R_C$ between two solutions  to the classical Maxwell equation with the same values of $E_0$.

In what follows, numerical results will be shown for calculations of $R_Q$ and other quantities such as $E(t)$, $\langle J_Q \rangle$, and $\langle N \rangle$ for scalar and spin-$\frac{1}{2}$ semiclassical electrodynamics for the asymptotically constant classical profile and then for the Sauter pulse classical profile. As stressed before, we mainly focus on the early-time behavior. In both cases it is assumed that the electric field and vector potential are initially zero.  As a result, for scalar fields the initial conditions for the mode functions are
\be
    f_k(t_0) = \frac{1}{\sqrt{2\omega}} \quad , \quad \dot{f}_k(t_0) = -i\sqrt{\frac{\omega}{2}} \quad .
\ee
For spin-$\frac{1}{2}$ fields the initial conditions are
\be
    h_k^{I}(t_0) = \sqrt{\frac{\omega-k}{2\omega}} \quad , \quad h_k^{II}(t_0) = -\sqrt{\frac{\omega+k}{2\omega}} \quad .
\ee

First, we discuss the mass dependence of the function $R_Q$ and its relation to the validity of the semiclassical approximation, with a focus on the asymptotically constant profile. Then, we show the results of our analysis for the most relevant case $E_0\sim E_{\rm crit}=m^2/q$ for both the asymptotically constant profile and the Sauter pulse.
As in Sec.~\ref{sec:energy}, for the numerical computations we use the dimensionless parameters %in~\eqref{scaled-parameters}
described therein.  However, in this section the electric field and the electric current are given in terms of $E/q$ and $J/q^2$ respectively.

Since we are considering multiple cases and subcases, a summary of all relevant information, including all cases and sub-cases with figure references, can be found in Table~\ref{tab:referencetable}.
\begin{table}[]
\begin{tabular}{|c|c|c|c|}
\hline
\textbf{Quantum Field} & \textbf{Classical Profile} & \textbf{Mass Cases} & \textbf{Figure Reference} \\ \hline
\multirow{4}{*}{Spin 1/2} & \multirow{2}{*}{\begin{tabular}[c]{@{}c@{}}Asymptotically\\ Constant\end{tabular}} & $m^2 \ll q E_0 \, \, (\textnormal{or} \, \, m\rightarrow 0)$ & 6, 7                     \\ \cline{3-4}
                               &                                                                                    & $m^2 \sim q E_0$                    & 5, 6, 7, 9               \\  \cline{2-4}
                                & \multirow{2}{*}{Sauter Pulse}                                                    & $m^2 \ll q E_0 (\textnormal{or} \, \, m\rightarrow 0)$ & N/A                       \\ \cline{3-4}
                                &                                                                                    & $m^2 \sim q E_0 $                   & 10                        \\ \hline
\multirow{4}{*}{Complex Scalar} & \multirow{2}{*}{\begin{tabular}[c]{@{}c@{}}Asymptotically\\ Constant\end{tabular}} & $m^2 \ll q E_0 \, \, (\textnormal{or} \, \, m\rightarrow 0)$ & 8                        \\ \cline{3-4}
                                &                                                                                    & $m^2 \sim q E_0 $                   & 8, 9
                            \\ \cline{2-4}
                               & \multirow{2}{*}{Sauter Pulse}                                                   & $m^2 \ll q E_0 \, \, (\textnormal{or} \, \, m\rightarrow 0)$ & N/A                       \\ \cline{3-4}
                               &                                                                                    & $m^2 \sim q E_0$                    & 10                        \\   \hline
\end{tabular}
\caption{A table organizing the various cases and subcases that are investigated in the paper. Included are figure references for ease of use. Note that cases with $m^2 \gg qE_0$ are not included; they are discussed in the main text on the basis of the decoupling mechanism.}
\label{tab:referencetable}
\end{table}

\subsection{Asymptotically constant classical profile}
\label{sec:validity-asym-class-profile}

\subsubsection{Massless spin-$\frac{1}{2}$ field}

As explained in Sec.~\ref{masslessLim}, for $m=0$ the mode equations \eqref{modeh1} and \eqref{modeh2} decouple, and
$\langle J_Q\rangle_{\rm ren}=-\frac{q^2}{\pi}A $. Thus the semiclassical Maxwell equation~\eqref{sb} reduces to
\bea \label{maxmassles}
\ddot A+\frac{q^2}{\pi}A=J_C \quad ,
\eea
which is the equation for a simple harmonic oscillator with frequency $\frac{|q|}{\sqrt{\pi}}$ and external source $J_C$.
In this case, the linear response equation is just
\bea
    \delta \ddot A+\frac{q^2}{\pi}\delta A=\delta J_C \quad .
\eea
Note that $\delta \langle J_Q \rangle_{\rm ren}=-\frac{q^2}{\pi}\delta A$ and also that the initial conditions for $\delta J_C$ can be arranged so that $\delta J_C \equiv \Delta J_C$.

For the asymptotically constant profile, $J_C$ is given in Eq.~\eqref{Jclass}.  With initial conditions $A(0)=0$ and $E(0)=0$, we immediately find
\bea
    A(t) &=& -\frac{E_0}{q}\Bigg[\cos(\frac{1+qt}{\sqrt{\pi}})\mathrm{Ci}(\pi^{-1/2})- \cos(\frac{1+qt}{\sqrt{\pi}})\mathrm{Ci}\Bigg(\frac{1+qt}{\sqrt{\pi}}\Bigg) \nonumber \\
    && \qquad \qquad + \sqrt{\pi}\sin(\frac{qt}{\sqrt \pi})+\sin(\frac{1+qt}{\sqrt{\pi}})\mathrm{Si}(\pi^{-1/2}) - \sin(\frac{1+qt}{\sqrt{\pi}})\mathrm{Si}\Bigg(\frac{1+qt}{\sqrt{\pi}}\Bigg) \Bigg] \quad , \nonumber \\  \label{A-soln-m-0}
\eea
where $\mathrm{Ci}(x)=-\int_{x}^{\infty}\frac{\cos(t)}{t} dt$ and $\mathrm{Si}(x)=\int_{0}^{x}\frac{\sin(t)}{t} dt$ are the cosine and the sine integral functions respectively.  Hence, we can conclude that for any two solutions $E_1(t)$ and $E_2(t)$ with $E_0 = E_{01}$ and $E_0 = E_{02}$ respectively, the relation
\bea \label{Rqmasslesf}
    R_Q(t)=\frac{|E_{Q1}(t)-E_{Q2}(t)|}{|E_{Q1}(t)|+|E_{Q2}(t)|}=\frac{|E_{01}-E_{02}|}{|E_{01}|+|E_{02}|}=R_C \quad ,
\eea
is always satisfied. Although this result was derived for the asymptotically constant profile~\eqref{Jclass}, it holds for any classical current of the form $J_C=-E_0 g(t)$.

\subsubsection{Massive spin-$\frac{1}{2}$ field}

We next study the relationship between the behavior of $R_Q$, the mass of the spin $\frac{1}{2}$ field, and the value of $E_0$ in Eq.~\eqref{Jclass}.  As illustrated in our numerical results below, the most important effect on $R_Q$ comes from the size of the dimensionless quantity $\frac{q E_0}{m^2}$.  We distinguish between three different cases: (i)   $\frac{q E_0}{m^2} \gg 1$ in which the mass is relatively small compared to the electric field and there is a lot of particle production,  (ii) the intermediate case $\frac{q E_0}{m^2} \sim 1$ where there is a significant amount of particle production, and (iii) $\frac{q E_0}{m^2} \ll 1$ in which the mass is relatively large compared to the electric field and there is very little particle production.

\begin{figure}[htbp]
\begin{center}
\begin{tabular}{c}
\hspace{-1.9cm}
\includegraphics[width=70mm]{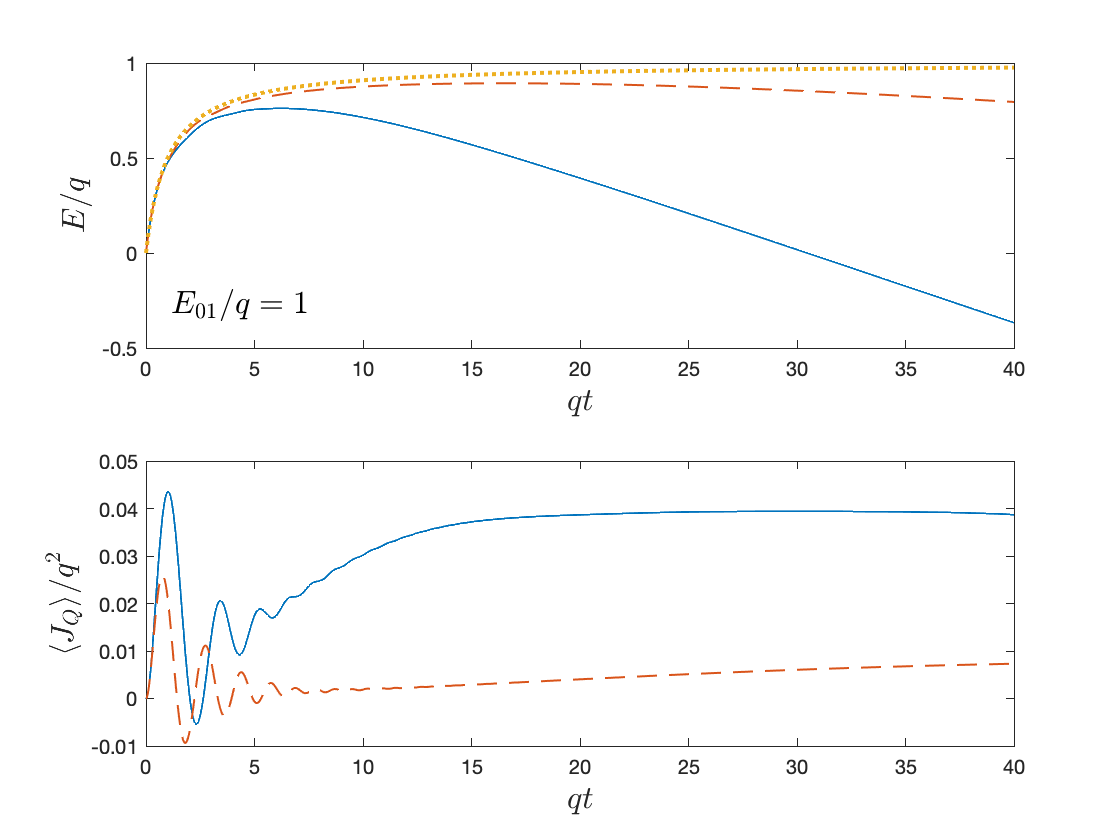}\hspace{-0.7cm}
\includegraphics[width=70mm]{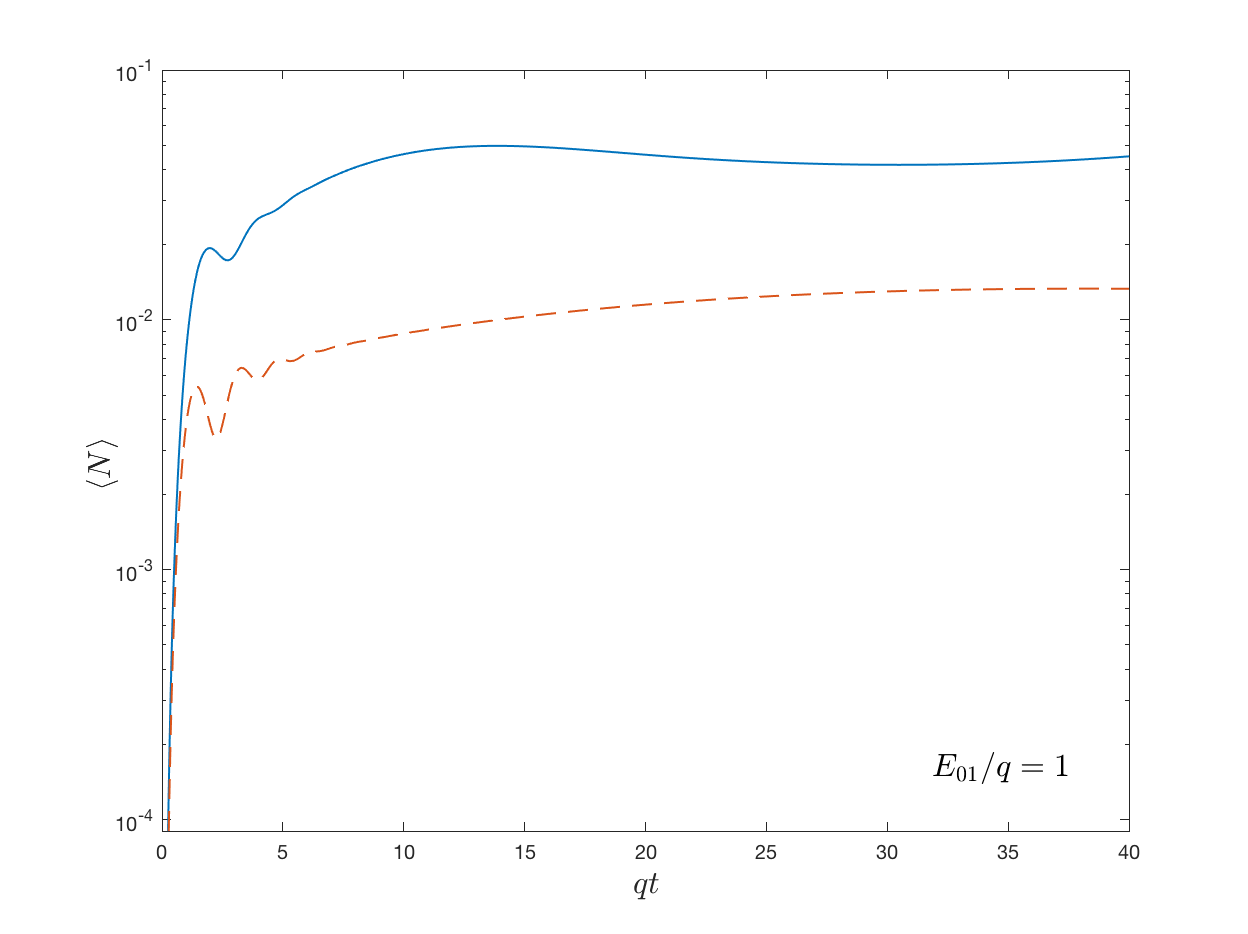}\hspace{-0.7cm}
\includegraphics[width=70mm]{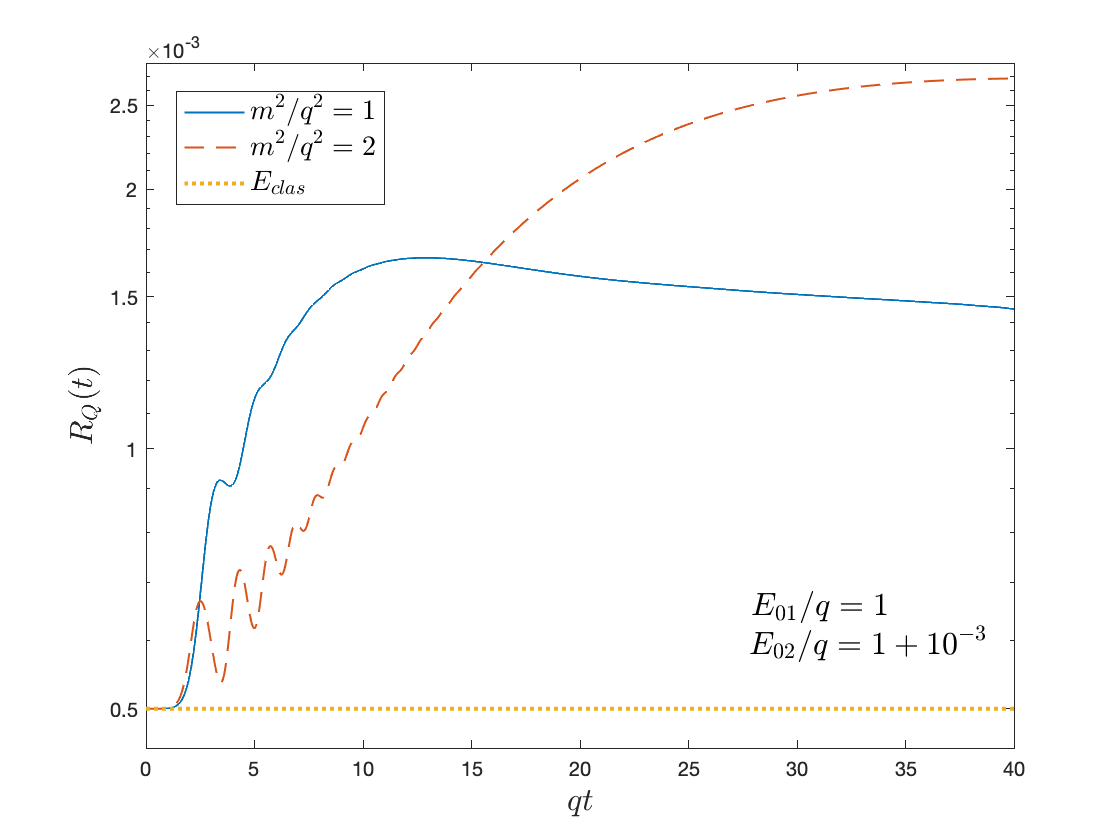}
\end{tabular}
\end{center}
\caption{\small{Results obtained from numerical solutions to the semiclassical backreaction equation for spin-$\frac{1}{2}$ fields and the asymptotically constant classical profile are shown for $E_0/q = 1$.  The masses are chosen so that $\frac{q E_0}{m^2}\le 1$. The electric field and the induced electric current $\langle J_Q \rangle/q^2$ for each case are plotted in the left panels. Plots for the corresponding number of particles, $\la N \ra$, are shown in the middle panel and plots of the quantity $R_Q$ appear in the right panel.  For the latter, the values $E_{01}/q=1$ and $E_{02}/q=1+10^{-3}$ have been chosen for the two solutions that are subtracted.  The values of $m^2/q^2$ for each case are shown along with the type of curve for that solution in the legend in the right panel.  }}
\label{fig:large-m-fermion}
\end{figure}

The beginning of the transition from intermediate to large effective masses is shown in Fig.~\ref{fig:large-m-fermion} where various quantities, such as the electric field, are plotted for $E_0/q=1$ and  $\frac{m^2}{q^2}=1$ and $\frac{m^2}{q^2}=2$.  As expected, the amount of particle production that occurs decreases significantly as $\frac{q E_0}{m^2}$ decreases and thus as the effective mass increases.  Note that the time scale on which backreaction effects occur increases significantly with an increase in the effective mass.

In the very-large-mass limit $\frac{q E_0}{m^2} \to 0$, the electric field will not have enough energy to create particles, so one expects that $\langle J_Q \rangle_{\rm ren} \to 0$ and $ E \to E_C$.  This is in agreement with the decoupling theorem in perturbative quantum field theory \cite{APtheorem}. Heavy masses decouple in the low-energy description of the theory, which in this case is purely classical electrodynamics for $m^2 \to \infty$, with $E_0$ fixed.

In the intermediate cases shown in Fig.~\ref{fig:large-m-fermion} where $\frac{q E_0}{m^2} \sim 1$, there is a significant amount of particle production and once enough particle production has occurred the value of $R_Q$ starts to increase rapidly, possibly exponentially for $\frac{q E_0}{m^2} = 1$.  This rapid rise continues until the backreaction of the particles on the background electric field is strong enough that the electric field has stopped increasing and has begun to noticeably decrease in size.  Thus in the intermediate case it appears that our criterion for the validity of the semiclassical approximation is not satisfied due the rapid and significant growth in $R_Q$ at relatively early times.

\begin{figure}[htbp]
\begin{center}
\begin{tabular}{c}
\hspace{-1.9cm}
\includegraphics[width=70mm]{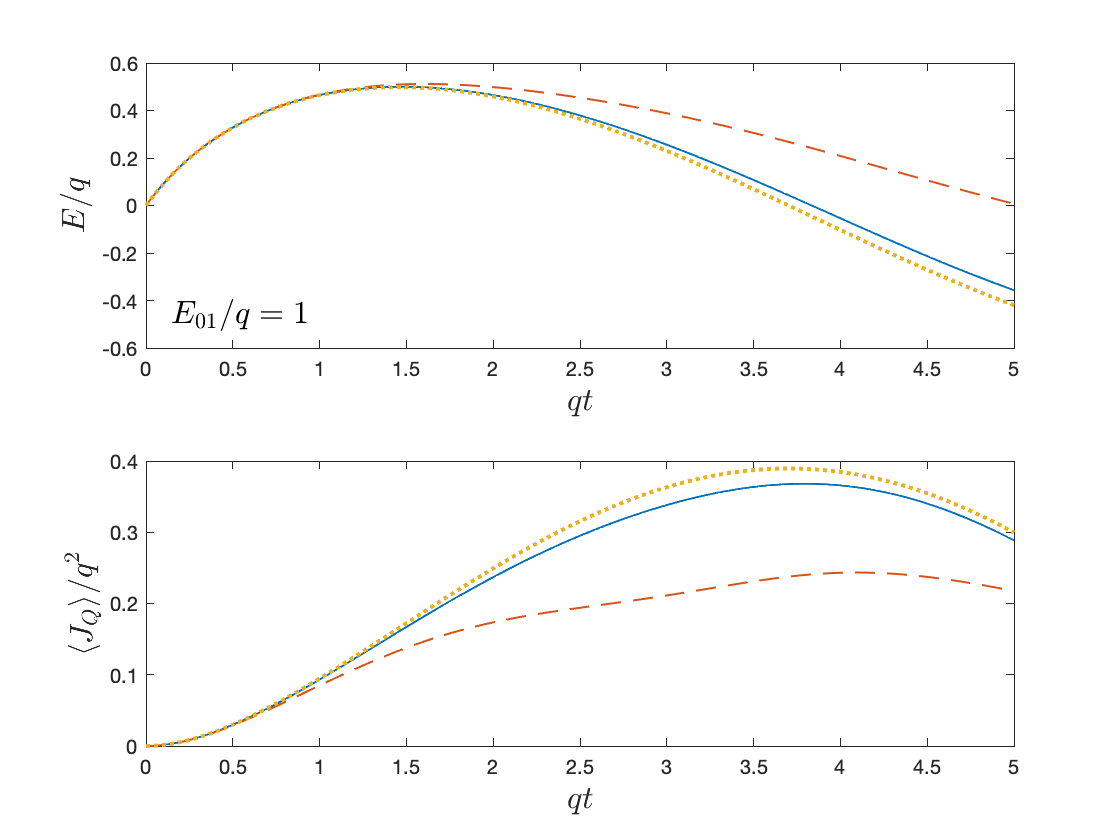}\hspace{-0.7cm}
\includegraphics[width=70mm]{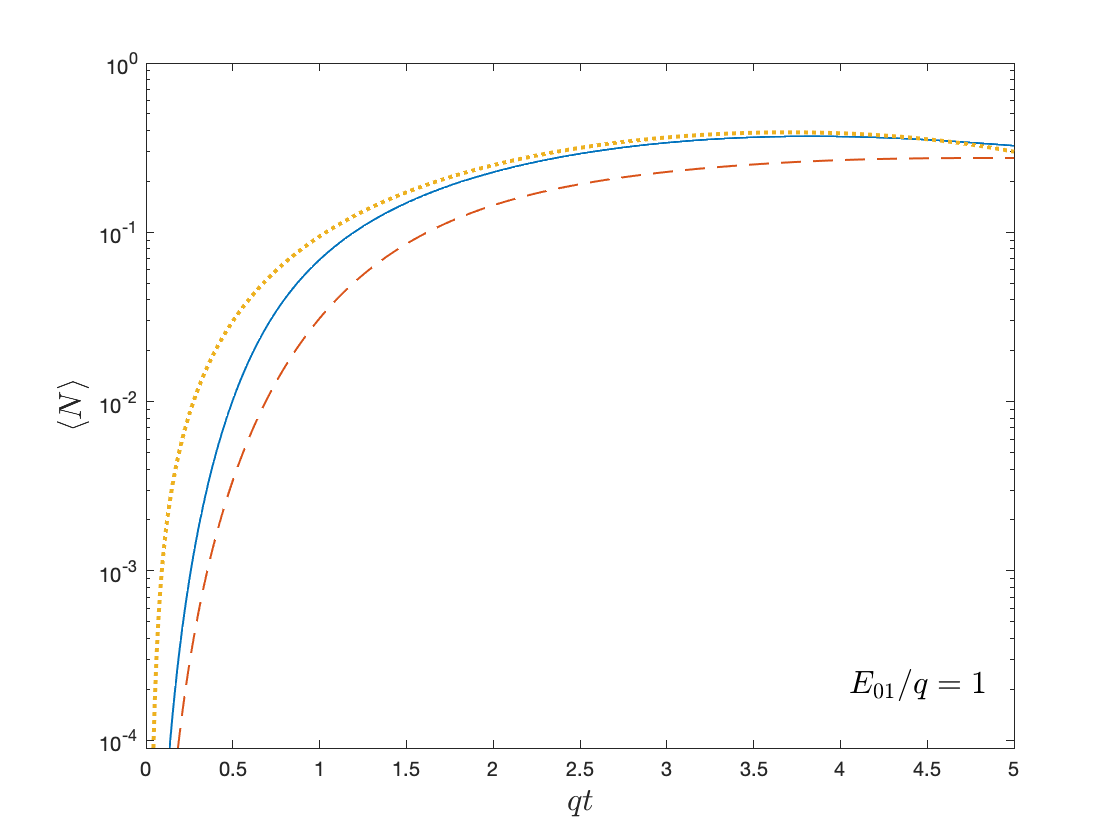}\hspace{-0.7cm}
\includegraphics[width=70mm]{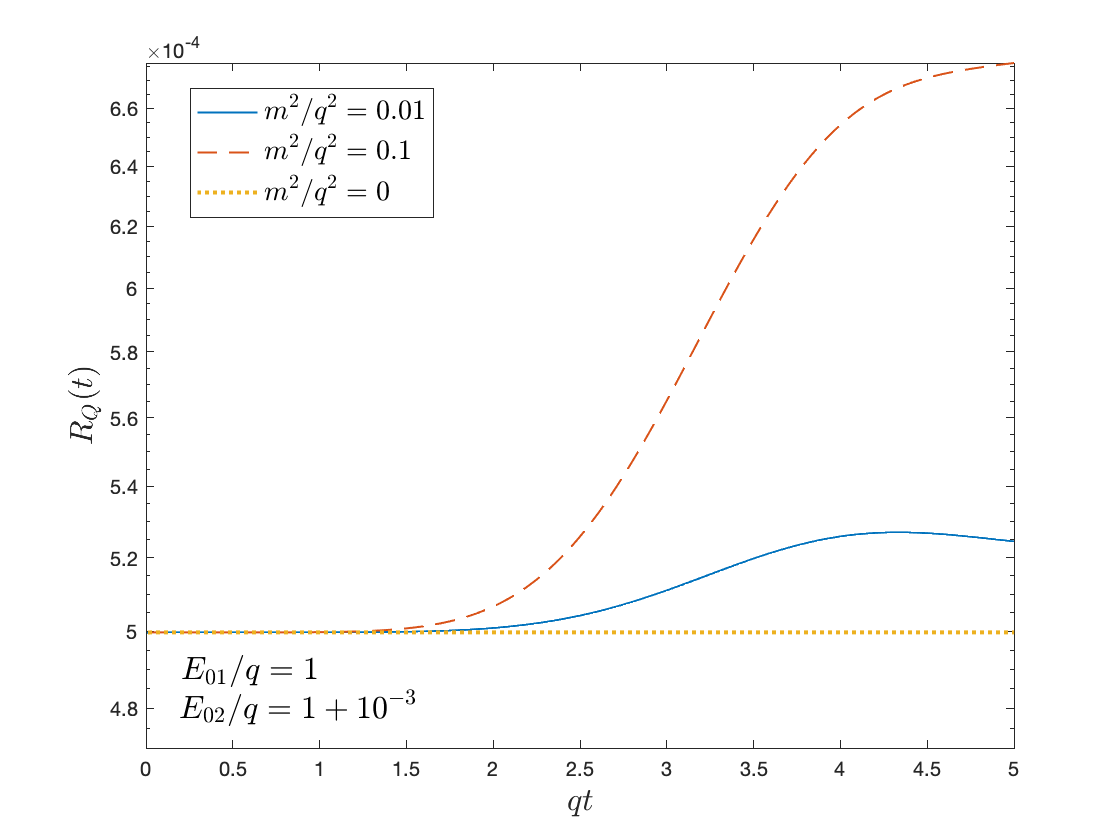}
\end{tabular}
\end{center}
\caption{\small{Results obtained from numerical solutions to the semiclassical backreaction equation for spin-$\frac{1}{2}$ fields and the asymptotically constant classical profile are shown for $E_0/q = 1$.  The masses are chosen so that $\frac{q E_0}{m^2} \ge 1$. he structure of the figure is the same as in Figure \ref{fig:large-m-fermion}. %The electric field and the induced electric current $\langle J_Q \rangle/q^2$ for each case are plotted in the left panels. Plots for the corresponding number of particles, $\la N \ra$, are shown in the middle panel and plots of the quantity $R_Q$ appear in the right panel.  For the latter, the values $E_{01}/q=1$ and $E_{02}/q=1+10^{-3}$ have been chosen for the two solutions that are subtracted.  The values of $m^2/q^2$ for each case are shown along with the type of curve for that solution in the legend in the right panel. 
}}
\label{fig:small-m-fermion}
\end{figure}

The transition from the intermediate case to the small-effective-mass case when $E_0/q = 1$ is shown in~Fig.~\ref{fig:small-m-fermion}. Comparison with Fig.~\ref{fig:large-m-fermion} shows that the intermediate case extends to
$\frac{m^2}{q^2}=0.1$, but not to $\frac{m^2}{q^2} = 0.01$ which has a qualitatively different behavior.  In particular for the relatively small-mass and zero-mass cases the particle production is more rapid and backreaction effects on the electric field are significant after a much smaller amount of time than for intermediate masses.  Examination of the behavior of $R_Q$ shows that it does not grow rapidly in time for the small-mass case and, as mentioned above, is constant in the massless case.  Thus our criterion for the validity of the semiclassical approximation is satisfied by the homogeneous approximate solutions that we consider in the relatively small-mass case.

\begin{figure}[htbp]
\begin{center}
\begin{tabular}{c}
\hspace{-1.9cm}
\includegraphics[width=70mm]{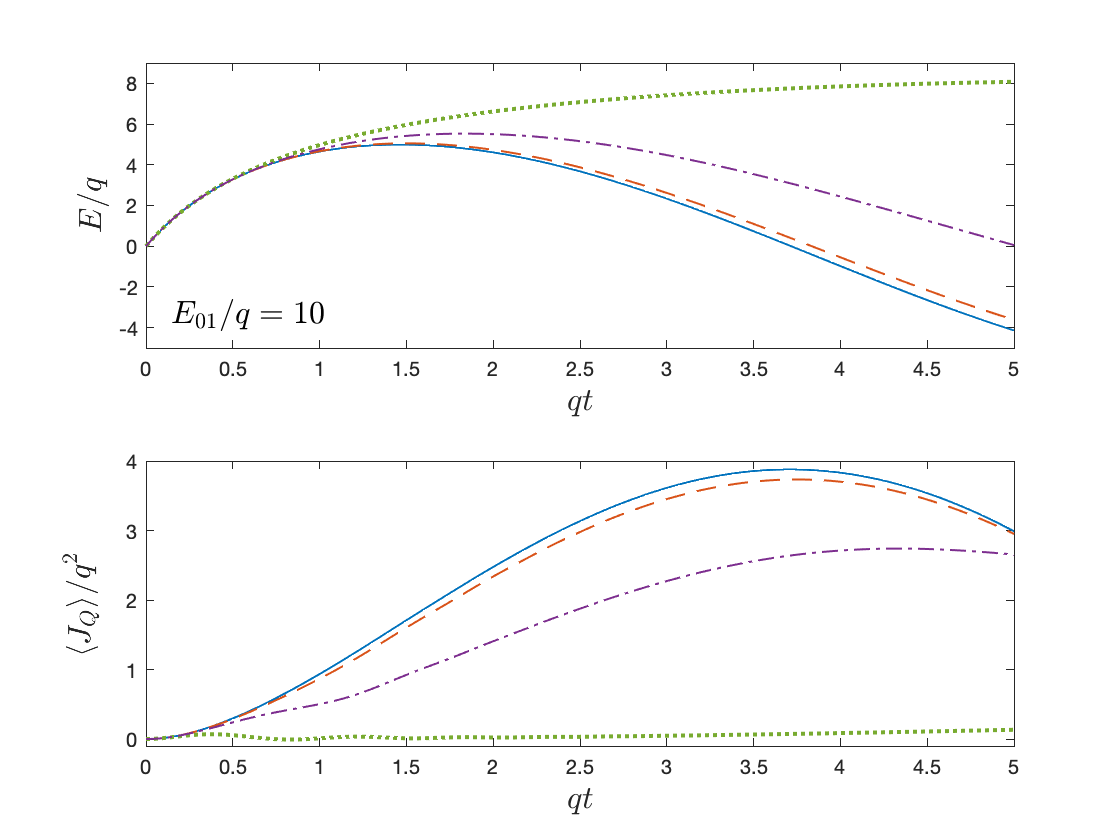}\hspace{-0.7cm}
\includegraphics[width=70mm]{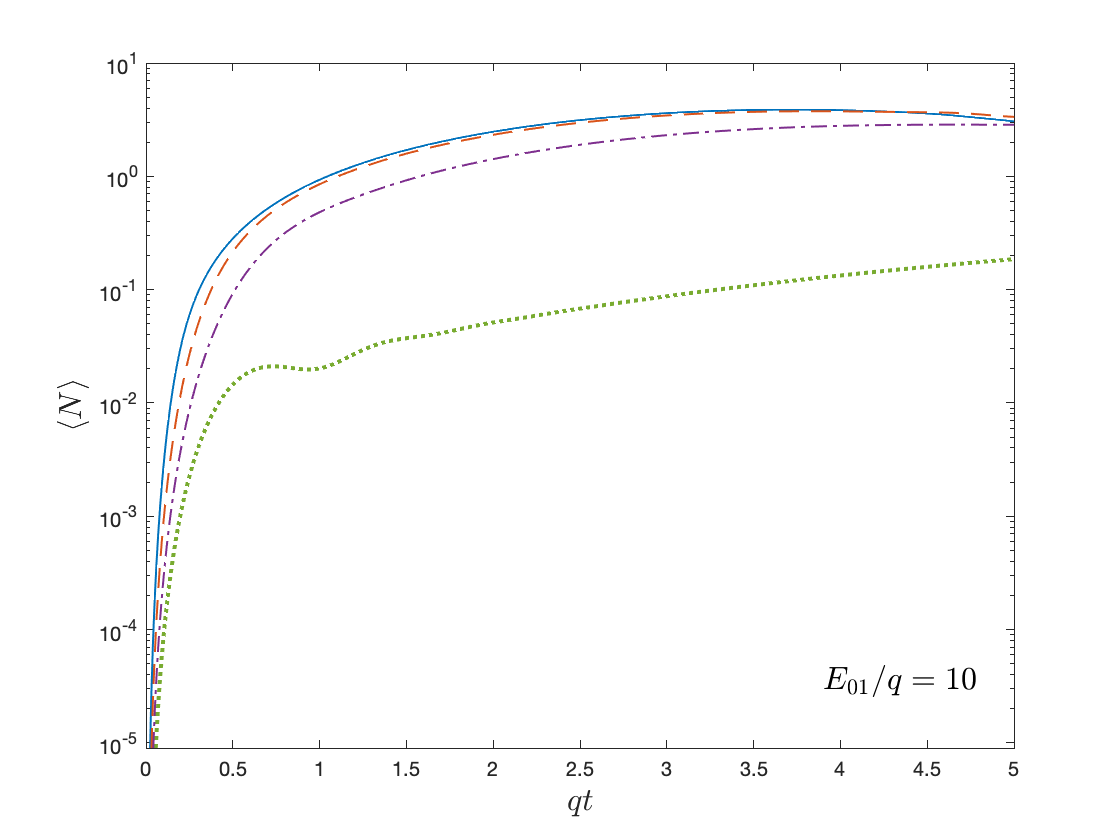}\hspace{-0.7cm}
\includegraphics[width=70mm]{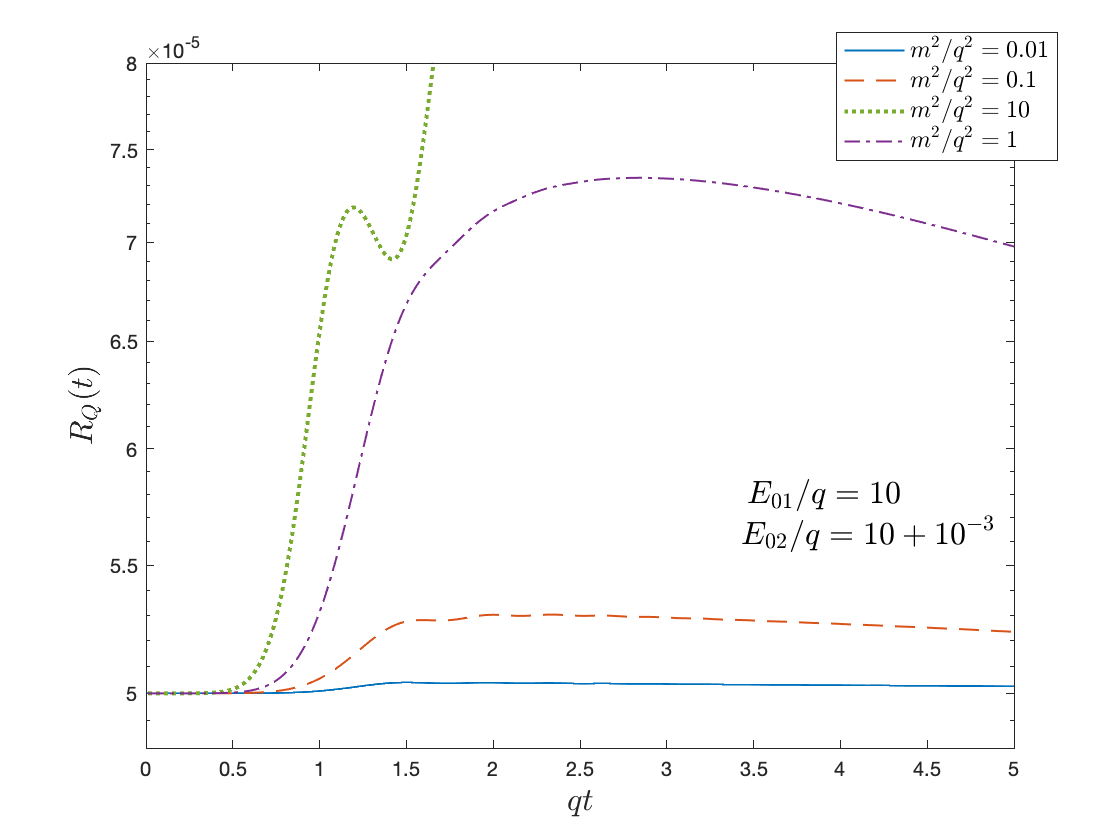}
\end{tabular}
\end{center}
\caption{\small{Results obtained from numerical solutions to the semiclassical backreaction equation for spin-$\frac{1}{2}$ fields and the asymptotically constant classical profile are shown for $E_0/q = 10$.  The masses are chosen so that $\frac{q E_0}{m^2} > 1$. The structure of the figure is the same as in Figure \ref{fig:large-m-fermion}. % The induced electric field and the induced electric current $\langle J_Q \rangle/q^2$ for each case are plotted in the left panels. Plots for the corresponding number of particles, $\la N \ra$, are shown in the middle panel and plots of the quantity $R_Q$ appear in the right panel.  For the latter, the values $E_{01}/q=10$ and $E_{02}/q=10+10^{-3}$ have been chosen for the two solutions that are subtracted.  The values of $m^2/q^2$ for each case are shown along with the type of curve for that solution in the legend in the right panel.
Here, the values $E_{01}/q=10$ and $E_{02}/q=10+10^{-3}$ have been chosen for the representation of the function $R_Q$. }}
\label{fig:small-m-fermion10}
\end{figure}

 In the above analysis the value of the ratio $\frac{q E_0}{m^2}$ has been shown to dictate the different types of qualitative behaviors the solutions have.  Of course one can change the values of $q E_0$ and $m^2$ in ways that keep the ratio fixed.  In Figs.~\ref{fig:large-m-fermion} and~\ref{fig:small-m-fermion}, $E_0/q = 1$.  In Fig.~\ref{fig:small-m-fermion10}, $E_0/q = 10$ is chosen along with several masses that lead to small and intermediate values of  $\frac{q E_0}{m^2}$.  Comparison with Fig.~\ref{fig:small-m-fermion} shows that while the details of the various curves are different, they are qualitatively the same when the ratio $\frac{q E_0}{m^2}$ is the same.

\subsubsection{Scalar field}

\begin{figure}[htbp]
\begin{center}
\begin{tabular}{c}
\hspace{-1.9cm}
\includegraphics[width=70mm]{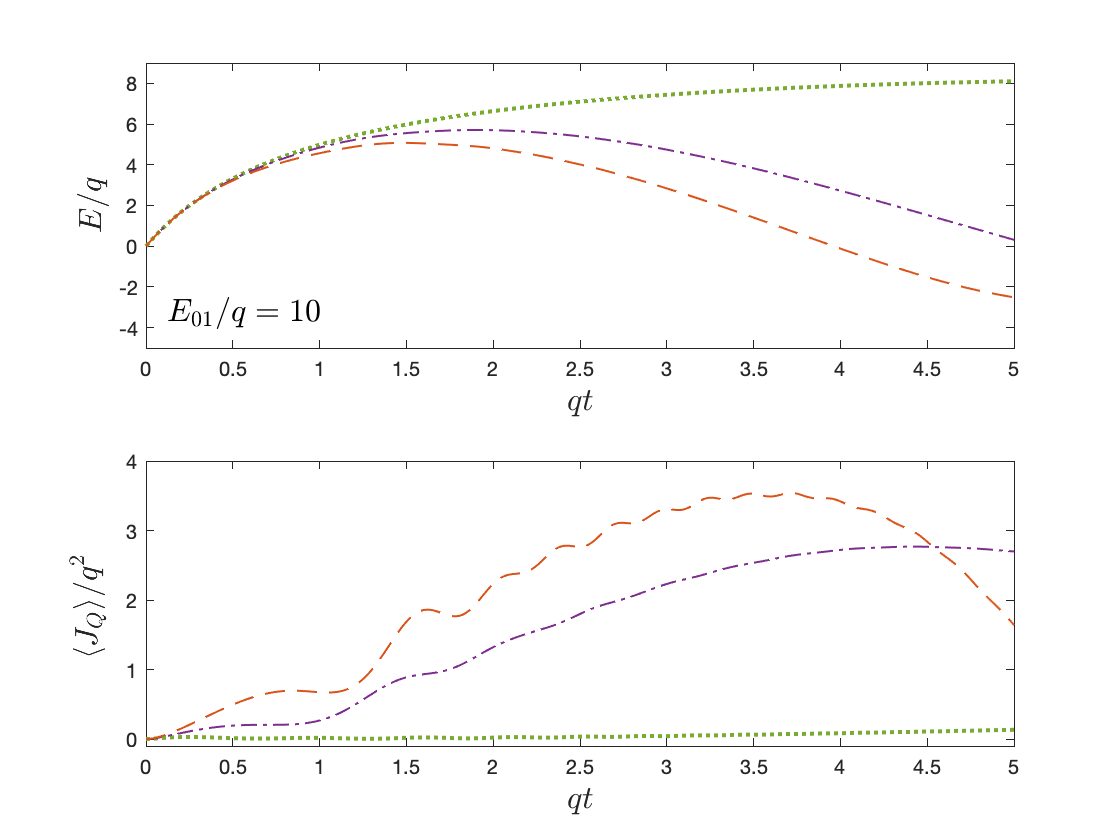}\hspace{-0.7cm}
\includegraphics[width=70mm]{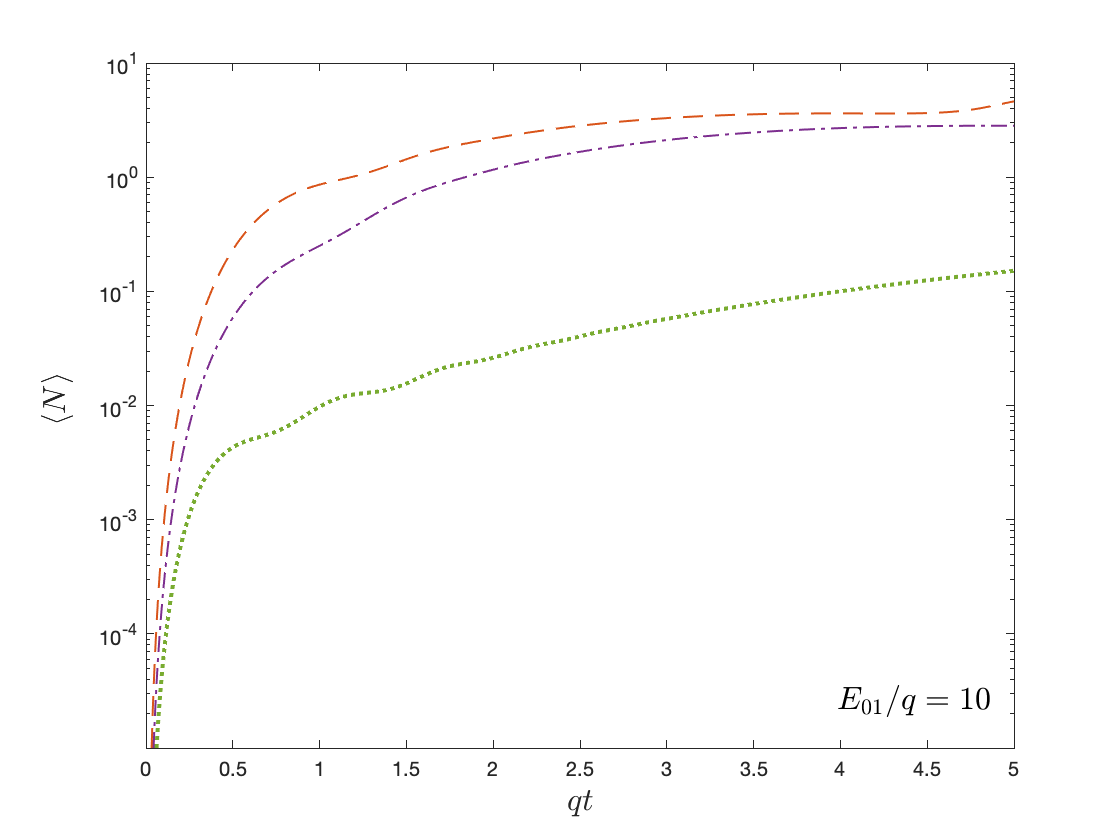}\hspace{-0.7cm}
\includegraphics[width=70mm]{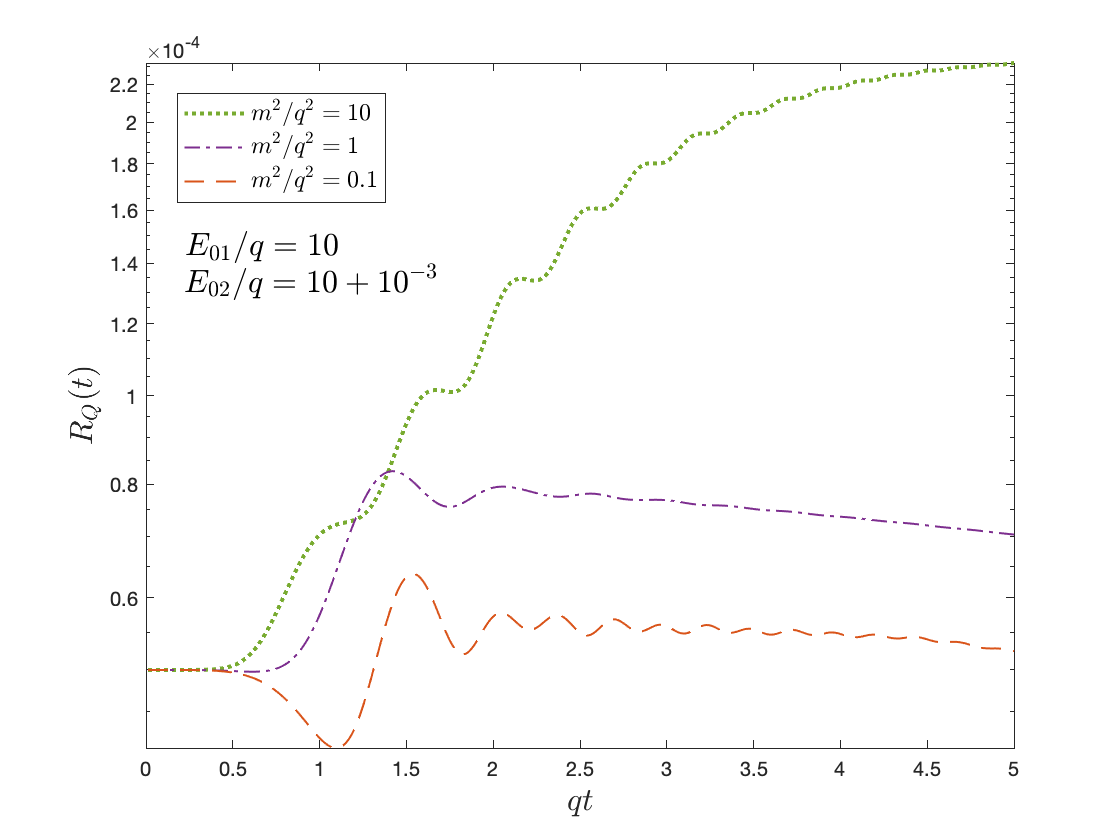}
\end{tabular}
\end{center}
\caption{\small{Results obtained from numerical solutions to the semiclassical backreaction equation for scalar fields and the asymptotically constant classical profile are shown for $E_0/q = 10$.  The masses are chosen so that $\frac{q E_0}{m^2} \ge 1$.   %The electric field and the induced electric current $\langle J_Q \rangle/q^2$ for each case are plotted in the left panels. Plots for the corresponding number of particles, $\la N \ra$, are shown in the middle panel and plots of the quantity $R_Q$ appear in the right panel.  For the latter, the values $E_{01}/q=10$ and $E_{02}/q=10+10^{-3}$ have been chosen for the two solutions that are subtracted.  The values of $m^2/q^2$ for each case are shown along with the type of curve for that solution in the legend in the right panel.
The structure of the figure is the same as in Figure \ref{fig:large-m-fermion}. We have chosen  $E_{01}/q=10$ and $E_{02}/q=10+10^{-3}$ to represent the function $R_Q$. }}
\label{fig:small-m-scalar}
\end{figure}

Unlike the case of the spin-$\frac{1}{2}$ field, there is
no clear limit that we have found as $m \to 0$ for a scalar field coupled to the electromagnetic field. However, our numerical results shown in Fig. \ref{fig:small-m-scalar} indicate that, as for the spin-$\frac{1}{2}$ field, $R_Q$ grows significantly at early times for $\frac{q E_0}{m^2} \sim 1$ but grows much less rapidly in time for larger values of $\frac{q E_0}{m^2}$. 
Thus our criterion is violated for $\frac{q E_0}{m^2} \sim 1$ but, at least for the homogeneous approximate solutions that we consider, it appears to be satisfied for $\frac{q E_0}{m^2} \gg 1$.

\begin{figure}[htbp]
\begin{center}
\begin{tabular}{c}
\hspace{-1.9cm}\includegraphics[width=70mm]{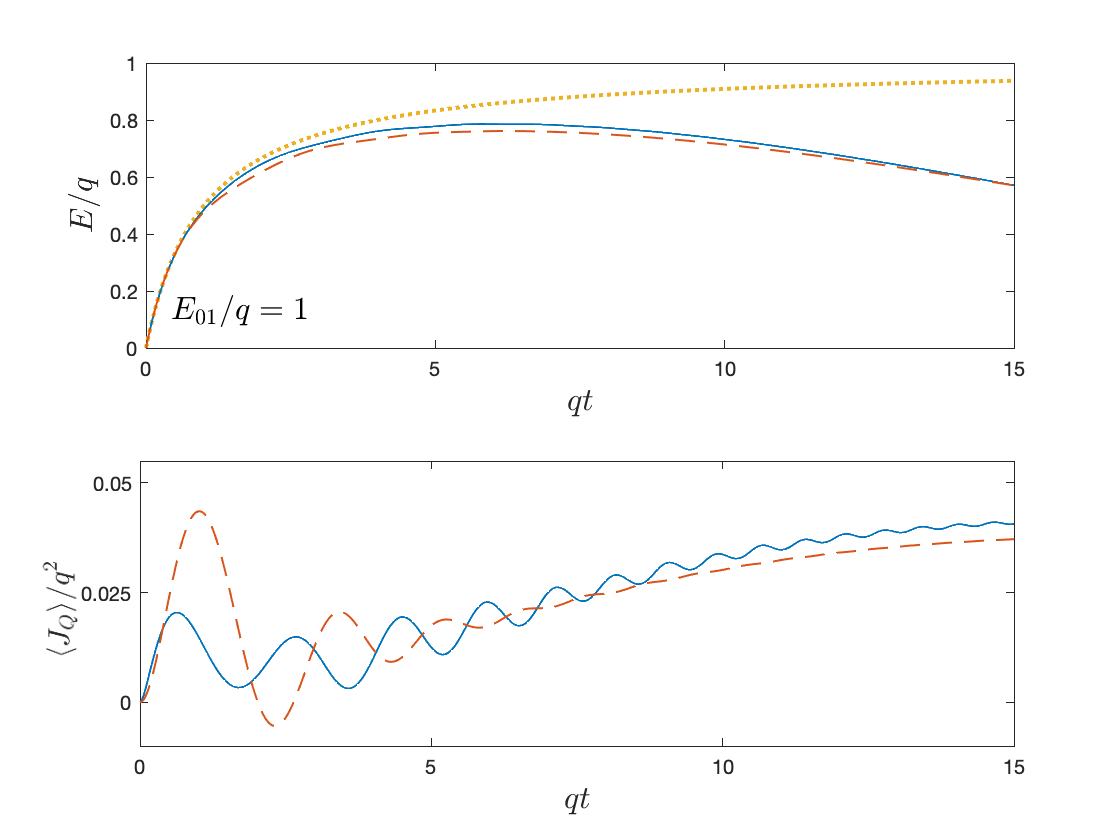}\hspace{-0.7cm}
\includegraphics[width=70mm]{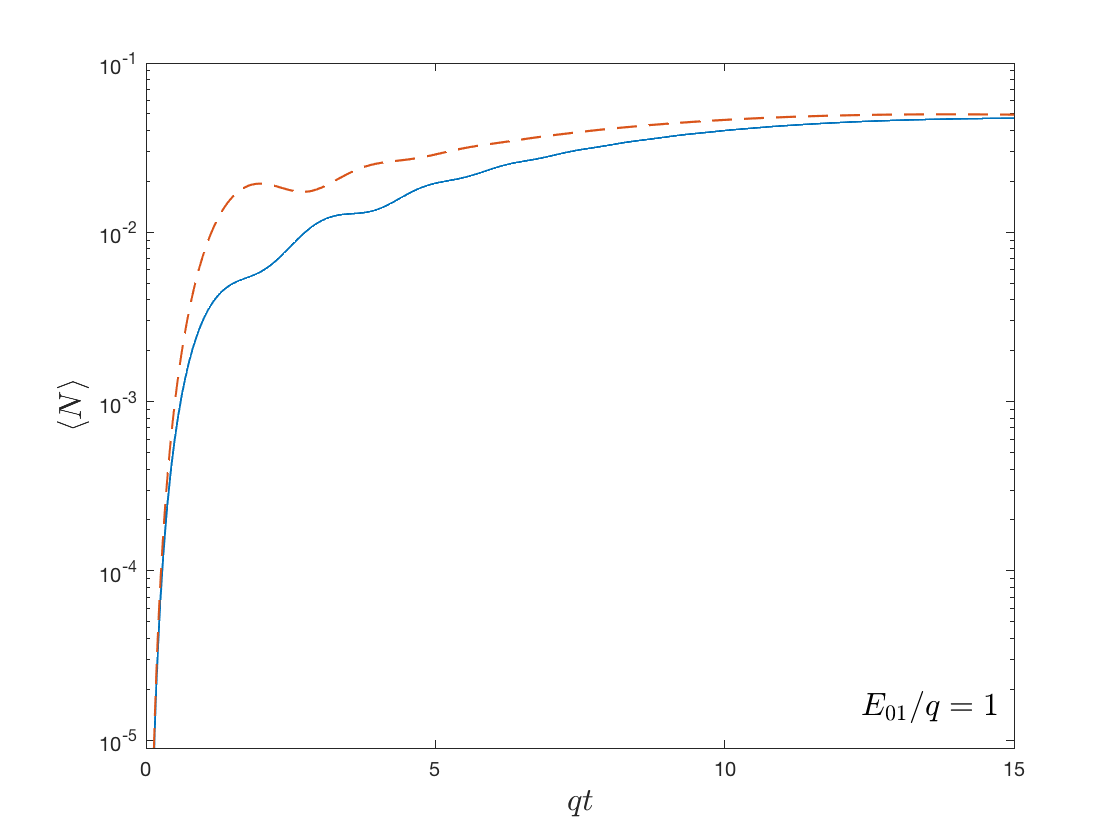}\hspace{-0.7cm}
\includegraphics[width=70mm]{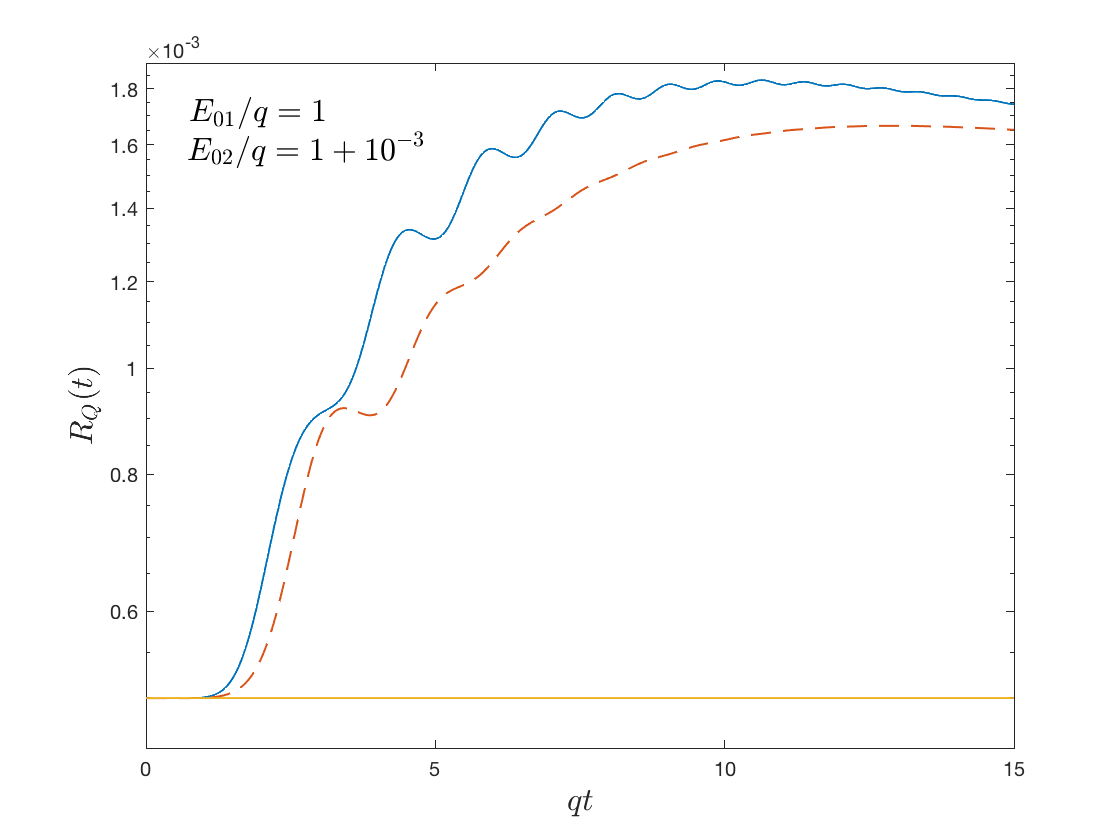}\\
\hspace{-1.9cm}\includegraphics[width=70mm]{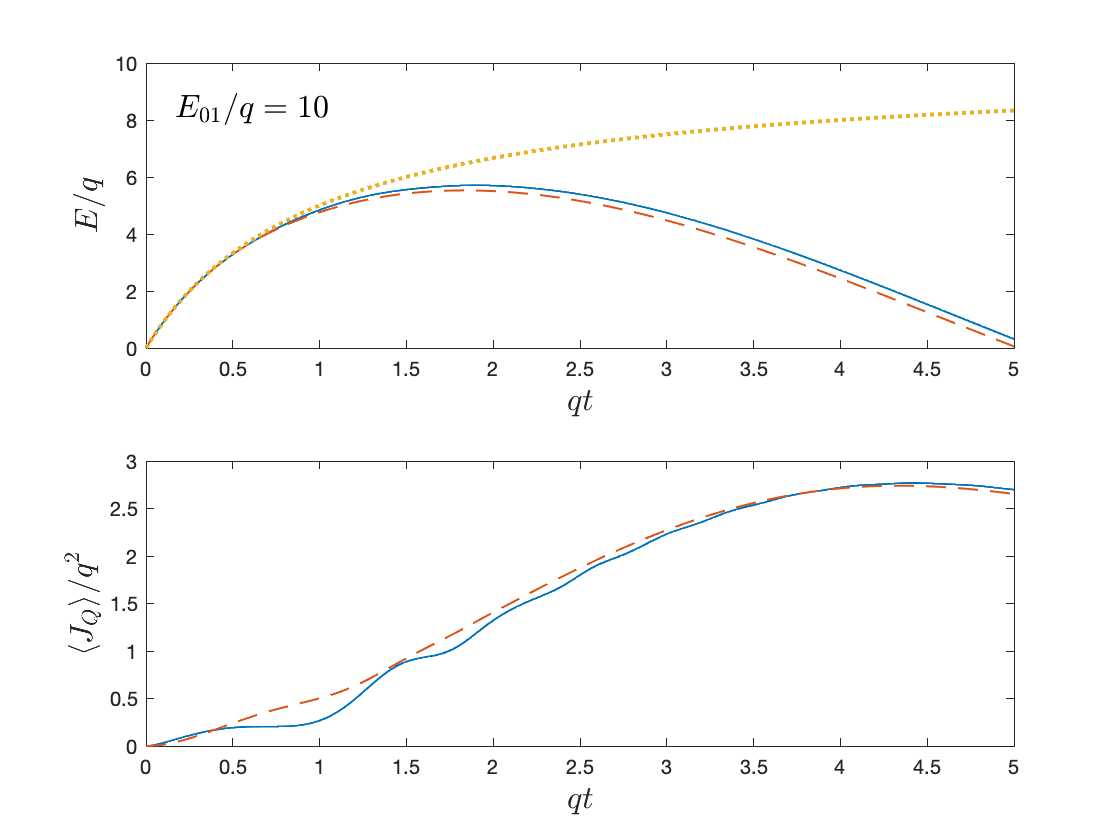}\hspace{-0.7cm}
\includegraphics[width=70mm]{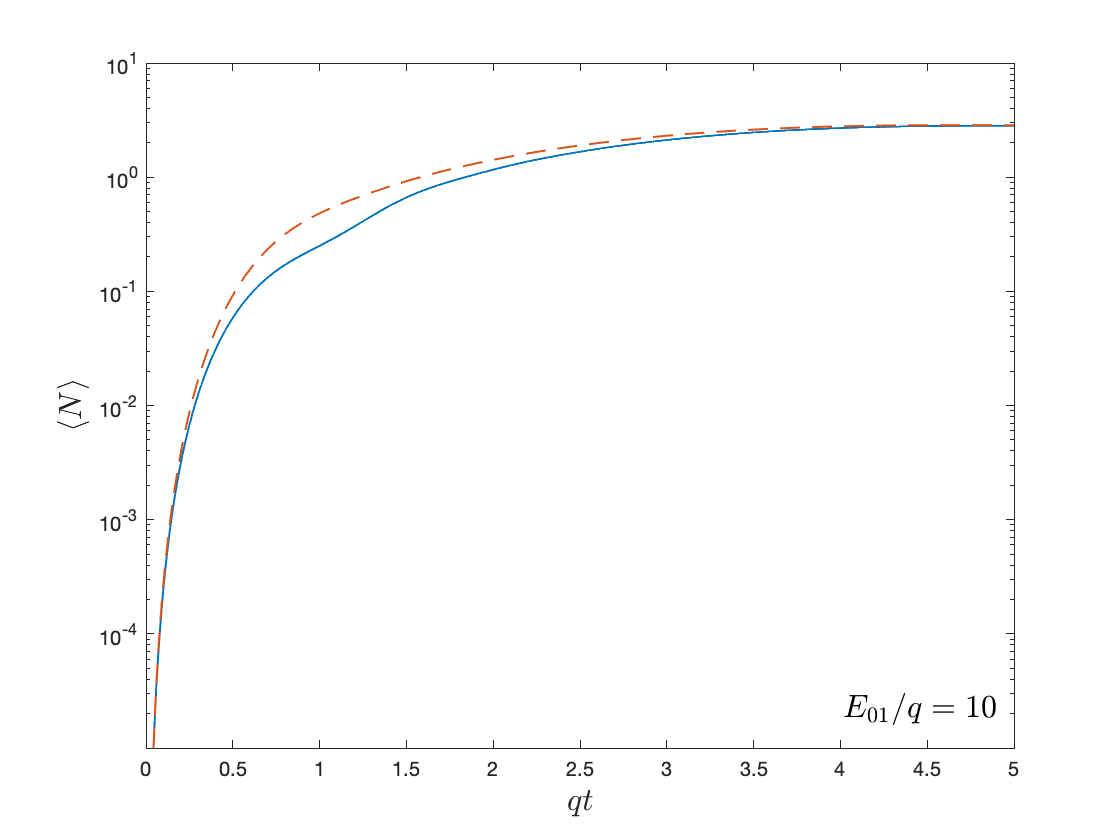}\hspace{-0.7cm}
\includegraphics[width=70mm]{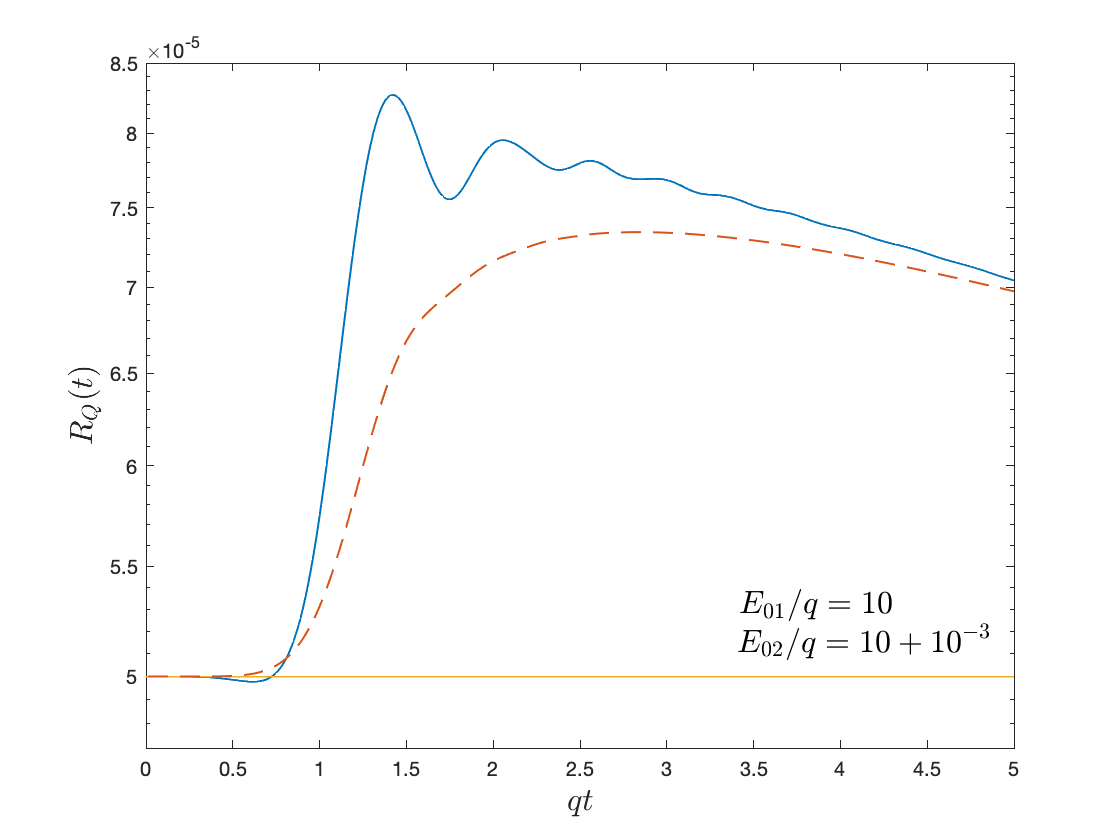}
\end{tabular}
\end{center}
\caption{\small{Results obtained from numerical solutions to the semiclassical backreaction equation for both spin-$\frac{1}{2}$ fields and scalar fields when the asymptotically constant classical profile is used are shown.  For all of the plots the solid curve (blue) corresponds to the scalar field, the dashed curve (orange) corresponds to the spin-$\frac{1}{2}$ field, and the dotted curve (yellow), when shown, corresponds to the classical solution when no quantum fields are coupled to the electromagnetic field.
In the upper tier $\frac{q E_{0}}{m^2} = 1$ and $m^2/q^2 = 1$ while in the lower tier $\frac{q E_{0}}{m^2} = 10$ and $m^2/q^2 = 1$.
For each tier the left panels show plots of the electric field and the induced electric current $\langle J_Q \rangle/q^2$, the middle panel shows plots of the number of particles $\la N \ra$, and  the right panel shows the quantity $R_Q$.  For the latter the values $E_{01}/q= E_0/q$ and $E_{02}/q=E_0/q+10^{-3}$ have been chosen for the two solutions that are subtracted.
}}
\label{fig:scalar-fermion}
\end{figure}

We have found that the behaviors of solutions to the semiclassical backreaction equation when a scalar field is present are in many ways qualitatively similar to the corresponding ones for the spin-$\frac{1}{2}$ field for cases in which the ratio $\frac{q E_0}{m^2}$ is not too large.  This is illustrated in Fig.~\ref{fig:scalar-fermion}
for $\frac{q E_0}{m^2} = 1$ and $10$.  The main difference occurs for the latter case where a larger ratio results in
more particle production for the scalar field than for the spin-$\frac{1}{2}$ field due to Pauli blocking.  Even in
that case the early-time behaviors of $R_Q$ are similar for the two fields.

For the large-mass limit, we expect that, as for the spin-$\frac{1}{2}$ case, the semiclassical approximation will approach the classical limit as $\frac{qE_0}{m^2}\to 0$.

\subsection{Sauter pulse classical profile}

While our results relating to the validity of the semiclassical approximation are the same for the scalar and spin-$\frac{1}{2}$ fields for the asymptotically constant classical profile, one might be concerned that
there could be significant differences for other classical profiles.  To test this we have also investigated the
validity of the semiclassical approximation for the Sauter pulse classical profile given in Eq.~\eqref{SauterE} with the classical current~\eqref{Sautercurrent}.  Unlike the asymptotically constant classical profile, the classical current in this case is a $C^\infty$ function so there is no extraneous particle production due to the sudden approximation.

We find for the Sauter pulse classical profile for both the scalar and spin-$\frac{1}{2}$ cases, that $R_Q$ grows significantly at early times for $\frac{q E_{0}}{m^2} \sim 1$, as it does for the asymptotically constant classical profile, and it is bounded for 
$\frac{q E_{0}}{m^2} \gg 1$.  Thus we find that our criterion for the validity of the semiclassical approximation is violated for $\frac{q E_{0}}{m^2} \sim 1$ while, for the approximate homogeneous solutions that we consider, our criterion appears to be satisfied for $\frac{q E_{0}}{m^2} \gg 1$.

\begin{figure}[htbp]
\begin{center}
\begin{tabular}{c}
\hspace{-1.9cm}\includegraphics[width=70mm]{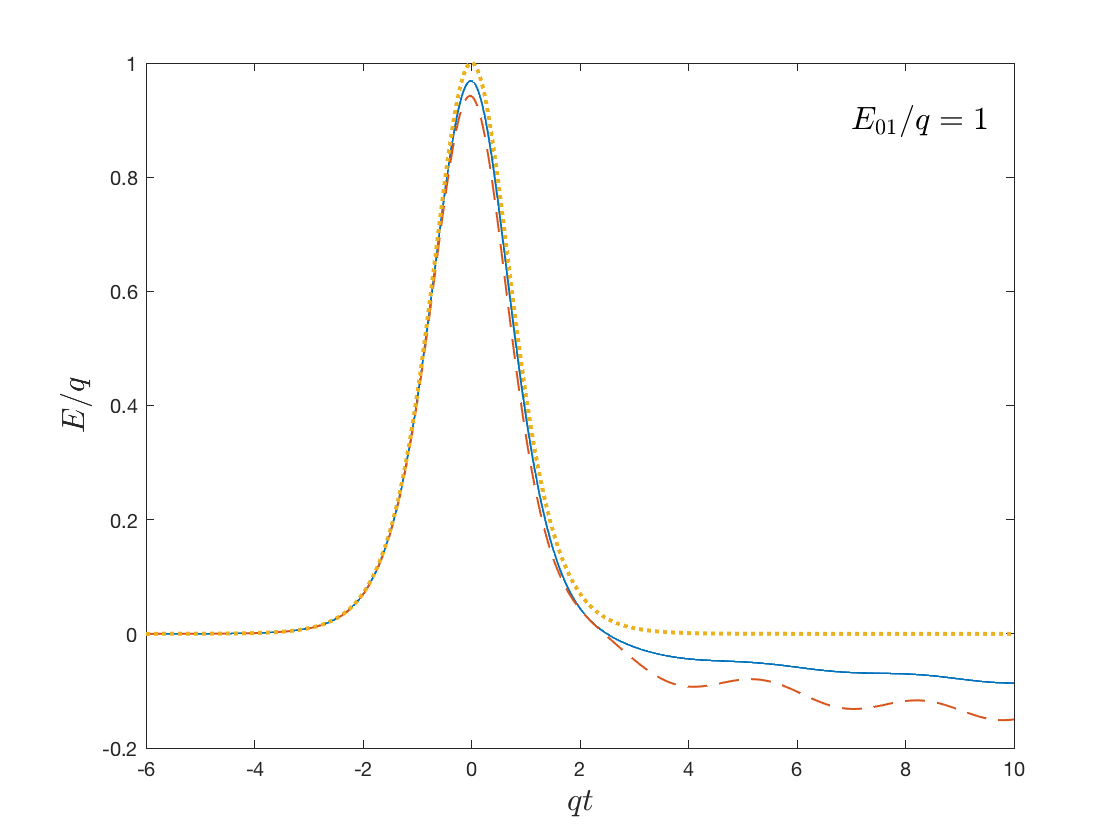}\hspace{-0.7cm}
\includegraphics[width=70mm]{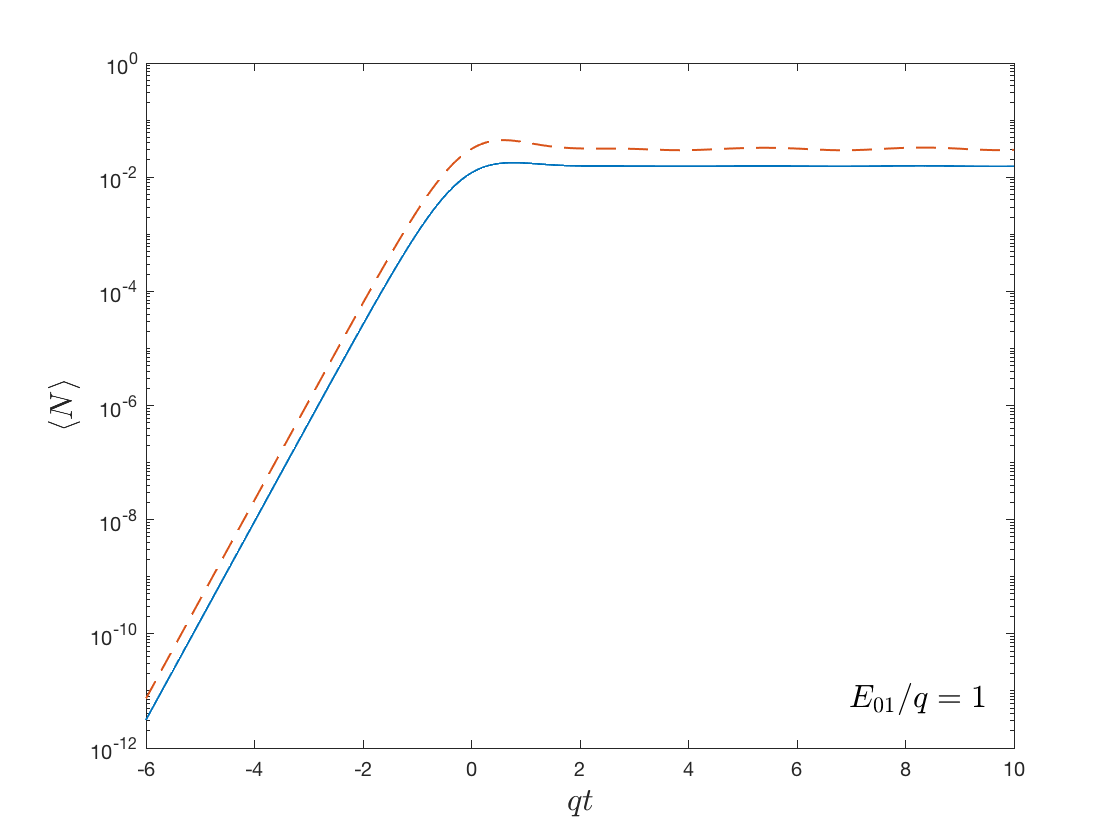}\hspace{-0.7cm}
\includegraphics[width=70mm]{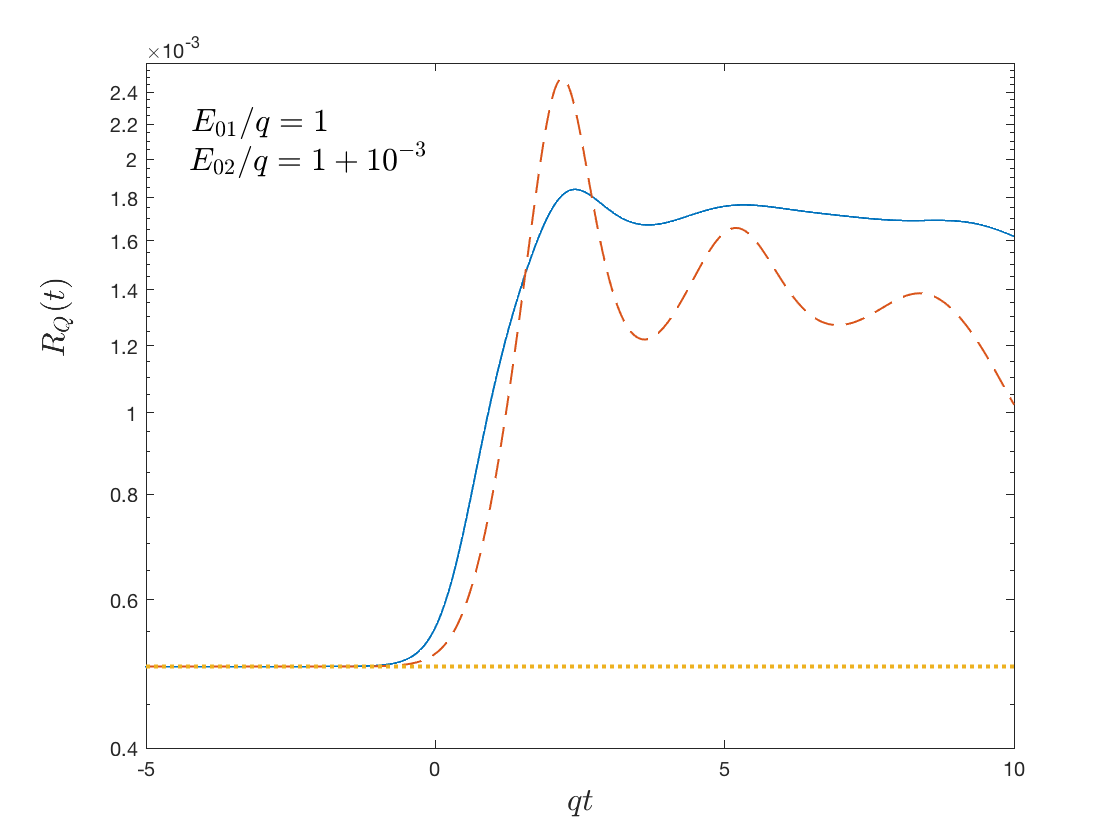}\\
\hspace{-1.9cm}\includegraphics[width=70mm]{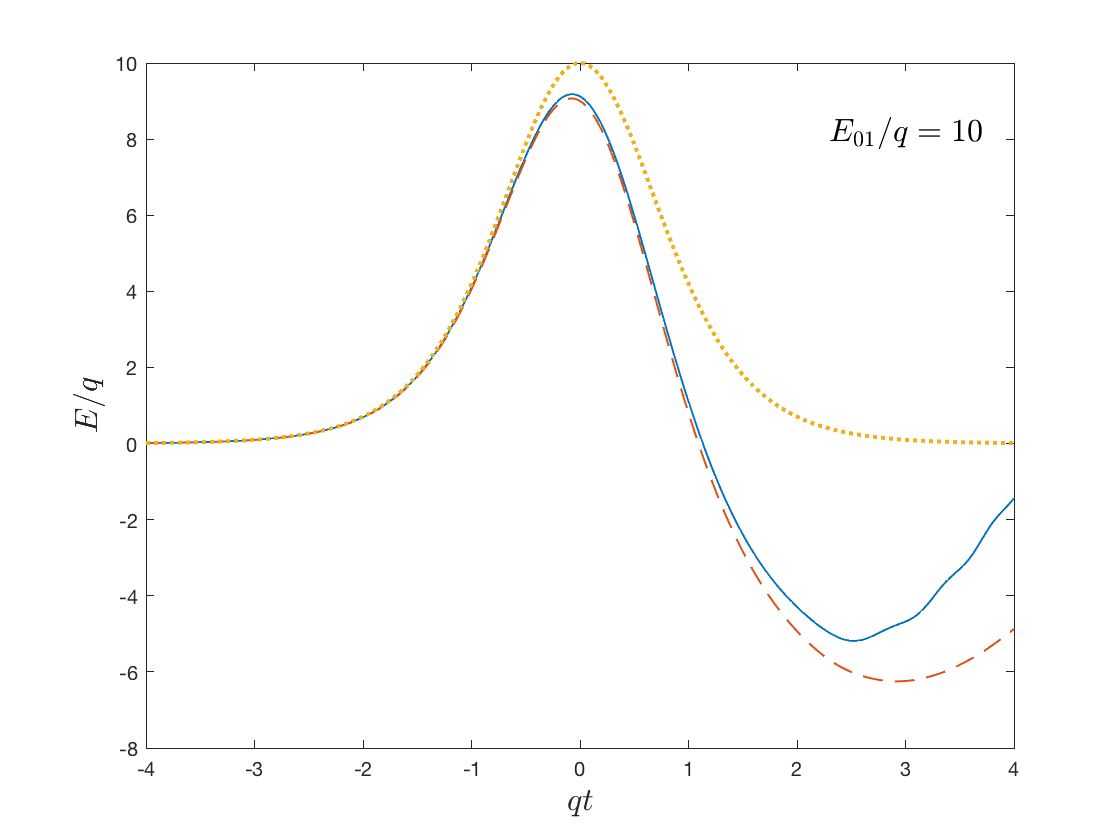}\hspace{-0.7cm}
\includegraphics[width=70mm]{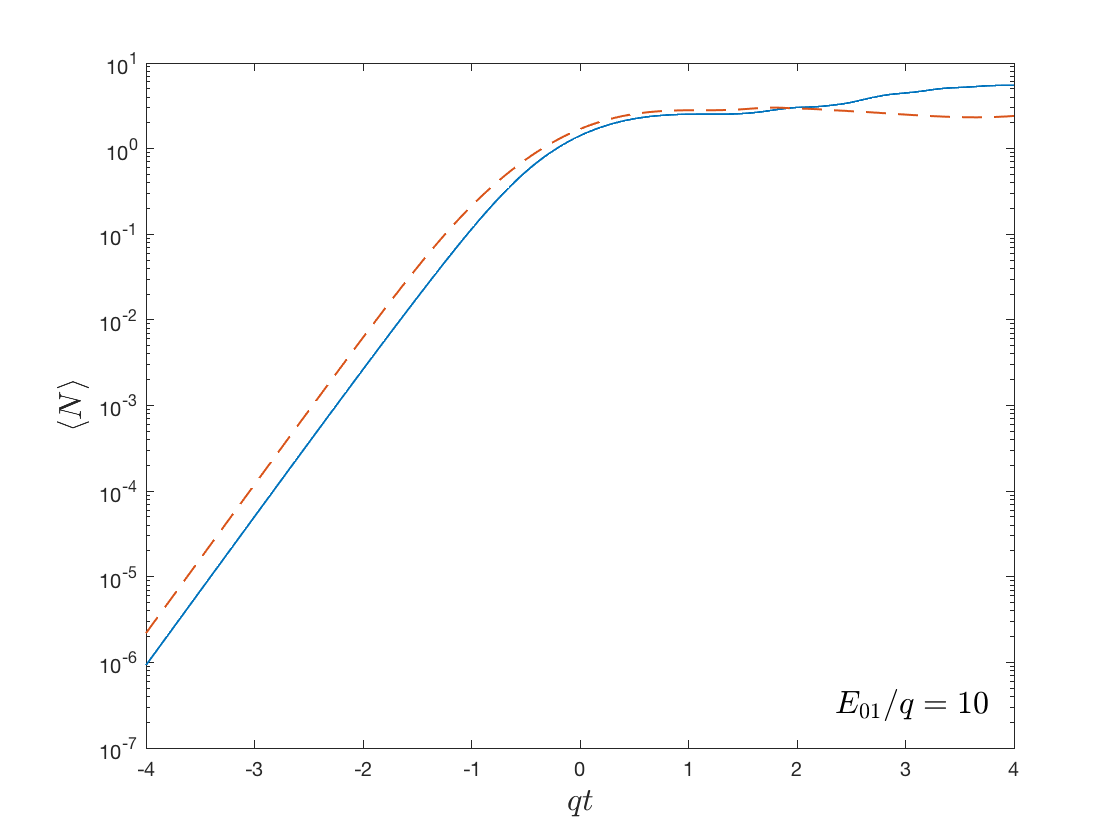}\hspace{-0.7cm}
\includegraphics[width=70mm]{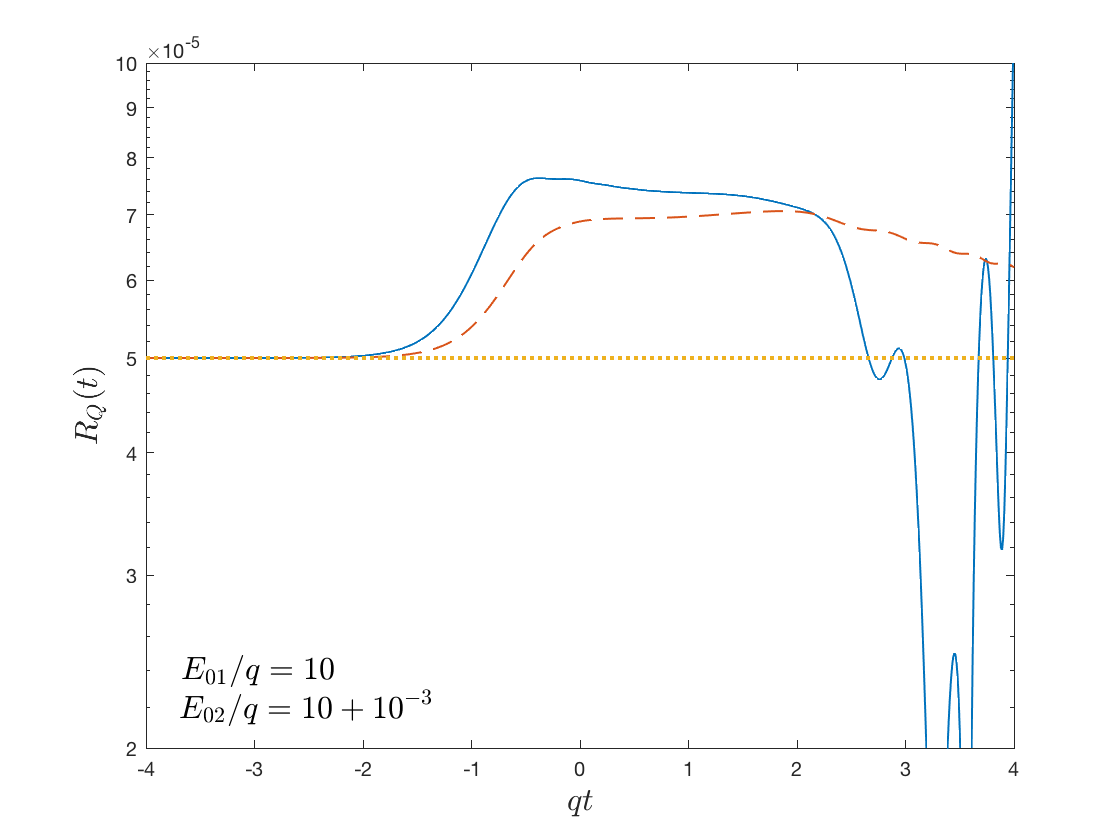}
\end{tabular}
\end{center}
\caption{\small{Results obtained from numerical solutions to the semiclassical backreaction equation for both spin-$\frac{1}{2}$ fields and scalar fields and the Sauter pulse classical profile are shown. The structure of the figure, including the initial values and the parameters of the fields, is the same than in Figure \ref{fig:scalar-fermion}. % For all of the plots the solid curve (blue) corresponds to the scalar field, the dashed curve (orange) corresponds to the spin $\frac{1}{2}$ field, and the dotted curve (yellow) when shown corresponds to the classical solution when no quantum fields are coupled to the electromagnetic field.
%In the upper tier $\frac{q E_{0}}{m^2} = 1$ and $m^2/q = 1$ while in the lower tier $\frac{q E_{0}}{m^2} = 10$ and $m^2/q = 1$. For each tier the left panel shows plots of the electric field, the middle panel shows plots of the number of particles $\la N \ra$, and  the right panel shows the quantity $R_Q$.  For the latter the values $E_{01}/q= E_0/q$ and $E_{02}/q=E_0/q+10^{-3}$ have been chosen for the two solutions that are subtracted.
}}
\label{fig:Sauter-scalar-fermi}
\end{figure}

Not surprisingly, given the difference between the Sauter pulse and asymptotically constant classical profiles, there are significant qualitative differences in the solutions for the electric field and in the time dependence of the number of particles that have been created.  These results are illustrated in Fig.~\ref{fig:Sauter-scalar-fermi} for both the scalar field and spin-$\frac{1}{2}$ field cases.
%From the plots in Fig.~\ref{fig:Sauter-scalar-fermi} 
It is clear from the plots that for the values $\frac{q E_0}{m^2} = 1$ and $10$ the backreaction effects start to be relevant before the classical pulse ends.  After the effect of the classical current subsides, plasma oscillations are expected to occur because of the current created by the produced particles.  There is evidence for
this in the plots of the electric field.  In the case $\frac{q E_0}{m^2} = 1$, backreaction effects are relatively weak and the particle creation essentially ceases once the pulse in the electric field has ended.  However, for $\frac{q E_0}{m^2} = 10$ the initial plasma oscillation is large enough that particles are created in the scalar field case after the pulse ends.
%As for the asymptotically constant profile when $\frac{q E_0}{m^2} \sim 1$ the quantity $R_Q$ has the same value as $R_C$ at early times but then grows rapidly for a short time.

\newpage

%%%%%%%%%%%%%%%%%%%%%%%%%%%%%%%%%%%%%%%%%%%%%%%%%%%%%%%%%%%%%%

\section{Conclusions and final comments}%SUMMARY, CONCLUSIONS, AND FINAL COMMENTS}
\label{sec:conclusions}

Numerical solutions to the semiclassical backreaction equation for quantum electrodynamics in 1+1D have been obtained for models of the Schwinger effect where particle production occurs due to the presence of a strong electric field.  The particle production results from the coupling of either a quantized massive charged scalar field or spin-$\frac{1}{2}$ field to a classical electric field. In each case the homogeneous electric field is zero initially, as it would be in a laboratory setting, and is generated by a classical current.  We have also used a renormalization scheme for the electric current and for the energy density of the quantum fields that is consistent with what would be used in a curved space background. This is different from previous backreaction calculations where the electric field was nonzero initially~\cite{Kluger91, Kluger92, CMRA}. % We have also used a renormalization scheme for the electric current and for the energy density of the quantum fields that is consistent with that which would be used in a curved space background.

In agreement with the previous backreaction calculations, it was found that if the electric field becomes large enough so that  $\frac{q E}{m^2} \gtrsim 1$ then a significant amount of particle production occurs. Subsequently, the produced particles create a current which generates an electric field in the opposite direction which begins to cancel the background electric field. After the initial damping of the background electric field, both the electric field and the current generated by the particles oscillate.  

 %We have studied in some detail the Schwinger particle production process in 1+1 dimensions with the backreaction of the created particles on the background homogeneous electric field included. Unlike previous calculations, the electric field is assumed to be zero initially and is turned on by a classical current.  %The assumed lower dimensional setting is a severe approximation, but it shares  with similar analysis of Hawking radiation in 2D gravity the advantage of possessing  a manageable set of semiclassical backreaction equations. 
 %We have expanded and improved on previous analyses in the literature \cite{Cooper-Mottola89, Kluger91, Kluger92, CMRA,and-mot-I,and-mot-san} in several crucial aspects. Our method for renormalizing the the electric current and the energy density has been done in a way consistent with gravity. 
 % Our detailed analysis of the time-dependent particle production has been displayed in a way  consistent with energy conservation (when the assumed classical source allows such a behaviour). We have also encountered for the first time multiple particle creation events as a consequence of  backreaction. ....    

The particle creation process has been discussed in detail for background electric fields in Refs.~\cite{and-mot-I,dunne-part-prod,and-mot-san}. %where time-dependent particle numbers were defined using various WKB approximations. If the field is turned off at late times then the time-dependent number becomes equal to the number of particles that would be obtained using a Bogolubov transformation.  
It was found that individual modes undergo a quasilocal particle creation event at roughly the time when $(k-qA)^2 \approx m^2$.  Here we have found that when backreaction effects are taken into account the same type of particle creation events occur.  What is different is that, because of the oscillations in the the vector potential at late times, there are modes that undergo multiple particle creation events.  Furthermore, once a given mode has undergone a particle creation event, it is possible for it to also undergo a particle destruction event although this does not always happen.  

The total number of particles was obtained using  the standard definition  of a time-dependent particle number \cite{and-mot-I, dunne-part-prod}.  For all three profiles considered it was found that the total particle number never decreases by any significant amount but that it is approximately constant for periods of time.  This is compatible with previous calculations of the total particle number when the electric field is turned on suddenly by a classical current that is proportional to $\delta(t)$ in $3+1$D using canonical quantization~\cite{Tanji} and in both $1+1$D and $3+1$D using lattice simulations~\cite{lattice-1,lattice-2}.  %Further, when there is an increase in the total number of particles after a period of being approximately constant,  that increase becomes smaller as time goes on so the particle number appears to approach an asymptotic value.  The process occurs more quickly in the case of a spin $\frac{1}{2}$ when the electric field is relatively strong due to the saturation of particle production caused by Pauli blocking.

The energy density of the quantum field was computed for a classical current that is proportional to $\delta(t)$ and is thus zero for $t >0$.  The total energy of the system is then constant and one can unambiguously track the transfer of energy between the particles and the electric field.  It was found that a significant amount of energy is permanently transferred to the particles during the first damping phase of the electric field.  More is then permanently transferred to the particles upon subsequent oscillations of the electric field.  This is also consistent with previous calculations
in $1+1$D using lattice simulations~\cite{lattice-1} and in $3+1$D using canonical quantization~\cite{Tanji} and classical statistical field theory techniques~\cite{stat-FT}.

Correlations between the energy density of the particles, the current due to the particles, and the total particle number were found.  In particular, times when the number of particles grows directly correspond to times when the current is changing, and times when the total number is not growing significantly correspond to times when the current is approximately constant.  However, the current keeps oscillating even after the particle number stops growing significantly.
%The exact relationship is less clear, but when the total number of particles increases significantly, energy transfer to the particles from the electric field is observed to occur.  However, there are times when energy is transferred to the particles and the particle number does not increase significantly.

%Furthermore, all the above analysis and results pave the way to a careful analysis of the validity of the previously assumed semiclassical framework. This is a very distinctive characteristic of this paper. To this end we have used a validity criterion, also inspired from gravitational physics, to go further and reconsider the validity range of the  semiclassical picture for the Schwinger effect.  The criterion proposed is based on the linear response equation. It allows us estimate the significance of the quantum fluctuations thought the behaviour of $R_Q$.     

Since semiclassical electrodynamics is an approximation to quantum electrodynamics, an important question is whether this approximation is a good one for a given solution to the semiclassical backreaction equation.  We have addressed this question by adapting a criterion developed for semiclassical gravity and modified for chaotic inflation models, that should be satisfied if the semiclassical approximation is valid.  It is therefore a necessary but not sufficient condition.  The condition is based upon the fact that the retarded two-point function for the current appears in the linear response equations for semiclassical electrodynamics.  If this correlation function grows significantly in time and therefore solutions to the linear response equation grow significantly, then one expects that quantum fluctuations are significant.  We have approximated homogeneous solutions to the linear response equation by taking two solutions to the semiclassical backreaction equation which are nearly the same at early times and plotting a relative difference between them which we call $R_Q$, defined in Eq.~\eqref{Rq-def}.  In cases where this difference grows significantly in time one expects that the corresponding solution to the linear response equation will also do so.   

%On physical grounds, and according with the decoupling theorem \cite{APtheorem}, one would expect that in the limit, $m^2 \to \infty$, the  quantized fields   decouple from the electromagnetic background, no particle production occurs, and the resulting theory behaves as a purely classical field theory. 

We have investigated the validity of the semiclassical approximation for both the scalar and spin-$\frac{1}{2}$ fields using two different classical current profiles which are shown along with the resulting electric field (if backreaction effects are ignored) in Fig.~\ref{electric_profiles2.0}. 
%If backreaction effects are ignored then the electric field in one case turns on gradually starting at $t = 0$ and asymptotically approaches the constant value $E_0$.  In the other case, which is the Sauter pulse, the electric field vanishes in the limits $t \to \pm \infty$ and has a single peak of height $E_0$ at time $t = 0$.         

In the zero-mass limit for the spin-$\frac{1}{2}$ field, the solutions to the semiclassical backreaction equations are completely determined by the axial anomaly.  In this case, there is no growth whatsoever in the relative difference $R_Q$, and thus, for the approximate homogeneous solutions to the linear response equation that we considered, our criterion appears to be satisfied.  We have investigated the behaviors of solutions in the small-mass case, i.e., $m^2 \ll qE_0 $, and found that they smoothly approach those found in the zero-mass limit.  Thus, for the same type of solutions to the linear response equation, our criterion appears to be satisfied in the small-mass limit as well.  Note that in this limit there is a great deal of particle production and backreaction effects are very strong (see Figs.~\ref{fig:small-m-fermion} and \ref{fig:small-m-fermion10}). Although there is no solvable massless limit for spin-$0$ field, we have also checked numerically that there is less growth in $R_Q$ with time  as we decrease the mass of the created particles (see Fig.~\ref{fig:small-m-scalar}).  

The intermediate case $m^2 \sim q E_0$ is very different.  In both the asymptotically constant and Sauter pulse models and for both the scalar and spin-$\frac{1}{2}$ fields, %the value of $R_Q$ appears to grow exponentially grows rapidly 
once the amount of particle production has become significant, there is a rapid and significant growth in the ratio $R_Q$.  %When backreaction effects begin to substantially modify the behavior of the electric field, the value of $R_Q$ levels off for awhile.  A similar behavior occurs for the total number of particles.
Thus in this case our criterion is not satisfied because of this growth.
This is similar to the breakdown of the semiclassical approximation found in Ref.~\cite{validitypreheating} for the preheating phase of chaotic inflation.% when quantized massless scalar fields are coupled to the inflaton field which is a classical massive scalar field.

In the large-mass limit where  $\frac{q E_0}{m^2} \to 0$, particle production does not occur and the behavior of the electric field can be predicted by classical electrodynamics. This is in agreement with the  decoupling theorem \cite{APtheorem}.

It is very likely that the first experimental verification of the Schwinger effect will be for the intermediate-mass case.  Thus it is worth examining the predictions for that case more carefully.  First, there is no observed growth in $R_Q$ at very early times before backreaction effects become significant.  Therefore our criterion appears to be initially satisfied.  However, given the difficulty in creating a strong enough electric field for the Schwinger effect to be observed in the laboratory (the field strength required being on the order of $E_{\rm crit} \sim 10^{18} $ V/m), the focus of the initial experiments is likely to be on the detection of particles rather than their backreaction effects.  Thus the semiclassical approximation should be able to give a good description of the particle production process at such early times. 
Second, once backreaction effects become significant, a relatively large number of particles is likely to have been created.  In previous work on the study of the validity of the semiclassical approximation for preheating in chaotic inflation~\cite{validitypreheating} it was found that in one case that could be compared there was good qualitative agreement with calculations that used a random-phase approximation~\cite{r-p-1,r-p-2,r-p-3} even though the semiclassical approximation broke down early in the process.  Similarly, the backreaction calculations in Ref.~\cite{lattice-1} using classical statistical field theory techniques in 1+1 D are in qualitative agreement with our calculations of the electric field, energy density, and total particle number.  Thus the semiclassical approximation can, at least in some cases, provide reasonable qualitative predictions even when its quantitative predictions cannot be trusted.

%Second, once backreaction effects become significant a relatively large number of particles is likely to have been created.  In this case, some type of random phase method might be used such as that developed for preheating in chaotic inflation~\cite{r-p-1,r-p-2,r-p-3}.  In~\cite{validitypreheating} it was noted that, even though the semiclassical approximation for preheating breaks down early in the process, the qualitative behaviors of solutions to the semiclassical backreaction equation are similar to those found using the random phase approximation in one case in which they could be compared.  It seems likely that a similar result may occur for the Schwinger effect in semiclassical electrodynamics.

%%%%%%%%%%%%%%%%%%%%%%%%%%%%%%%%%%%%%%%%%%%%%%%%%%%%%%%%%%%%%%%%%%%%%%%%%%%%%%%%%%%%%%%%%%

\section*{Acknowledgments}
I.M.N. would like to thank Eric Carlson and William Kerr for helpful conversations. J.N.-S. and S.P. would like to thank Antonio Ferreiro and Pau Beltran-Palau for useful discussions.  P. R. A. would like to thank Fred Cooper and Emil Mottola for useful discussions and we would also like to than Emil Mottola for helpful comments on the manuscript.  Most of the numerical work has been performed with the {\it MATLAB} software. The authors also acknowledge the Distributed Environment for Academic Computing (DEAC) at Wake Forest University for providing HPC resources that have contributed to the research results reported within this paper. This work was supported in part by Spanish Ministerio de  Economia,  Industria  y  Competitividad  Grants  No. FIS2017-84440-C2-1-P (MINECO/FEDER, EU) and No.  FIS2017-91161-EXP, the project PROMETEO/2020/079 (Generalitat
Valenciana), and by
National Science Foundation Grants No. PHY-1505875 and PHY-1912584 to Wake Forest University.
S. P.  is supported by a Ph.D. fellowship, Grant No. FPU16/05287.

%%%%%%%%%%%%%%%%%%%%%%%%%%%%%%%%%%%%%%%%%%%%%%%%%%%%%%%%%%%%%%%%%%%%%%%%%%%%%%%%%%%%%%%%%%
\appendix

\section{Derivation of the linear response equation}

\subsection{Scalar field}

The mode equation for a massive complex scalar field can be obtained by substituting Eq.~\eqref{phi1} into Eq.~\eqref{mode3}  with the result
\be
    \left[-\partial_{t}^{2}+\partial_{x}^{2}-2iqA(t)\partial_{x}-q^{2}A^{2}(t)-m^{2}\right]U_{k}(t,x)=0 \quad . \label{modeU}
\ee
If one perturbs the vector potential about some solution to the semiclassical backreaction equation $A(t)$ such that
$ A(t) \to A(t) + \delta A(t)$ and writes for the mode function $U_k(t,x) \to U_k(t,x) + \delta U_k(t,x)$, then to leading order
\be
\left[-\partial_{t}^{2}+\partial_{x}^{2}-2iqA(t)\partial_{x}-q^{2}A^{2}(t)-m^{2}\right]\delta U_{k}(t,x) = 2 q \delta A(t) \bigg( i \partial_x U_k (t,x) + q A(t) U_k (t,x) \bigg) \quad . \label{deltaU-1}
\ee

For a massive scalar field, the retarded Green's function
\cite{birrell-davies}
\be
    G_{R}(t,x;t',x')= i \theta(t-t')\la [\phi(t,x),\phi^{\dagger}(t',x')]\ra \quad , \label{scalarGreen}
\ee
is a solution to the inhomogeneous equation
\be
    \bigg[-\partial_{t}^{2}+\partial_{x}^{2}-2iqA(t)\partial_{x}-q^{2}A(t)^{2}-m^{2}\bigg]G_{R}(t,x;t',x') =  -\delta(t-t')\delta(x-x') \quad .
\ee
Thus the solutions to Eq.~\eqref{deltaU-1} can be written in the form
\be
    \delta U_{k}(t,x)=\delta U_{k}^{H}(t,x)-2q\int_{-\infty}^{\infty} dt'\int_{-\infty}^{\infty} dx' G_{R}(t,x;t',x')\left[i\,\partial_{x'}+q\,A(t')\right]U_{k}(t',x')\delta A(t') \quad , \label{deltaU-2}
\ee
where $\delta U_{k}^{H}(t,x)$ is a solution to the homogeneous part of Eq.~\eqref{deltaU-1}.

The explicit form of $G_{R}(x,x^{'})$ can be found using Eq.~\eqref{scalarGreen} with Eq.~\eqref{phi3} evaluated in the vacuum state, which yields
\be
    G_{R}(t,x;t'x')=\frac{i}{2\pi}\theta(t-t^{'})\int_{-\infty}^{\infty} dk\bigg[f_{k}(t)f_{k}^{*}(t^{'})-f_{k}(t^{'})f_{k}^{*}(t)\bigg]e^{ik(x-x^{'})} \quad . \label{GR-f}
\ee
Restricting attention to spatially homogeneous perturbations we have %and using~\eqref{func}, gives
\be \delta U_{k}(t,x)=\delta f_{k}(t) e^{ikx} \;.  \label{deltaf-def} \ee
Substituting Eqs.~\eqref{GR-f} and \eqref{deltaf-def} into Eq.~\eqref{deltaU-2} and integrating %first over $x^{'}$and then over $k$
yields
\be
    \delta f_{k}(t)= \delta f_{k}^{H}(t)+2\,i\,q \int_{-\infty}^{t}dt'\bigg(k-q A(t')\bigg)\bigg[f_{k}(t)f_{k}^{*}(t')-f_{k}(t')f_{k}^{*}(t)\bigg] f_{k}(t')\delta A(t') \quad . \label{deltamodes}
\ee
The perturbation of the renormalized current \eqref{Jrenorm} yields
\be
    \delta \la J_{Q} \ra_{\rm ren} = \frac{q}{\pi} \int_{-\infty}^{\infty} dk \, \bigg[ \bigg(k-qA(t)\bigg)\bigg[f_{k}(t)\delta f_{k}^{*}(t)+\delta f_{k}(t)f_{k}^{*}(t)\bigg] -q|f_{k}(t)|^{2}\delta A(t) + \frac{q\, m^{2}}{2\omega^{3}}\delta A(t) \bigg] \quad . \label{deltaJrenorm}
\ee
Substituting Eq.~\eqref{deltamodes} and its complex conjugate into Eq.~\eqref{deltaJrenorm}  yields
\bea
    \delta \la J_{Q}\ra _{\rm ren} &=&  \frac{q}{\pi}\int_{-\infty}^{\infty} dk \, \left\{\bigg(k-qA(t)\bigg)\bigg[f_{k}(t)\delta f_{k}^{*H}(t)+f_{k}^{*}(t)\delta f_{k}^{H}(t)\bigg]-\bigg[|f_{k}(t)|^{2}- \frac{m^{2}}{2\omega^{3}}\bigg]q \,\delta A(t) \right\} \nonumber \\ \nonumber \\
    &&\qquad  - \frac{4q^{2}}{\pi}\int_{-\infty}^{\infty} dk \int_{-\infty}^{t}dt'\bigg(k-qA(t)\bigg)\bigg(k-qA(t')\bigg)\textnormal{Im}\bigg\{f_{k}(t)^{2}f_{k}^{*}(t')^{2}\bigg\}\delta A(t') \quad . \label{deltaJb}
\eea

Our goal is to show that the above linear response equation can be written in terms of the two-point correlation function for the current, $\la [J_{Q}(t,x),J_{Q}(t',x')] \ra$.
To accomplish this we next calculate the two-point correlation function using the symmetrized current density \eqref{jsymm1} and the scalar field mode expansion \eqref{phi3} evaluated in the vacuum state. After integrating over the spatial coordinate one finds
\be
    \int_{-\infty}^{\infty} dx' \langle [J_{Q}(t,x),J_{Q}(t',x')] \rangle =\frac{4 \, i \, q^{2}}{\pi}  \int_{-\infty}^{\infty} dk \bigg(k-qA(t)\bigg)\bigg(k-qA(t')\bigg)\textnormal{Im}\bigg\{f_{k}(t)^{2}f_{k}^{*}(t')^{2}\bigg\} \quad . \label{Itpc}
\ee
Comparing Eqs.~\eqref{deltaJb} and \eqref{Itpc}, it is clear that Eq.~\eqref{deltaJb} can be written in the form
\bea
    \delta \la J_{Q}\ra _{\rm ren} &=&  \frac{q}{\pi}\int_{-\infty}^{\infty} dk \, \bigg[\bigg(k-qA(t)\bigg)\bigg[f_{k}(t)\delta f_{k}^{*H}(t)+f_{k}^{*}(t)\delta f_{k}^{H}(t)\bigg]-\bigg[|f_{k}(t)|^{2}- \frac{m^{2}}{2\omega^{3}}\bigg]q \,\delta A(t)\bigg] \nonumber \\ \nonumber \\
    && \qquad  \qquad \qquad  \qquad \qquad  \qquad  + \, i \int_{-\infty}^{\infty} dx' \int_{-\infty}^{t}dt' \, \la [J_{Q}(t,x),J_{Q}(t',x')] \ra \delta A(t') \quad . \label{deltaJfinal}
\eea
Thus $\delta \la J_{Q} \ra_{\rm ren}$ for a scalar field has been cast in terms of the current-current two-point correlation function. Note that $\delta f_{k}^{H}(t)$ corresponds to a change of state of the quantum field.
For the cases considered in this paper the vector potential and its first time derivative are zero initially so the perturbations do not cause a change in the state of the field so $\delta f_{k}^{H}(t) = 0$.
Then the linear response equation~\eqref{LRE} becomes
\bea
    \frac{d^{2}}{dt^{2}}\delta A(t) &=& -\frac{d}{dt}\delta E(t) = \delta J_{C}- \frac{q^2}{\pi} \delta A(t) \int_{-\infty}^{\infty} dk \,
    \left[ |f_{k}(t)|^{2} - \frac{m^{2}}{2\omega^{3}} \right] \nonumber \\
    && \qquad \qquad \qquad \qquad \qquad \qquad \qquad \qquad \qquad + \, i \int_{-\infty}^{\infty} dx' \int_{-\infty}^{t}dt' \, \la [J_{Q}(t,x),J_{Q}(t',x')] \ra \delta A(t') \quad . \nonumber \\
\eea

%%%%%%%%%%%%%%%%%%%%%%%%%%%%%%%%%%%%%%%%%%%%%%%

\subsection{Spin-$\frac{1}{2}$ field}

The mode equation for a massive charged spin $\frac{1}{2}$ field can be obtained by substituting Eq.~\eqref{psi1} into Eq.~\eqref{modefermi} with the result
\be
    \bigg[i \, \gamma^{t}\partial_{t}+i \, \gamma^{x}\partial_{x}+q \, \gamma^{x}A(t)-m \bigg]u_{k}(t,x)=0 \quad . \label{modeuk}
\ee
If one perturbs the vector potential about some solution to the semiclassical backreaction equation $A(t)$ such that $A(t) \to A(t) + \delta A(t)$ and writes for the mode function $u_k(t,x) \to u_k(t,x) + \delta u_k(t,x)$ then to leading order
\be
\bigg[i \, \gamma^{t}\partial_{t}+i \, \gamma^{x}\partial_{x}+q \, \gamma^{x}A(t)-m \bigg]\delta u_{k}(t,x)= - q \gamma^x u_k(t,x) \delta A(t) \quad . \label{delta-uk-eq}
\ee
For a massive spin-$\frac{1}{2}$ field the retarded Green's function
\be
    G_{R}(t,x;t',x') = i \, \theta(t-t') \la \{\psi(t,x),\Bar{\psi}(t',x')\}\ra \quad , \label{greenDirac}
\ee
is a solution to the inhomogeneous equation
\be
    \bigg[i \, \gamma^{t}\partial_{t}+i \, \gamma^{x}\partial_{x}+q \, \gamma^{x}A(t)-m \bigg]G_{R}(t,x;t',x') =  - \mathbbm{1} \, \delta(t-t')\delta(x-x') \quad ,
\ee
where $\mathbbm{1}$ is the identity matrix.
Thus the solution to Eq.~\eqref{delta-uk-eq} can be written in the form
\be
    \delta u_{k}(t,x) = \delta u_{k}^{H}(t,x)+q\int_{-\infty}^{\infty}dt'\int_{-\infty}^{\infty}dx' \, G_{R}(t,x;t',x') \, \gamma^{x} \, \delta A(t^{'}) \, u_{k}(t',x') \quad . \label{deltauk}
\ee
where $H$ represents the homogeneous solution. The explicit form of $G_{R}(x,x^{'})$ can be found using Eq.~\eqref{greenDirac} with the Dirac field expansion \eqref{psi1} in terms of spinor solutions \eqref{v} evaluated in the vacuum state, which yields
\be
    G_{R}(t,x;t',x') = \frac{i \, \theta(t-t^{'})}{2\pi}\int dk \, e^{ik(x-x^{'})}
    \begin{bmatrix}
    -h_{k}^{I}(t)h_{k}^{II*}(t^{'})+h_{k}^{II*}(t)h_{k}^{I}(t^{'}) & h_{k}^{I}(t)h_{k}^{I*}(t^{'})+h_{k}^{II*}(t)h_{k}^{II}(t^{'}) \\  \\
    h_{k}^{II}(t)h_{k}^{II*}(t^{'})+h_{k}^{I*}(t)h_{k}^{I}(t^{'}) & -h_{k}^{II}(t)h_{k}^{I*}(t^{'})+h_{k}^{I*}(t)h_{k}^{II}(t^{'})
\end{bmatrix} \quad . \label{greendiracfinal}
\ee
Restricting attention to spatially homogeneous perturbations and using Eq.~\eqref{v} gives
\be
    \delta u_{k}(t,x)=\frac{e^{ikx}}{\sqrt{2\pi}}
    \begin{bmatrix}
    \delta h_{k}^{I}(t) \\
    -\delta h_{k}^{II}(t)
    \end{bmatrix} \quad . \label{deltahomok}
\ee
%Substituting Eqs.~\eqref{deltahomok} and \eqref{greendiracfinal} in %for
%Eq.~\eqref{deltauk} and integrating %over the spatial $x^{'}$ variable and then the momentum $k$ variable yields 
%for the perturbation of the mode functions

Changing the integration variable to $k'$ in \eqref{greendiracfinal}, substituting the result along with \eqref{modeuk} and \eqref{deltahomok} into \eqref{deltauk}, and integrating first over $x'$ and then over $k'$ gives

\bea
\begin{bmatrix}
\delta h_{k}^{I}(t) \\ \\
-\delta h_{k}^{II}(t)
\end{bmatrix}
&=&
\begin{bmatrix}
\delta h_{k}^{I H}(t) \\ \\
-\delta h_{k}^{II H}(t)
\end{bmatrix}
-i\, q \int_{-\infty}^{t}dt^{'}
\begin{bmatrix}
h_{k}^{I}(t)\bigg(|h_{k}^{I}(t^{'})|^{2}-|h_{k}^{II}(t^{'})|^{2}\bigg)+2h_{k}^{II*}(t)h_{k}^{I}(t^{'})h_{k}^{II}(t^{'}) \\ \\
h_{k}^{II}(t)\bigg(|h_{k}^{II}(t^{'})|^{2}-|h_{k}^{I}(t^{'})|^{2}\bigg)+2h_{k}^{I*}(t)h_{k}^{I}(t^{'})h_{k}^{II}(t^{'})
\end{bmatrix}\delta A(t^{'}) \quad . \label{h1h2} \nonumber \\
\eea
The perturbation of the renormalized current \eqref{Jrenorm2} yields
\be
    \delta \la J_{Q} \ra_{\textnormal{ren}} = \frac{q}{2\pi}\int_{-\infty}^{\infty} dk \, \bigg[ h_{k}^{I*}(t)\delta h_{k}^{I}(t)+h_{k}^{I}(t)\delta h_{k}^{I*}(t)-h_{k}^{II*}(t)\delta h_{k}^{II}(t)-h_{k}^{II}(t)\delta h_{k}^{II*}(t) - \frac{q\, m^{2}}{\omega^{3}}\delta A(t) \bigg] \quad . \label{jdiracrenorm}
\ee
Equation~\eqref{h1h2} and its complex conjugate can be substituted into Eq.~\eqref{jdiracrenorm} to yield
\bea
    \delta \la J_{Q}\ra_{\rm ren} &=& \frac{q}{2\pi}\int_{-\infty}^{\infty} dk \, \bigg[ h_{k}^{I *}(t)\, \delta h_{k}^{I, H}(t) + h_{k}^{I}(t) \, \delta h_{k}^{I *H}(t) - h_{k}^{II *}(t)\, \delta h_{k}^{II,H}(t) - h_{k}^{II}(t)\, \delta h_{k}^{II *H}(t) -\frac{q \, m^{2}}{\omega^{3}}\delta A(t)\bigg] \nonumber \\ \nonumber \\
    && \qquad \qquad \qquad \qquad \qquad \qquad -\frac{4\,q^{2}}{\pi}\int_{-\infty}^{\infty}dk \int_{-\infty}^{t}dt^{'} \textnormal{Im}\bigg\{ h^{I}(t)h^{II}(t)h^{I *}(t^{'})h^{II *}(t^{'}) \bigg\} \delta A(t^{'}) \quad . \nonumber \\ \label{deltaJdirac}
\eea

As in the scalar field case, an explicit expression for the two-point correlation function is needed. To calculate the two-point correlation function we begin by utilizing the antisymmetrized current density \eqref{JsymDirac} with the fermion field mode expansion \eqref{psi1} evaluated in the vacuum state. Integrating over the spatial coordinate gives
\be
\int_{-\infty}^{\infty}dx^{'} \la [ J_{Q}(t,x) , J_{Q}(t',x') ] \ra = \frac{4 i \, q^{2}}{\pi}\int_{-\infty}^{\infty}dk \, \textnormal{Im}\bigg\{h_{k}^{I}(t)h_{k}^{II}(t)h_{k}^{I*}(t^{'})h_{k}^{II*}(t^{'}) \bigg\} \quad . \label{JJDiracFinal}
\ee
Comparing Eqs.~\eqref{deltaJdirac} and \eqref{JJDiracFinal}, it is clear that Eq.~\eqref{deltaJdirac} can be written in the form
\bea
\delta \la J_{Q} \ra_{\textnormal{ren}} &=&  \frac{q}{2\pi}\int_{-\infty}^{\infty} dk \, \bigg[ h_{k}^{I*}(t)\, \delta h_{k}^{I, H}(t) + h_{k}^{I}(t) \, \delta h_{k}^{I*H}(t) - h_{k}^{II*}(t)\, \delta h_{k}^{II, H}(t) - h_{k}^{II}(t)\, \delta h_{k}^{II*H}(t) \nonumber  \\ \nonumber \\ & & \qquad \qquad \qquad -\frac{q \, m^{2}}{\omega^{3}}\delta A(t)\bigg]
+ \, i\int_{-\infty}^{\infty}dx^{'}\int_{-\infty}^{t}dt^{'} \la [ J_{Q}(t,x) , J_{Q}(t',x') ] \ra \, \delta A(t^{'}) \quad . \label{deltaJfermionfinal}
\eea
Thus $\delta \la J_{Q} \ra_{\textnormal{ren}}$ for spin-$\frac{1}{2}$ particle production has been cast in terms of the current-current two-point correlation function. Note that $\delta h_{k}^{(I,II) H}(t)$ corresponds to a change of state of the quantum field. As mentioned above, for the cases considered in this paper the vector potential and its first time derivative are zero initially so the perturbations do not cause a change in the state of the field so $\delta h_{k}^{(I,II) H}(t) = 0$.
Then the linear response equation \eqref{LRE} becomes
\be
    \frac{d^{2}}{dt^{2}}\delta A(t)=-\frac{d}{dt}\delta E(t) = \delta J_{C} - \frac{q^2 m^2}{2\pi} \delta A(t) \int_{-\infty}^{\infty} \frac{dk}{\omega^{3}} + i\int_{-\infty}^{\infty}dx^{'}\int_{-\infty}^{t}dt^{'} \la [ J_{Q}(t,x) , J_{Q}(t',x') ] \ra \, \delta A(t^{'}) \quad .
\ee

\newpage

%%%%%%%%%%%%%%%%%%%%%%%%%%%%%%%%%%%%%%%%%%%%%%%%%%%%%%%%%%%%%%%%%%%%%%%%%%%%%%%%%%%%%%%%%%


\begin{thebibliography}{9}

\bibitem{schwinger}
J. Schwinger,
%\textit{On Gauge Invariance and Vacuum Polarization}.
Phys. Rev. {\bf82}, 664 (1951).

\bibitem{parker66} L. Parker, {\it The creation of particles in an expanding universe}, Ph.D. thesis, Harvard University, 1966;  Dissexpress.umi.com, Publication
No.  7331244; %{\it Phys.~Rev.~Lett.}
 Phys.~Rev.~Lett. {\bf 21}, 562 (1968); %{\it Phys.~Rev.~D}
 Phys.~Rev. {\bf 183}, 1057 (1969); %{\it Phys.~Rev.~D}
% {\it Phys.~Rev.~D}
Phys.~Rev.~D {\bf 3}, 346 (1971).

\bibitem{hawking}
S.W. Hawking,
%\textit{Particle creation by black holes}.
Commun. Math. Phys. {\bf 43}, 199 (1975).

\bibitem{parker-toms}L.~Parker and D.~J.~Toms, {\it Quantum Field Theory in Curved Spacetime: Quantized Fields
and Gravity} (Cambridge University Press, Cambridge, England, 2009).

\bibitem{birrell-davies} N.~D.~Birrell  and P.~C.~W.~Davies, {\it Quantum Fields in Curved Space} (Cambridge University Press, Cambridge, England, 1982).


\bibitem{qftbook} M. D. Schwartz, {\it Quantum Field Theory and the Standard Model} (Cambridge University Press, Cambridge, England, 2014).

\bibitem{Kluger91} Y. Kluger, J. M. Eisenberg, B. Svetitsky, F. Cooper and E. Mottola, %{\it Phys. Rev. Lett.}
Phys. Rev. Lett. {\bf 67}, 2427 (1991).
\bibitem{Kluger92}Y. Kluger, J. M. Eisenberg, B. Svetitsky, F. Cooper and E. Mottola, %{\it Phys. Rev. D}
Phys. Rev. D {\bf 45}, 4659 (1992).

\bibitem{Kluger93}Y. Kluger, J. M. Eisenberg, and B. Svetitsky, Int. J. Mod.Phys. E {\bf02}, 333 (1993).
\bibitem{Tanji} N. Tanji, Ann. Phys. (Amsterdam) {\bf 324}, 1691 (2009).


\bibitem{stat-FT} F. Gelis and N. Tanji, Phys. Rev. D {\bf 87}, 125035 (2013).

\bibitem{Bloch} J. C. R. Bloch, V. A. Mizerny, A. V. Prozorkevich, C. D. Roberts, S. M. Schmidt, S. A. Smolyansky, and D. V. Vinnik, Phys. Rev. D {\bf 60}, 116011 (1999).

\bibitem{lattice-1} F. Hebenstreit, J. Berges, and D. Gelfand, Phys. Rev. D {\bf87}, 105006 (2013).
\bibitem{lattice-2} V. Kasper, F. Hebenstreit, and J. Berges, Phys. Rev. D {\bf90}, 025016 (2014).

\bibitem{Sauter} F. Sauter, %{\it Z. Phys.}
Z. Phys. {\bf 69}, 742 (1931).


\bibitem{and-mot-I}	P. R. Anderson and E. Mottola, % {\it Phys. Rev. D}
Phys. Rev. D {\bf 89}, 104038 (2014).

\bibitem{dunne-part-prod} R. Dabrowski and G. V. Dunne, % {\it Phys. Rev. D}
Phys. Rev. D {\bf 90}, 025021 (2014).

\bibitem{and-mot-san} P. R. Anderson, E. Mottola, and D. H. Sanders, %{\it Phys. Rev. D}
Phys. Rev. D {\bf 97}, 065016 (2018).

\bibitem{Schwingermass} J. Schwinger, {\it Phys. Rev.} {\bf 128}, 2425 (1962).

\bibitem{validitypreheating}
P. R. Anderson, C. Molina-Paris, and D. H. Sanders,
%\textit{Breakdown of the semiclassical approximation during the early stages of preheating}. {\it Phys. Rev. D}
Phys. Rev. D {\bf 92}, 083522 (2015).

\bibitem{validitygravity}
P. R. Anderson, C. Molina-Paris, and E. Mottola,
%\textit{Linear response, validity of semiclassical gravity, and the stability of flat space}.
Phys. Rev. D {\bf 67}, 024026 (2003).

\bibitem{validitydeSitter}
P. R. Anderson, C. Molina-Paris, and E. Mottola,
%\textit{Cosmological horizon modes and linear response in de Sitter spacetime}
Phys. Rev. D {\bf 80}, 084005 (2009).

\bibitem{FN} A. Ferreiro and J. Navarro-Salas, %{\it Phys. Rev. D}
Phys. Rev. D {\bf 97}, 125012   (2018).

\bibitem{BNP} P. Beltr\'an-Palau, J. Navarro-Salas, and S. Pla,  %{\it Phys. Rev. D}
Phys. Rev. D {\bf 99}, 105008 (2019).

%\bibitem{large-N-1994} F. Cooper, S. Habib, Y. Kluger, E. Mottola, J. P. Paz, and P. R. Anderson,
%\textit{Nonequilibrium quantum fields in the large-N expansion}
%Phys. Rev. D {\bf 50}, 2848 (1994).

%\bibitem{hu-verdaguer} B.-L. B. Hu and E. Verdaguer, {\it Semiclassical and Stocastic Gravity}, Cambridge University Press, Cambridge, England, (2020).
%(**)

\bibitem{parker-fulling} L.~Parker and S.~A.~Fulling, % {\it Phys.~Rev.~D}
 Phys.~Rev.~D {\bf 9}, 341 (1974); S. A. Fulling and L. Parker, %{\it Ann.~Phys.}
Ann.~Phys. (N.Y.) {\bf 87}, 176 (1974).
\bibitem{Birrell78} N. D. Birrell,  %{\it Proc. R. Soc.  B}
Proc. R. Soc.  B {\bf 361}, 513 (1978).

\bibitem{Anderson-Parker}  P.~R.~Anderson and L.~Parker, % {\it Phys.~Rev.~D}
 Phys.~Rev.~D {\bf 36}, 2963 (1987).

\bibitem{rio1} A. Landete, J. Navarro-Salas, and F. Torrenti, %{\it Phys. Rev. D}
Phys. Rev. D {\bf 88}, 061501 (2013).
\bibitem{rio2} A. Landete, J. Navarro-Salas, and F. Torrenti,
Phys. Rev. D {\bf 89}, 044030 (2014).
\bibitem{rio3} A. del Rio, J. Navarro-Salas, and F. Torrenti,
Phys. Rev. D {\bf 90}, 084017 (2014).

\bibitem{ghosh1} S.~Ghosh, %{\it Phys.~Rev.~D}
Phys.~Rev.~D {\bf 91}, 124075 (2015).
\bibitem{ghosh2} S. Ghosh, %{\it Phys.~Rev.~D}
 Phys.~Rev.~D {\bf 93},   044032 (2016).


\bibitem{BFNV}J. F. Barbero G., A. Ferreiro, J. Navarro-Salas, and E. J. S. Villase\~nor, %{\it Phys. Rev. D}
Phys. Rev. D {\bf 98}, 025016 (2018).



\bibitem{CMRA} F. Cooper, E. Mottola, B. Rogers and P. Anderson,  Pair production from an external electric field, {\it in Proceedings of the Workshop on Intermittency in High-Energy Collisions}, (Santa Fe HE Coll, 1990) pp. 399-414.

 \bibitem{Cooper-Mottola89} F. Cooper and E. Mottola, %{\it Phys. Rev. D}
  Phys. Rev. D {\bf 40}, 456 (1989).



\bibitem{FNP} A. Ferreiro, J. Navarro-Salas, and S. Pla, %{\it Phys. Rev. D}
Phys. Rev. D {\bf 98}, 045015 (2018).

\bibitem{FNP2}  A. Ferreiro, J. Navarro-Salas, and S. Pla, Pair creation in electric fields, renormalization, and backreaction, arXiv:1903.11425; in  \textit{Proceedings of the 15$^{th}$ Marcel Grossmann Meeting, Rome, 2018}.



\bibitem {BNP20} P. Beltr\'an-Palau, J. Navarro-Salas, and S. Pla, % {\it Phys. Rev. D}
Phys. Rev. D {\bf 101}, 105014 (2020).

\bibitem{Heisenberg1936} W.~Heisenberg and H.~Euler, %``Consequences of Dirac's theory of positrons,''{\it  Z.\ Phys.}
Z.\ Phys. {\bf 98}, 714 (1936).




% \bibitem{kim} S. P. Kim and D.N. Page, %{\it Phys. Rev. D}
% Phys. Rev. D {\bf 65}, 105002 (2002).


\bibitem{dunne} G. Dunne,  Heisenberg-Euler effective Lagrangians: Basics and extensions. %hep-th/0406216. 
Proceedings of the I. Kogan Memorial, in {\it From Fields to Strings: Circumnavigating Theoretical Physics}, edited by M. Shifman, A. Vainshtein and J. Wheater (World Scientific, 2005).

\bibitem{BFNP} P. Beltr\'an-Palau, A. Ferreiro, J. Navarro-Salas, and S. Pla %{\it Phys. Rev. D}
 Phys. Rev. D {\bf 100}, 085014 (2019).
 
\bibitem{Dash} A. Dash,  {\it Field Theory, a Path Integral Approach,} (World Scientific, Singapore, 2019).

\bibitem{coleman-1} S. R. Coleman, R. Jackiw, and L. Susskind, Ann. Phys. (N.Y) {\bf 93}, 267 (1975).
 
\bibitem{gross} D.J. Gross,  I. R. Klebanov, A. V. Matytsin, and A. V. Smilga, {\it Nucl.Phys.} {\bf B461}, 109 (1996)
 
 \bibitem{large-N-1994} F. Cooper, S. Habib, Y. Kluger, E. Mottola, J. P. Paz, and P. R. Anderson,
%\textit{Nonequilibrium quantum fields in the large-N expansion}
Phys. Rev. D {\bf 50}, 2848 (1994).

\bibitem{hu-verdaguer} B.-L. B. Hu and E. Verdaguer, {\it Semiclassical and Stocastic Gravity}  (Cambridge University Press, Cambridge, England, 2020).
%(**)
 
 %(***)

\bibitem{fred1} B. Mihaila, T. Athan, F. Cooper, J. Dawson, and S. Habib,
Phys. Rev. D {\bf 62}, 125015 (2000).

\bibitem{fred-emil-private} F. Cooper and E. Mottola (private communication).

\bibitem{fred2} B. Mihaila, J. F. Dawson, and F. Cooper,
Phys. Rev. D {\bf 63}, 096003 (2001).

\bibitem{fred3} F. Cooper, J. F. Dawson, and B. Mihaila,
Phys. Rev. D {\bf 67}, 056003 (2003).

\bibitem{wu-ford} C.-H. Wu and L. H. Ford,
Phys. Rev. D {\bf 60}, 104013 (1999).

\bibitem{phillips-hu} N. G. Phillips and B. L. Hu,
Phys. Rev. D {\bf 62}, 084017 (2000).

\bibitem{kubostuff}
A.L. Fetter and J.D. Walecka.
\textit{Quantum Theory of Many-Particle Systems} (McGraw-Hill, New York, 1971).

\bibitem{APtheorem} T. Appelquist and J. Carazzone, {\it Phys. Rev. D} {\bf 11}, 2856 (1975).

%\bibitem{ford-tensor-pert}  J.-T. Hsiang, L. H. Ford, D.-S. Lee, and H.-L. Yu, %{\it Phys. Rev.D} Phys. Rev.D {\bf 83}, 084027 (2011).



\bibitem{r-p-1} T. Prokopec and T. G. Roos, Phys. Rev. D {\bf 55}, 3768
(1997).
\bibitem{r-p-2} S. Khlebnikov and I. I. Tkachev, Phys. Rev. Lett. {\bf 79}, 1607
(1997).

\bibitem{r-p-3}  G. Felder and I. Tkachev, Comput. Phys. Commun. {\bf 178},
929 (2008).

%(*revisar BNP*)














 %\bibitem{cghs} J. A. Harvey and A. Strominger, {\it Quantum Aspects of Black Holes}, hep-th/9209055; C. G. Callan, S. B. Giddings, J. A. Harvey, and A. Strominger, %{\it Phys.Rev.D}Phys.Rev.D {\bf 45}, 1005, (1992).

%\bibitem{rst} J. G. Russo, L. Susskind, and L. Thorlacius, %{\it Phys.Rev.D} Phys.Rev.D {\bf46}, 3444 (1992).

%\bibitem{bpp} S. Bose, L. Parker, and Y. Peleg,  %{\it Phys.Rev.D}Phys.Rev.D {\bf 52}, 3512 (1995).

%\bibitem{fabbri-navarro} A. Fabbri and J. Navarro-Salas, {\it Modeling Black Hole evaporation}, ICP-World Scientific, London (2005).



%\bibitem{BFNP} P. Beltr\'an-Palau, A. Ferreiro, J. Navarro-Salas, and S. Pla %{\it Phys. Rev. D} Phys. Rev. D {\bf 100}, 085014 (2019).

%{\color{red}
%\bibitem{Schwingermass} J. Schwinger, {\it Phys. Rev.} {\bf 128}, 2425 (1962).

%\bibitem{Dash} A. Dash,  {\it Field Theory, a Path Integral Approach,} World Scientific, Singapore (2019).

%\bibitem{coleman-1} S. R. Coleman, R. Jackiw, L. Susskind, Annals Phys. {\bf 93} 267 (1975).
 
%\bibitem{gross} D.J. Gross,  I. R. Klebanov, A. V. Matytsin, A. V. Smilga, {\it Nucl.Phys.} {\bf B 461}, 109 (1996)

%{\color{red} References suggested by the referee:

%\bibitem{Tanji} N. Tanji, Ann. Phys. {\bf 324}, 1691 (2009).



%\bibitem{Bloch} J. C. R. Bloch, V. A. Mizerny, A. V. Prozorkevich, C. D. Roberts, S. M. Schmidt, S. A. Smolyansky, and D. V. Vinnik, Phys. Rev. D {\bf 60}, 116011 (1999).

%\bibitem{stat-FT} F. Gelis and N. Tanji, Phys. Rev. D {\bf 87}, 125035 (2013).

%\bibitem{lattice-1} F. Hebenstreit, J. Berges, and D. Gelfand, Phys. Rev. D {\bf87}, 105006 (2013).
%\bibitem{lattice-2} V. Kasper, F. Hebenstreit, and J. Berges, Phys. Rev. D {\bf90}, 025016 (2014).


%\bibitem{lattice-3} F. Hebenstreit, D. Banerjee, M. Hornung, F.-J. Jiang, F. Schranz, and U.-J. Wiese, Phys. Rev. B {\bf92}, 035116 (2015).

%}

%\bibitem{PSbook} M. Peskin and D. Schroeder, {\it An Introduction to Quantum Field Theory}, Addison-Wesley, Reading MA, (1995).

%\bibitem{kubostuff}A.L. Fetter and J.D. Walecka.\textit{Quantum theory of many-particle systems}.McGraw-Hill (1971).






\end{thebibliography}
\end{document}